\documentclass[cernpreprint,atlasdraft=true,texlive=2016,UKenglish,texmf]{atlasdoc}
 
\usepackage{atlaspackage}
\usepackage{atlasbiblatex}
 
\usepackage{atlasphysics}
 
\graphicspath{{logos/}{figures/}}
 
\usepackage{ANA-STDM-2017-27-PAPER-defs}
\usepackage{rotating}

\AtlasTitle{
Differential cross-section measurements for the electroweak production of dijets in association with a $Z$ boson in proton--proton collisions at ATLAS}
 
\PreprintIdNumber{CERN-EP-2020-045}
\AtlasAbstract{
Differential cross-section measurements are presented for the electroweak production of two jets in association with a $Z$ boson. These measurements are sensitive to the vector-boson fusion production mechanism and provide a fundamental test of the gauge structure of the Standard Model. The analysis is performed using proton--proton collision data collected by ATLAS at $\sqrt{s}=13$~\TeV{} and with an integrated luminosity of 139~fb$^{-1}$. The differential cross-sections are measured in the $Z\rightarrow \ell^+\ell^-$ decay channel ($\ell=e,\mu$) as a function of four observables: the dijet invariant mass, the rapidity interval spanned by the two jets, the signed azimuthal angle between the two jets, and the transverse momentum of the dilepton pair. The data are corrected for the effects of detector inefficiency and resolution and are sufficiently precise to distinguish between different state-of-the-art theoretical predictions calculated using \textsc{Powheg+Pythia8}, \textsc{Herwig7+Vbfnlo} and \textsc{Sherpa}~2.2. The differential cross-sections are used to search for anomalous weak-boson self-interactions using a dimension-six effective field theory. The measurement of the signed azimuthal angle between the two jets is found to be particularly sensitive to the interference between the Standard Model and  dimension-six scattering amplitudes and provides a direct test of charge-conjugation and parity invariance in the weak-boson self-interactions.
}
 
\author{The ATLAS Collaboration}
 
\AtlasRefCode{STDM-2017-27}

\AtlasJournal{Eur.Phys.J. C}
\AtlasJournalRef{Eur. Phys. J. C 81 (2021) 163}
\AtlasDOI{10.1140/epjc/s10052-020-08734-w}

\hypersetup{pdftitle={ATLAS document},pdfauthor={The ATLAS Collaboration}}
 
\begin{document}
 
\maketitle
 
\tableofcontents

\section{Introduction}
\label{sec:intro}
\newcommand{\AtlasCoordFootnote}{
ATLAS uses a right-handed coordinate system with its origin at the nominal interaction point (IP)
in the centre of the detector and the \(z\)-axis along the beam pipe.
The \(x\)-axis points from the IP to the centre of the LHC ring,
and the \(y\)-axis points upwards.
Cylindrical coordinates \((r,\phi)\) are used in the transverse plane,
\(\phi\) being the azimuthal angle around the \(z\)-axis.
The pseudorapidity is defined in terms of the polar angle \(\theta\) as \(\eta \equiv -\ln \tan(\theta/2)\), and is equal to the rapidity $y \equiv 0.5\ln{\left((E+p_{z})/(E-p_{z})\right)}$
in the relativistic limit. Angular distance is measured in units of \(\Delta R \equiv \sqrt{(\Delta y)^{2} + (\Delta\phi)^{2}}\).}
 
Measurements that exploit the weak vector-boson scattering (VBS) and weak vector-boson fusion (VBF) processes have become increasingly prevalent at the Large Hadron Collider (LHC) in the last few years.
In the Higgs sector, measurements of Higgs boson production via VBF have been used to determine the strength, charge-conjugation (C) and parity (P) properties of the Higgs boson's interactions with weak bosons~\cite{Khachatryan:2016vau,Aaboud:2018xdt,Aaboud:2017vzb, Sirunyan:2018koj,Aad:2016nal,Sirunyan:2019nbs,Sirunyan:2019twz}.
These measurements have recently been augmented by the observation of the electroweak production of two jets in association with a weak-boson pair \cite{Sirunyan:2017ret,Aaboud:2019nmv,Aaboud:2018ddq,Aad:2020zbq, Sirunyan:2020gyx}, which
is extremely sensitive to the VBS production mechanism and provides a stringent test of the gauge structure of the Standard Model of particle physics (SM).
In the search for physics beyond the SM, the VBF and VBS production mechanisms have been used to search for dark matter~\cite{Aaboud:2018sfi,Sirunyan:2018owy}, heavy-vector triplets~\cite{Aaboud:2018bun}, Higgs-boson pair production~\cite{Aad:2020kub}, and signatures of warped extra dimensions~\cite{Aad:2020ddw}.
 
All of these measurements and searches rely on theoretical predictions to accurately model the electroweak processes that are sensitive to the VBF and VBS production mechanisms.
Specifically, Monte Carlo (MC) event generators are used to optimise the event selection and to extract the electroweak signal from the dominant background, with the signal extraction typically performed using fits to kinematic spectra.
However, it is known that the theoretical predictions from different event generators do not agree, both in the overall production rate~\cite{Aaboud:2019nmv} as well as in the kinematic properties of the final state~\cite{ATL-PHYS-PUB-2019-004}.
Model-independent measurements that directly probe the kinematic properties of VBF and VBS are therefore crucial, to determine which event generators can be used reliably in  physics analysis at the LHC experiments.
 
\begin{figure}[t]
\begin{center}
\includegraphics[width=0.7\textwidth]{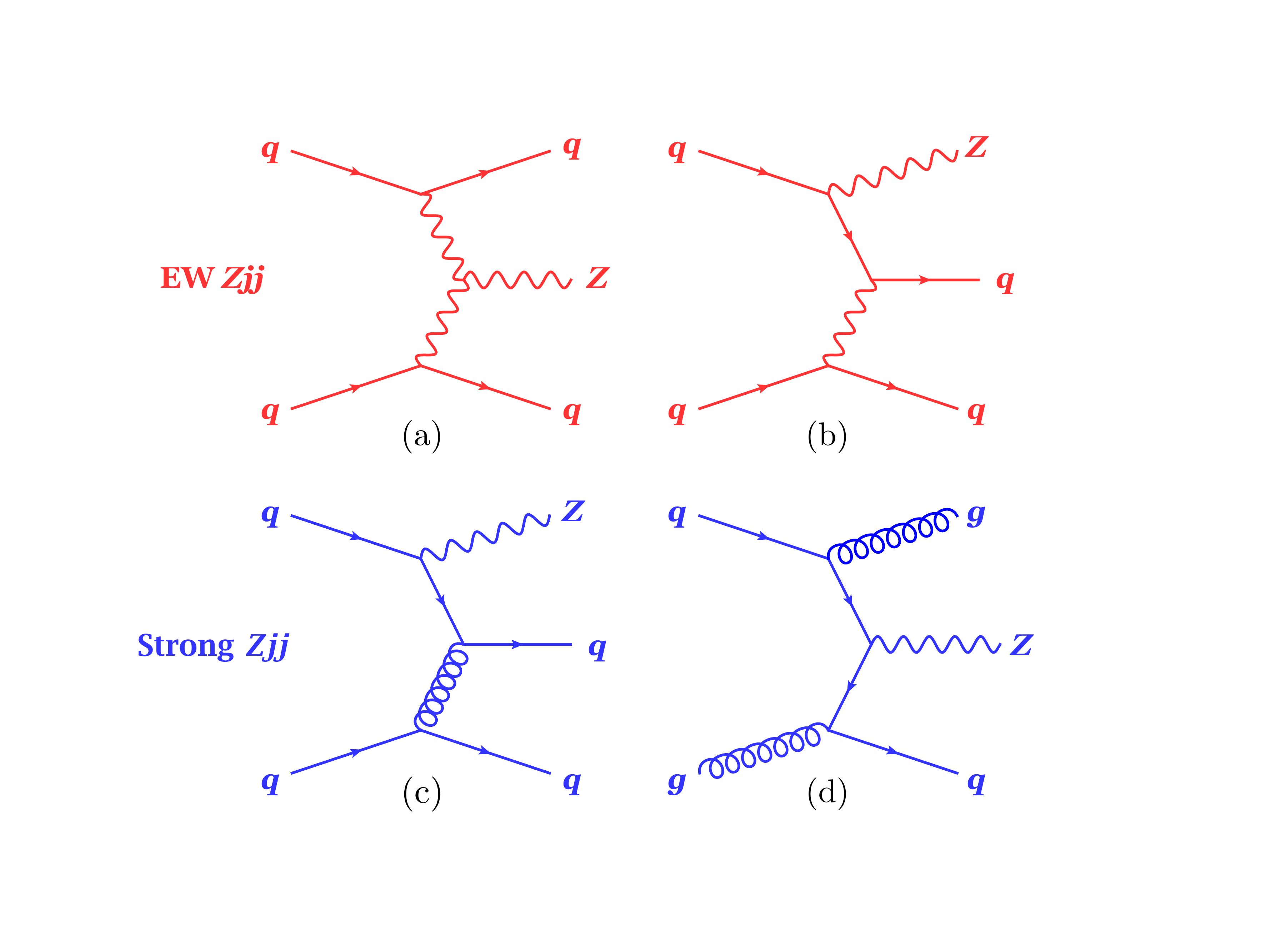}
\end{center}
\caption{Representative Feynman diagrams for EW $Zjj$ production (a,b) and strong $Zjj$ production (c,d). The electroweak $Zjj$ process is defined by the $t$-channel exchange of a weak boson and at tree level is calculated at $O(\alpha_\text{EW}^4)$ when including the decay of the $Z$ boson. The strong $Zjj$ process has no weak boson exchanged in the $t$-channel and at tree level is calculated at $O(\alpha_\text{EW}^2\alphas{}^2)$ when including the decay of the $Z$ boson.}
\label{fig:Zjj}
\end{figure}

This article presents differential cross-section measurements for the electroweak production of dijets in association with a $Z$ boson (referred to as EW $Zjj$ production).  The EW $Zjj$ process is defined by the $t$-channel exchange of a weak vector boson, as shown in Figures~\ref{fig:Zjj}(a) and \ref{fig:Zjj}(b), and is very sensitive to the VBF production mechanism. Previous measurements of EW $Zjj$ production by ATLAS~\cite{Aad:2014dta,Aaboud:2017emo} and CMS~\cite{Chatrchyan:2013jya,Khachatryan:2014dea,Sirunyan:2017jej} have focused on measuring only an integrated fiducial cross-section in a VBF-enhanced topology. The analysis presented in this article measures differential cross-sections of EW $Zjj$ production in the $Z\rightarrow \ell^+\ell^-$ decay channel ($\ell=e,\mu$) and as a function of four observables; the transverse momentum of the dilepton pair ($p_{{\mathrm T},\ell\ell}$), the dijet invariant mass ($m_{jj}$), the absolute rapidity\footnote{\AtlasCoordFootnote} separation of the two jets (\Dyjj), and the signed azimuthal angle between the two jets ($\Delta\phi_{jj}$). The $\Delta\phi_{jj}$ variable is
defined as $\Delta\phi_{jj} = \phi_f -\phi_b$, where the two highest transverse-momentum jets are ordered such that $y_f>y_b$~\cite{Hankele:2006ma}. Collectively, these four observables probe the important kinematic properties of the VBF and VBS production mechanisms. The measurements are performed using proton--proton collision data collected by the ATLAS experiment at a centre-of-mass energy of $\sqrt{s}=13$~\TeV{} and with an integrated luminosity of 139~fb$^{-1}$.
 
The EW $Zjj$ differential cross-section measurements presented here are sufficiently precise that they can be used to probe a diverse range of physical phenomena. First, under the assumption of no beyond-the-SM physics contributions to the EW $Zjj$ process, the measurements can be used to distinguish between the SM EW $Zjj$ predictions produced by different event generators or by different parameter choices within each event generators. In the short term, the measurements will therefore help determine which event generator predictions can be used reliably in analyses that seek to exploit VBF and VBS at the LHC. In the longer term, the measurements will provide crucial input if the theoretical predictions are to be improved. Second, and more generally, the measurements provide a new avenue to search for signatures of physics beyond the SM. The differential cross-section as a function of  $\Delta\phi_{jj}$, for example, is found to be particularly sensitive to anomalous weak-boson self-interactions that arise from CP-even and CP-odd operators in a dimension-six effective field theory. This parity-odd observable has been proposed as a method to search for CP-violating effects in Higgs boson production~\cite{Hankele:2006ma}, but has not yet been measured in a final state sensitive to anomalous weak-boson self-interactions.
 
The layout of the article is as follows. The ATLAS detector is briefly described in Section~\ref{sec:detector}. The signal and background simulations used in the analysis are described in Section~\ref{sec:MC}. The event reconstruction and selection are described in Section~\ref{sec:evsel}.  The method used to extract the electroweak component is described in Section~\ref{sec:ewextraction}. This includes a data-driven constraint on the dominant background process in which the jets that are produced in association with the $Z$ boson arise from the strong interaction (strong $Zjj$ production) as shown in Figures~\ref{fig:Zjj}(c) and \ref{fig:Zjj}(d). The corrections applied to remove the impact of detector resolution and inefficiency are described in Section~\ref{sec:unfolding}. The experimental and theoretical systematic uncertainties are presented in Section~\ref{sec:systematics}. Finally, the differential cross-sections for EW $Zjj$ production are presented in Section~\ref{sec:results}. Differential cross-sections for inclusive $Zjj$ production are also presented in Section~\ref{sec:results} for the signal and control regions used to extract the electroweak component. The EW $Zjj$ differential cross-sections are used in Section~\ref{sec:eft} to search for anomalous weak-boson self-interactions. A brief summary of the analysis is given in Section~\ref{sec:conclusion}.
 

\section{ATLAS detector}
\label{sec:detector}
The ATLAS detector~\cite{PERF-2007-01} at the LHC covers nearly the entire solid angle around the collision point.
It consists of an inner tracking detector surrounded by a thin superconducting solenoid, electromagnetic and hadronic calorimeters,
and a muon spectrometer incorporating three large superconducting toroidal magnets.
 
The inner-detector system is immersed in a \SI{2}{\tesla} axial magnetic field
and provides charged-particle tracking in the range \(|\eta| < 2.5\). The high-granularity silicon pixel detector covers the vertex region and typically provides four measurements per track,
the first hit normally being in the insertable B-layer (IBL) installed before the start of  Run~2~\cite{ATLAS-TDR-2010-19,PIX-2018-001}.
The IBL is followed by the silicon microstrip tracker which usually provides eight measurements per track.
These silicon detectors are complemented by the transition radiation tracker (TRT),
which enables radially extended track reconstruction up to \(|\eta| = 2.0\).
The TRT also provides electron identification information
based on the fraction of hits (typically 30 in total) above a higher energy-deposit threshold corresponding to transition radiation.
 
The calorimeter system covers the pseudorapidity range \(|\eta| < 4.9\).
Within the region \(|\eta|< 3.2\), electromagnetic calorimetry is provided by barrel and
endcap high-granularity lead/liquid-argon (LAr) calorimeters,
with an additional thin LAr presampler covering \(|\eta| < 1.8\),
to correct for energy loss in material upstream of the calorimeters.
Hadronic calorimetry is provided by the steel/scintillator-tile calorimeter,
segmented into three barrel structures within \(|\eta| < 1.7\), and two copper/LAr hadronic endcap calorimeters.
The solid angle coverage is completed with forward copper/LAr and tungsten/LAr calorimeter modules
optimised for electromagnetic and hadronic measurements respectively.
 
The muon spectrometer comprises separate trigger and
high-precision tracking chambers measuring the deflection of muons in a magnetic field generated by the superconducting air-core toroids.
The field integral of the toroids ranges between \num{2.0} and \SI{6.0}{\tesla\metre}
across most of the detector.
A set of precision chambers covers the region \(|\eta| < 2.7\) with three layers of monitored drift tubes,
complemented by cathode-strip chambers in the forward region, where the background is highest.
The muon trigger system covers the range \(|\eta| < 2.4\) with resistive-plate chambers in the barrel, and thin-gap chambers in the endcap regions.
 
Interesting events are selected for further analysis by the level-one (L1) trigger system, which is implemented in custom hardware. The selections are further refined by algorithms implemented in software in the high-level trigger (HLT)~\cite{TRIG-2016-01}. The L1 trigger selects events from the \SI{40}{\MHz} bunch crossings at a rate below \SI{100}{\kHz}. The HLT further reduces the rate in order to write events to disk at about \SI{1}{\kHz}.
 
\section{Dataset and Monte Carlo event simulation}
\label{sec:MC}
The analysis is performed on proton--proton collision data at a centre-of-mass energy of $\sqrt{s}=13$~\TeV{}. The data were recorded between 2015 and 2018 and correspond to an integrated luminosity of 139 fb$^{-1}$.
 
Monte Carlo event generators are used to simulate the signal and background events produced in the proton--proton collisions. These samples are used to optimise the analysis, evaluate systematic uncertainties, and correct the data for detector inefficiency and resolution. A summary of the event generators is presented in Table~\ref{tab:samples} and further details of each generator are given below.

Electroweak $Zjj$ production was simulated using three MC event generators. The default EW $Zjj$ sample was produced with  \POWHEGBOXV{v1}~\cite{powheg1,powheg2,powheg3} using the CT10nlo~\cite{Alwall:2014hca} parton distribution functions (PDF) and is accurate to next-to-leading order (NLO) in perturbative QCD. The sample was produced with the `VBF approximation', which requires a $t$-channel colour-singlet exchange to remove overlap with diboson topologies~\cite{Jager:2012xk}. The parton-level events were passed to \PYTHIAV{8.186} to add parton-showering, hadronisation and underlying-event activity, using the AZNLO~\cite{Aad:2014xaa} set of tuned parameters. The \evtgen{}  program~\cite{LANGE2001152} was used for the properties of the bottom and charm hadron decays. This sample is referred to as \powpy{} EW $Zjj$ production.

The second EW $Zjj$ sample was produced in the VBF approximation with \herwig{}.1.5~\cite{Bahr:2008pv,Bellm:2015jjp}. The samples were produced at NLO accuracy in the strong coupling using \vbfnlo{} v3.0.0~\cite{Arnold:2011wj} as the loop-amplitude provider. The MMHT2014LO PDF set~\cite{Harland-Lang:2014zoa} was used along with the default set of tuned parameters for parton showering, hadronisation and underlying event. EvtGen was used for the properties of the bottom and charm hadron decays.
This sample is referred to as \hervbf{} EW $Zjj$ production.
 
The third EW $Zjj$ sample was produced in the VBF approximation with the \sherpa{}~2.2.1 event generator~\cite{Bothmann:2019yzt}. The samples were produced using leading-order (LO) matrix elements with up to two additional parton emissions. The NNPDF3.0nnlo PDFs~\cite{Ball:2014uwa} were used and the matrix elements were merged with the \sherpa{} parton shower using the MEPS@LO prescription~\cite{Catani:2001cc}. Hadronisation and underlying-event algorithms were used to construct the fully hadronic final state using the set of tuned parameters developed by the \sherpa{} authors.
This sample is referred to as \sherpa{} EW $Zjj$ production.
 
\begin{table}[tb]
\centering
\resizebox{\textwidth}{!}{
\def\arraystretch{1.5}
\newcolumntype{L}[1]{>{\raggedright\let\newline\\\arraybackslash}p{#1}}
\begin{tabular}{L{0.12\textwidth}L{0.27\textwidth}L{0.23\textwidth}L{0.18\textwidth}L{0.20\textwidth}L{0.12\textwidth}}
\toprule
Process & Generator & ME accuracy & PDF & Shower and \newline hadronisation & Parameter set  \\
\midrule
EW $Zjj$ & \POWHEGBOXV{v1} & NLO & CT10nlo & \pythia{} + \evtgen & AZNLO \\
& \herwig{} $+$ \vbfnlo & NLO & MMHT2014lo  & \herwig{} + \evtgen & default \\
& \sherpa~2.2.1 & LO (2--4j) & NNPDF3.0nnlo & \sherpa & default \\
\hline
Strong $Zjj$ & \sherpa~2.2.1 & NLO (0--2j), LO (3--4j) & NNPDF3.0nnlo & \sherpa{} & default \\
& \fxfx & NLO (0--2j), LO (3--4j) & NNPDF2.3nlo & \pythia{} + \evtgen & A14 \\
& \mg & LO (0--4j) & NNPDF3.0lo & \pythia{} + \evtgen & A14 \\
\hline
$VV$ & \sherpa & NLO (0--1j), LO (2--3j) & NNPDF3.0nnlo & \sherpa{} & default \\
$t\bar{t}$ & \POWHEGBOXV{v2} hvq & NLO & NNPDF3.0nnlo & \pythia{} + \evtgen & A14 \\
$VVV$ & \sherpa & LO (0--1j) & NNPDF3.0nnlo & \sherpa & default \\
$W$+jets & \sherpa & NLO (0--2j), LO (3--4j) & NNPDF3.0nnlo & \sherpa{} & default \\
\bottomrule
\end{tabular}
}
\caption{Summary of generators used for simulation. The details and the corresponding references are provided in the body of the text. In the final column, `default' refers to the default set of tuned parameters provided with the event generator. \label{tab:samples}}
\end{table}
 
The dominant background arises from $Zjj$ final states in which the two jets are produced from the strong interaction, as shown in Figures~\ref{fig:Zjj}(c) and \ref{fig:Zjj}(d). This is referred to as the strong $Zjj$ background and was simulated using three different MC event generators. \sherpa{}~2.2.1 was used to produce $Z$+$n$-parton predictions ($n = 0,1,2,3,4$), at NLO accuracy for up to two partons in the final state and at LO accuracy for three or four partons in the final state, using the Comix~\cite{Gleisberg:2008fv} and OpenLoops~\cite{Denner:2016kdg,Cascioli:2011va} libraries. The different final-state topologies were merged into an inclusive sample using an improved CKKW matching procedure~\cite{Catani:2001cc,Hoeche:2009rj}, which has been extended to NLO accuracy using the MEPS@NLO prescription~\cite{Hoeche:2012yf}. The \sherpa{} prediction was produced using the NNPDF3.0nnlo PDFs and normalised to a next-to-next-to-leading-order (NNLO) prediction for inclusive $Z$-boson production~\cite{Anastasiou:2003ds}. The default set of tuned parameters in \sherpa{} was used for hadronisation and underlying-event activity. This sample is referred to as \sherpa{} strong $Zjj$ production.

The second strong $Zjj$ sample was produced using the \fxfx{} event generator \cite{Alwall:2014hca} and is accurate to NLO in the strong coupling for up to two partons in the final state. The NNPDF2.3nlo PDF set~\cite{Ball:2012cx} was used in the calculation. The \fxfx{} generator was interfaced to \PYTHIAV{8.186} to provide parton showering, hadronisation and underlying-event activity, using the A14 set of tuned parameters. To remove overlap between the matrix element and the parton shower, the different jet multiplicities were merged using the FxFx prescription~\cite{Frederix:2012ps}. \evtgen{} was used for the properties of the bottom and charm hadron decays. The sample is normalised to the same NNLO prediction as for the \sherpa{} sample and is referred to as \mgnlo{} strong $Zjj$ production.
 
The third strong $Zjj$ sample was also produced with \fxfx{}, but with the  $Z$+$n$-parton matrix-elements produced at LO accuracy for up to four partons in the final state. The NNPDF3.0lo PDFs were used in the calculation. The parton-level events were passed to \PYTHIAV{8.186} to provide parton-showering, hadronisation and underlying-event activity, using the A14 set of tuned parameters~\cite{ATL-PHYS-PUB-2014-021}. To remove overlap between the matrix element and the parton shower, the CKKW-L merging procedure~\cite{Lonnblad:2001iq,Lonnblad:2011xx} was applied.  \evtgen{} was used for properties of the bottom and charm hadron decays. The sample is normalised to the same NNLO prediction as for the \sherpa{} sample and is referred to as \mglo{} strong $Zjj$ production.
 
Production of diboson ($VV$) final states were simulated using \sherpa{} at NLO accuracy for up to one parton in the final state, and at LO accuracy for two or three partons in the final state. The NNPDF3.0nnlo PDF set was used in the calculation. The virtual corrections were taken from OpenLoops and the different topologies were merged using the MEPS@NLO algorithm. The default set of tuned parameters in \sherpa{} was used for hadronisation and underlying-event activity.
 
Backgrounds from events containing a single top quark or a top--antitop ($t\bar{t}$) pair were estimated at NLO accuracy, using the hvq program \cite{Frixione:2007nw} in \POWHEGBOXV{v2}. The parton-level events were passed to \PYTHIAV{8.230} to provide the parton showering, hadronisation and underlying-event activity using the A14 set of tuned parameters. \evtgen{} was used for the properties of the bottom and charm hadron decays. The NNPDF3.0nnlo PDF set was used and the $h_{\mathrm{damp}}$ parameter in the \powbox{} was set to $1.5\,m_{\mathrm{top}}$. The background from the $W$+jets final state was estimated using \sherpa{}, with the same set-up as for the $Z$+jets final state. The small contribution from triboson events ($VVV$ production) was estimated using \sherpa{} at LO accuracy for up to one parton in the final state. The MEPS@LO prescription was used to merge the samples. The samples were produced using the NNPDF3.0nnlo PDF and the \sherpa{} authors' default parameterisation was used for hadronisation and underlying-event activity.

The signal and background events were passed through the \GEANT{}4~\cite{Agostinelli2003250} simulation of the ATLAS detector~\cite{Aad:2010ah} and reconstructed using the same algorithms as used for the data (except for the \hervbf{} and \mgnlo{} samples, which were produced only at particle level). Differences in lepton trigger, reconstruction and isolation efficiencies between simulation and data are corrected on an event-by-event basis using $\pt$- and $\eta$- dependent scale factors for each lepton~\cite{Aad:2016jkr,Aaboud:2019ynx}. The effect of multiple proton--proton interactions (pile-up) in the same or nearby bunch crossings is accounted for using inelastic proton--proton interactions generated by \pythia~\cite{Sjostrand:2007gs}, with the A3 tune~\cite{ATL-PHYS-PUB-2016-017} and the NNPDF2.3LO PDF  set~\cite{Ball:2012cx}. These inelastic proton--proton interactions were added to the signal and background samples and weighted such that the distribution of the average number of proton--proton interactions in simulation matches that observed in the data.
 
An approximate detector-level prediction for \mgnlo{} is obtained by reweighting the strong $Zjj$ simulation produced by \mglo{} such that the kinematic distributions match \mgnlo{} at particle level.
This is referred to as \mgnloR{}. Similarly, an approximate  detector-level prediction for \hervbf{} is obtained by reweighting the EW $Zjj$ simulation produced by \powpy{} to match \hervbf{} at particle level. This is referred to as \hervbfR{}.
 
\section{Event reconstruction and selection}
\label{sec:evsel}
Events are required to pass unprescaled dilepton triggers with transverse momentum thresholds that depend on the lepton flavour and running periods. In 2015, the dielectron triggers retained events with two electron candidates that had $\pt>12 $~\GeV{}, whereas the dimuon triggers selected events with leading (subleading) muon candidates having $\pt>18\,(8) $~\GeV{}. The transverse momentum thresholds for the lepton candidates were gradually increased during data taking, such that both electron candidates had $\pt>24 $~\GeV{} in 2018, whereas the leading muon threshold was increased to 22~\GeV{} in the same running period.
 
Events are used in the analysis if they were recorded during stable beam conditions and if they satisfy detector and data-quality requirements~\cite{Aad:2019hmz}. The positions of the proton--proton interactions are reconstructed using tracking information from the inner detector, with each associated vertex required to have at least two  tracks with $\pt>0.5$~\GeV{}. The primary hard-scatter vertex is defined as the one with the largest value of the sum of squared track transverse momenta.
 
Muons are identified by matching tracks reconstructed in the muon spectrometer to tracks reconstructed in the inner detector. Each muon is then required to satisfy the `medium’ identification criteria and the `Gradient’ isolation working point~\cite{Aad:2016jkr}. Muons are required to be associated with the primary hard-scatter vertex by satisfying $|d_0/\sigma_{d_0}| < 3$ and $|z_0 \times \mathrm{sin}\theta| < 0.5$~mm,
where $d_0$ is the transverse impact parameter calculated with respect to the measured beam-line position, $\sigma_{d_0}$ is its uncertainty, and $z_0$ is the longitudinal difference between the point at which $d_0$ is measured and the primary vertex.
Reconstructed muons are used in the analysis if they have $\pt>25$~\GeV{} and $|\eta|<2.4$.
 
Electrons are reconstructed from topological clusters of energy deposited in the electromagnetic calorimeter that are matched to a reconstructed track~\cite{Aaboud:2019ynx}. They are calibrated using $Z\to ee$ data~\cite{Aaboud:2018ugz}.
Each electron is required to satisfy the `medium' likelihood identification criteria~\cite{Aaboud:2019ynx}, as well as the same isolation working point as for muons. Electrons are required to be associated with the primary hard-scatter vertex by satisfying $|d_0/\sigma_{d_0}| < 5$ and $|z_0 \times \mathrm{sin}\theta| < 0.5$~mm. Reconstructed electrons are used in the analysis if they have $\pt>\SI{25}{\GeV}$ and $|\eta|<2.47$, but excluding the transition region between the barrel and end-cap calorimeters ($1.37<|\eta|<1.52$).
 
Jets are reconstructed with the anti-$k_t$ algorithm~\cite{Cacciari:2008gp,Cacciari:2011ma} using a radius parameter of $R=0.4$. The inputs to the algorithm are clusters of energy deposited in the electromagnetic and hadronic calorimeters. The jets are initially calibrated by applying energy- and pseudorapidity- dependent correction factors derived from simulation in the `EM+JES’ scheme~\cite{Aaboud:2017jcu}, and then further calibrated using data-driven correction factors derived from the transverse momentum balance of jets in $\gamma$+jet, $Z$+jet and multijet topologies. Jets are used in the analysis if they have $\pt >\SI{25}{\GeV}$ and $|y|<4.4$. As all high-$\pt$ electrons pass the above requirements, jets are required to not overlap with a reconstructed electron (i.e. $\Delta R(j,e)>0.2$). Jets with $\pt<\SI{120}{\GeV}$ and $|\eta|<2.4$ are also required to be consistent with originating from the primary hard-scatter vertex using the `medium' working point of the jet vertex tagger ($\mathrm{JVT}>0.59$)~\cite{Aad:2015ina}.
 
Following jet reconstruction, an additional quality requirement is placed on the events, by removing events containing jets that originate from noise bursts in the calorimeter. This removes 0.4\% of the events in data.
 
Events are then selected if they have a topology consistent with EW $Zjj$ production. A $Z$-boson candidate is reconstructed by requiring that each event contains exactly two charged leptons ($\ell=e,\mu$) that are opposite in charge and of the same flavour. These leptons are required to be well separated from jets by imposing $\Delta R(\ell,j)>0.4$. The invariant mass and transverse momentum of the dilepton system is required to fulfil $m_{\ell\ell}\in(81,101)$~\GeV{} and $\pTll>20$~\GeV. Events are required to contain two or more jets, with the leading and subleading jets satisfying $\pt > \SI{85}{\GeV}$ and $\pt > \SI{80}{\GeV}$, respectively. The dijet system is then constructed from the two leading jets and is required to fulfil $\mjj>\SI{1}{\TeV}$ and $\Dyjj>2.0$. The $Z$ boson is required to be centrally produced relative to the dijet system by imposing $\xi_Z<1.0$; the quantity $\xi_Z$ is defined as $\xi_Z = |y_{\ell\ell} - 0.5(y_{j1}+ y_{j2})|\,/\,\Dyjj$, where $y_{\ell\ell}$, $y_{j1}$ and $y_{j2}$ are the rapidities of the dilepton system, the leading jet, and the subleading jet, respectively.   Finally, to reduce the impact of jets that originate from pile-up interactions and that survive the JVT selection criteria, the $Z$-boson candidate and the dijet system are required to be approximately balanced in transverse momentum, by requiring that $p_{\mathrm{T}}^{\mathrm{bal}}<0.15$, where $p_{\mathrm{T}}^{\mathrm{bal}}=|\Sigma_i \vec{p}_{\mathrm{T},i}|\,/\,\Sigma_i {p}_{\mathrm{T},i}$ and the summation includes the dilepton system, the dijet system, and the highest transverse-momentum additional jet reconstructed in the rapidity interval spanned by the dijet system.
 
\begin{table} [tp]
\centering
\begin{tabular}{l|r@{$~\pm~$}r@{$~\pm~$}r|r@{$~\pm~$}r@{$~\pm~$}r}\hline
Sample & \multicolumn{3}{c|}{$Z \to ee$} & \multicolumn{3}{c}{$Z\to\mu\mu$}\\ \hline\hline
Data                  & \multicolumn{3}{c}{10\,870} & \multicolumn{3}{c}{12\,125} \\ \hline
EW $Zjj$ (\powpy{})       &  2670 &  120 &  280 &  2740 & 120 &  290 \\
EW $Zjj$ (\sherpa{})       &  1280 &   60 &  140 &  1350 &  60 &  150 \\
EW $Zjj$ (\hervbfR)         &  2290 &  100 &  210 &  2350 & 100 &  220 \\ \hline
Strong $Zjj$ (\sherpa{})   & 13\,500 &  600 & 4500 & 15\,100 & 600 & 5000 \\
Strong $Zjj$ (\mglo{})       & 13\,140 &  480 &  N/A & 14\,810 & 540 &  N/A \\
Strong $Zjj$ (\mgnloR{})     &  8800 &  300 & 1000 & 10\,000 & 400 & 1200 \\ \hline
$ZV$ $(V\to jj)$               &    179 &    8 &    6 &    178 &   8 &  6\\ \hline
Other $VV$           &     45 &    2 &    2 &   45 &  2 &  2 \\ \hline
$t\bar{t}$, single top     &     92 &    8 &    6 &   98 &  8 & 6\\ \hline
$W(\to\ell\nu)$+jets, \,$Z(\to\tau\tau)$+jets & \multicolumn{3}{c|}{negligible} & \multicolumn{3}{c}{negligible} \\ \hline\hline
\end{tabular}
\caption{Observed and expected event yields in the dielectron and dimuon decay channels following the event selection described in Section~\ref{sec:evsel}. The first (second) uncertainty quoted for each generator is the experimental (theoretical) systematic uncertainty. The experimental systematic uncertainties are shown for each prediction. Theoretical uncertainties are calculated for all predictions except for
\mglo{} strong $Zjj$, which is denoted `N/A' in the table. The statistical uncertainty on each prediction is negligible.}
\label{tab:eventYields}
\end{table}

\begin{figure}[tp]
\includegraphics[width=0.48\textwidth]{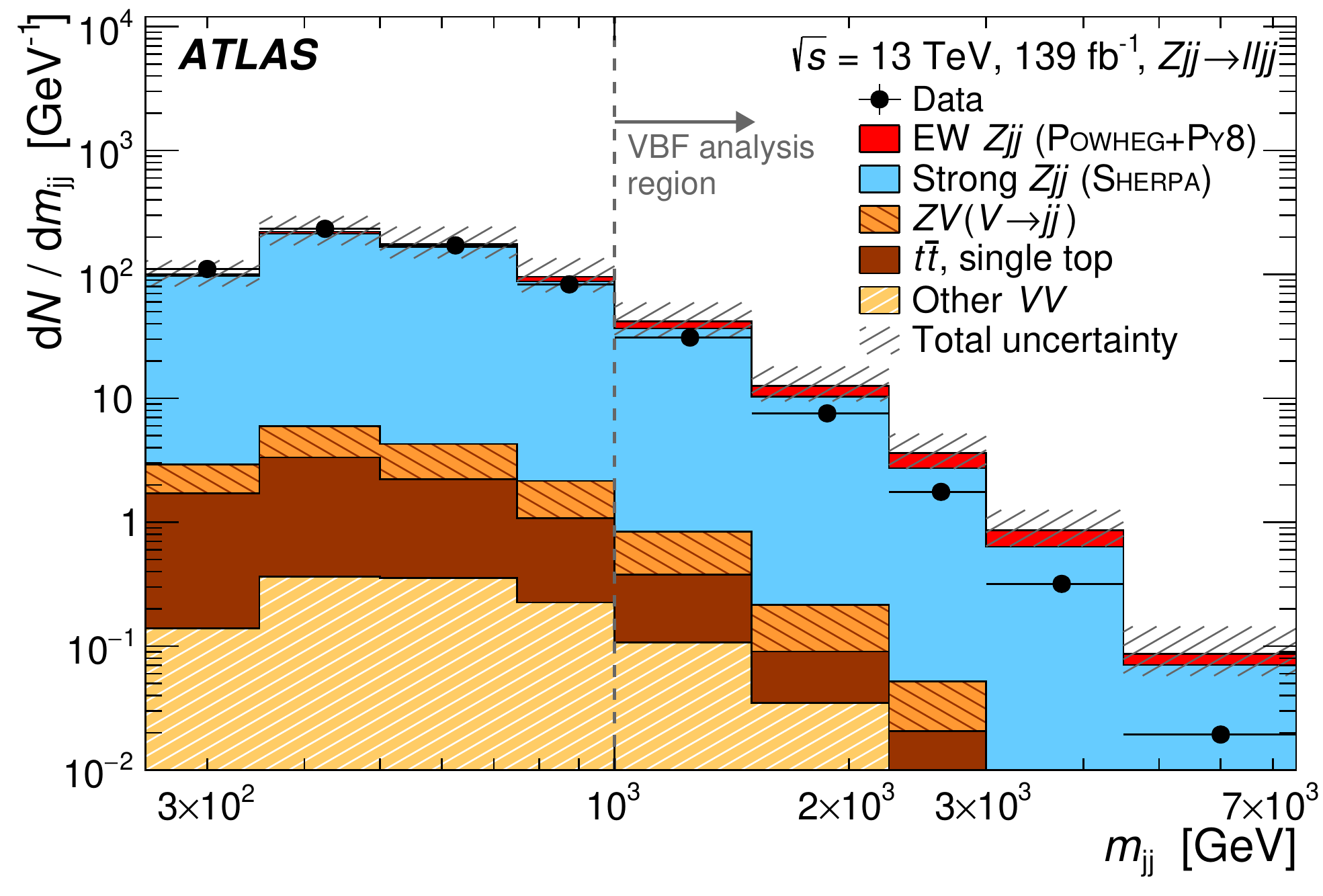}
\includegraphics[width=0.48\textwidth]{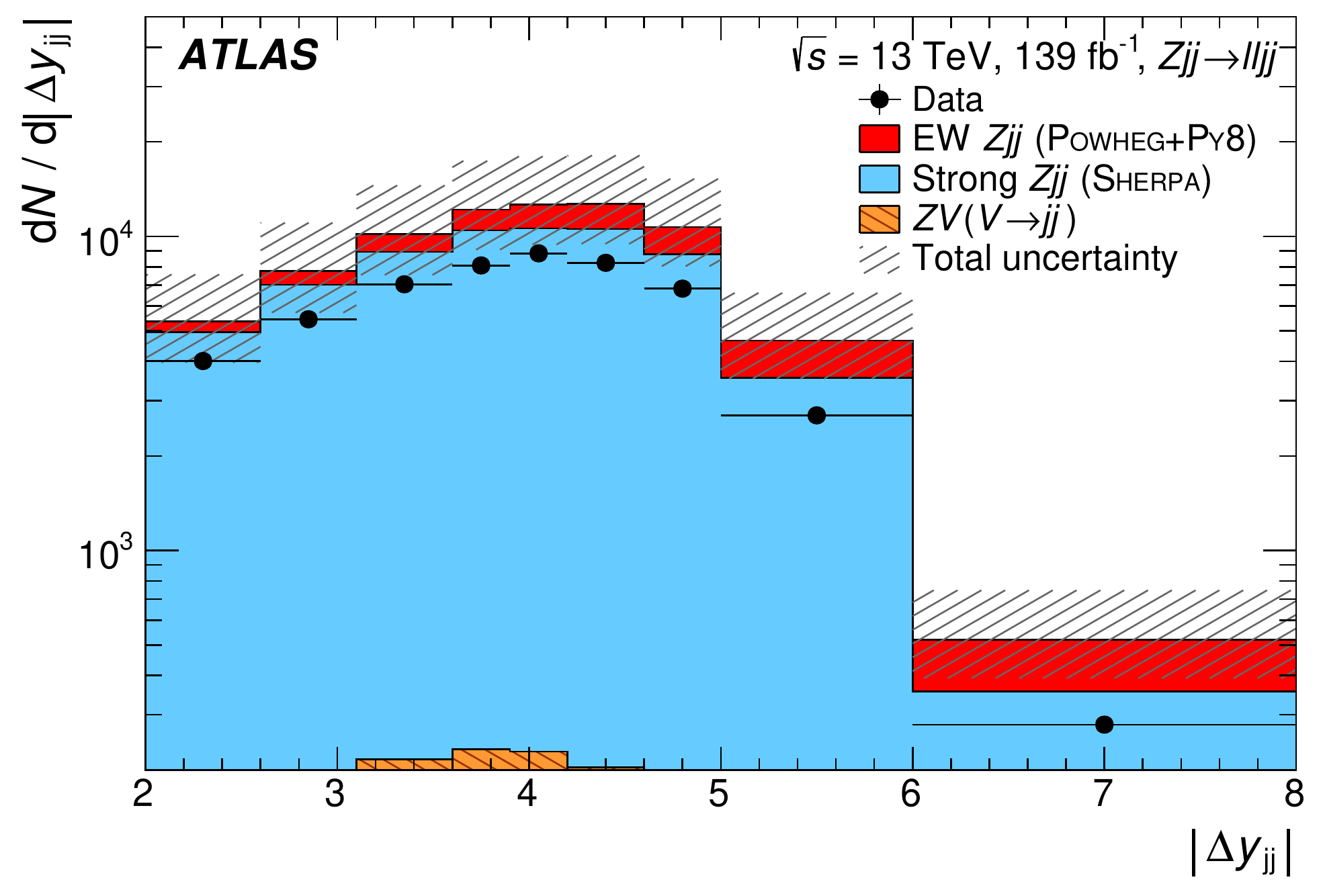}\\
\includegraphics[width=0.48\textwidth]{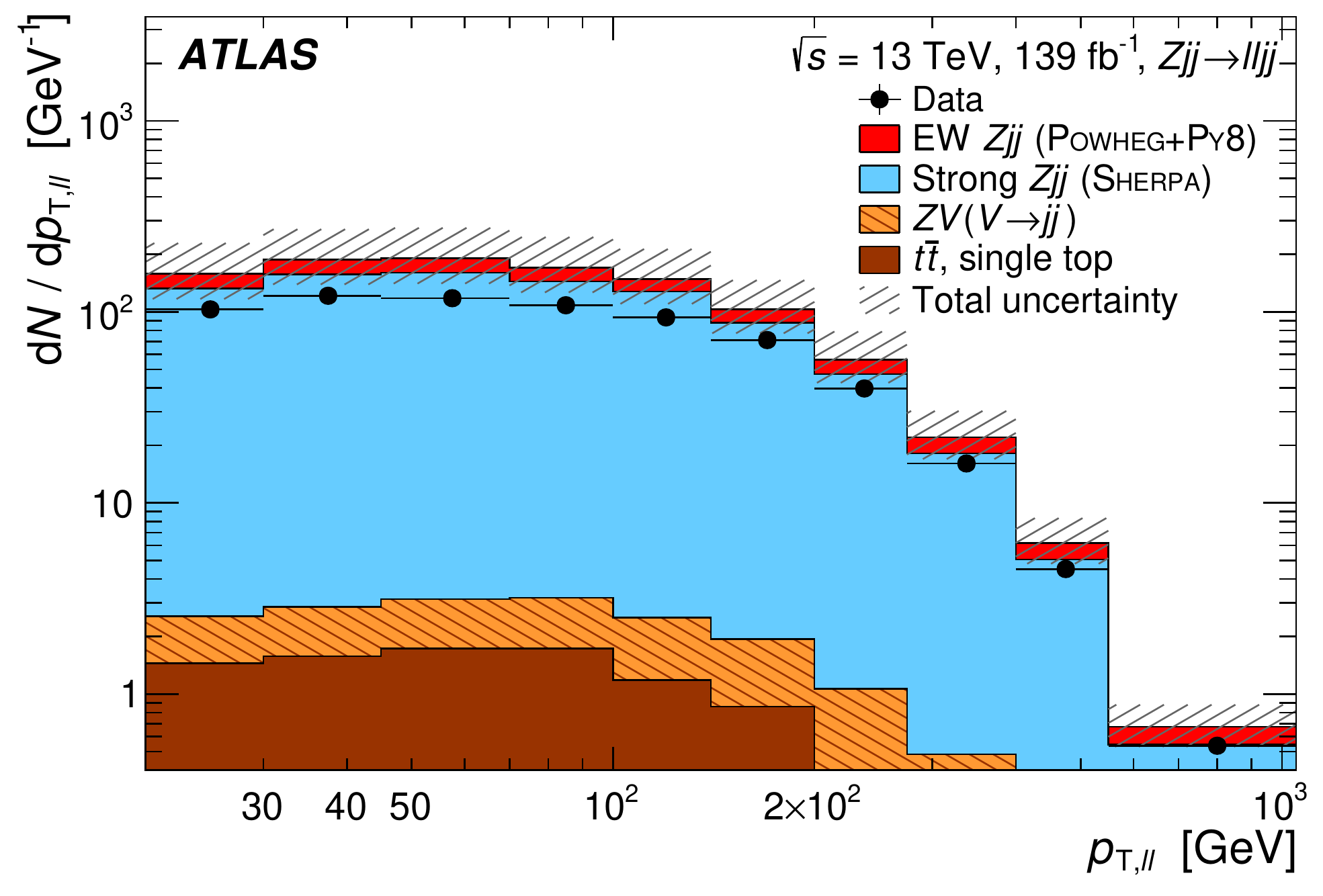}
\includegraphics[width=0.48\textwidth]{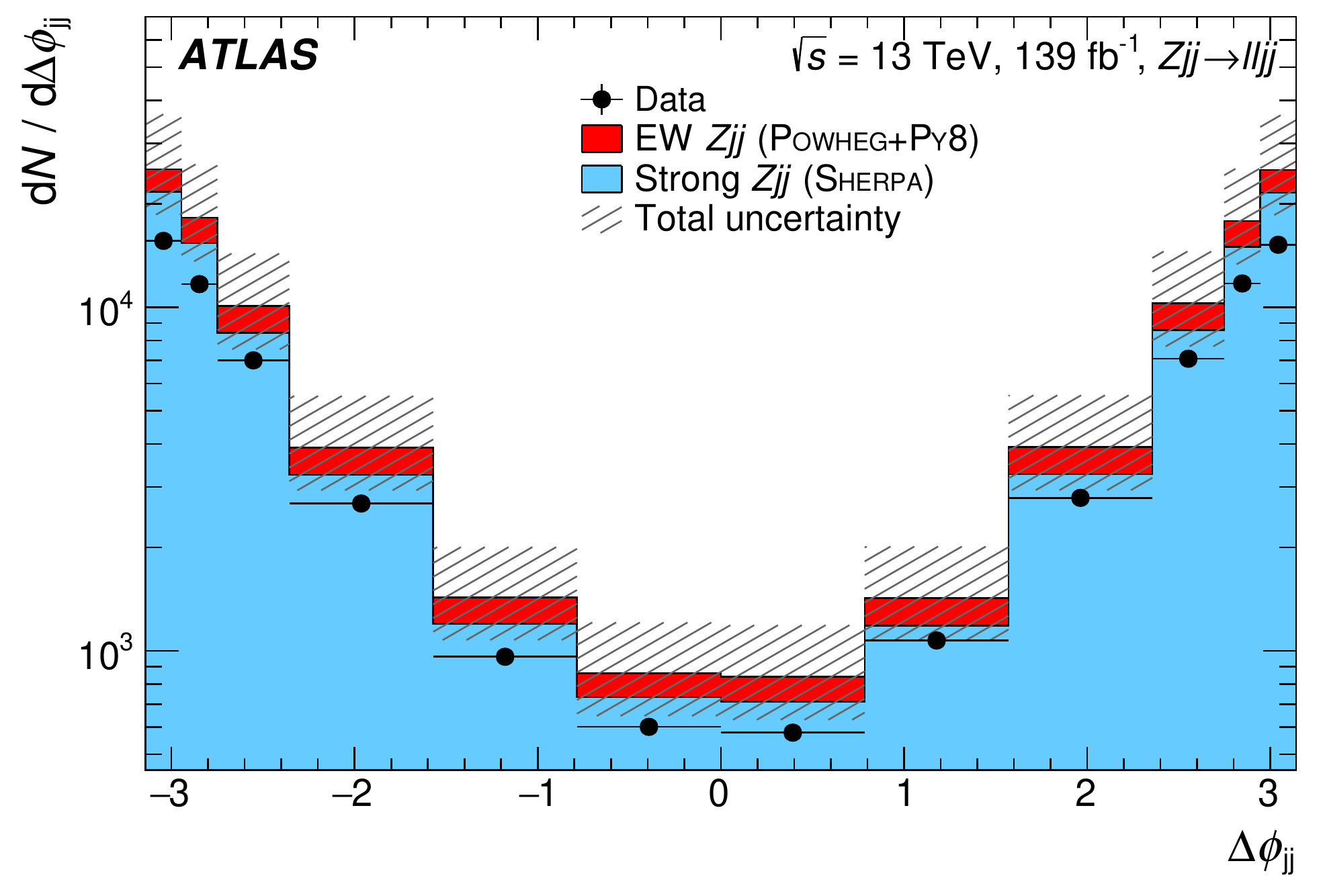}
\caption{Event yields as a function of \mjj{} (top left), \Dyjj{} (top right), \pTll{} (bottom left) and \Dphijj{} (bottom right) in data and simulation, measured after the event selection described in Section~\ref{sec:evsel}. The data are represented as black points and the associated error bar includes only statistical uncertainties. The \mjj{} spectrum is shown starting from 250~\GeV{}, and hence includes more events than the other plots that use the default $\mjj > \SI{1000}{\GeV}$ criterion.}
\label{fig:dataMCcomparison}
\end{figure}
 
The number of events in data that pass these selection requirements is shown in Table \ref{tab:eventYields}. The predicted event yield for each MC simulation is also presented. There is a large spread of EW $Zjj$ event yields predicted by the different event generators. Furthermore, the predicted strong $Zjj$ event yield also has significant uncertainties, with large theory uncertainties in each prediction and a large difference between the predictions of the different event generators. The contribution of the other processes amounts to about 3\%.

The disagreement between data and simulation is not just observed in the total event yield. Figure \ref{fig:dataMCcomparison} shows the data and predicted event yield as a function of \mjj{}, \Dyjj{}, \pTll{}, and \Dphijj{}, with \sherpa{} used to model the strong $Zjj$ process and \powpy{} used for the EW $Zjj$ process. The level of agreement between data and simulation depends on the kinematic properties of the event, with  agreement at large \mjj{} being particularly poor for this configuration of MC simulations.

 
\section{Extraction of electroweak component}
\label{sec:ewextraction}
 
\begin{figure}[t]
\includegraphics[width=0.48\textwidth]{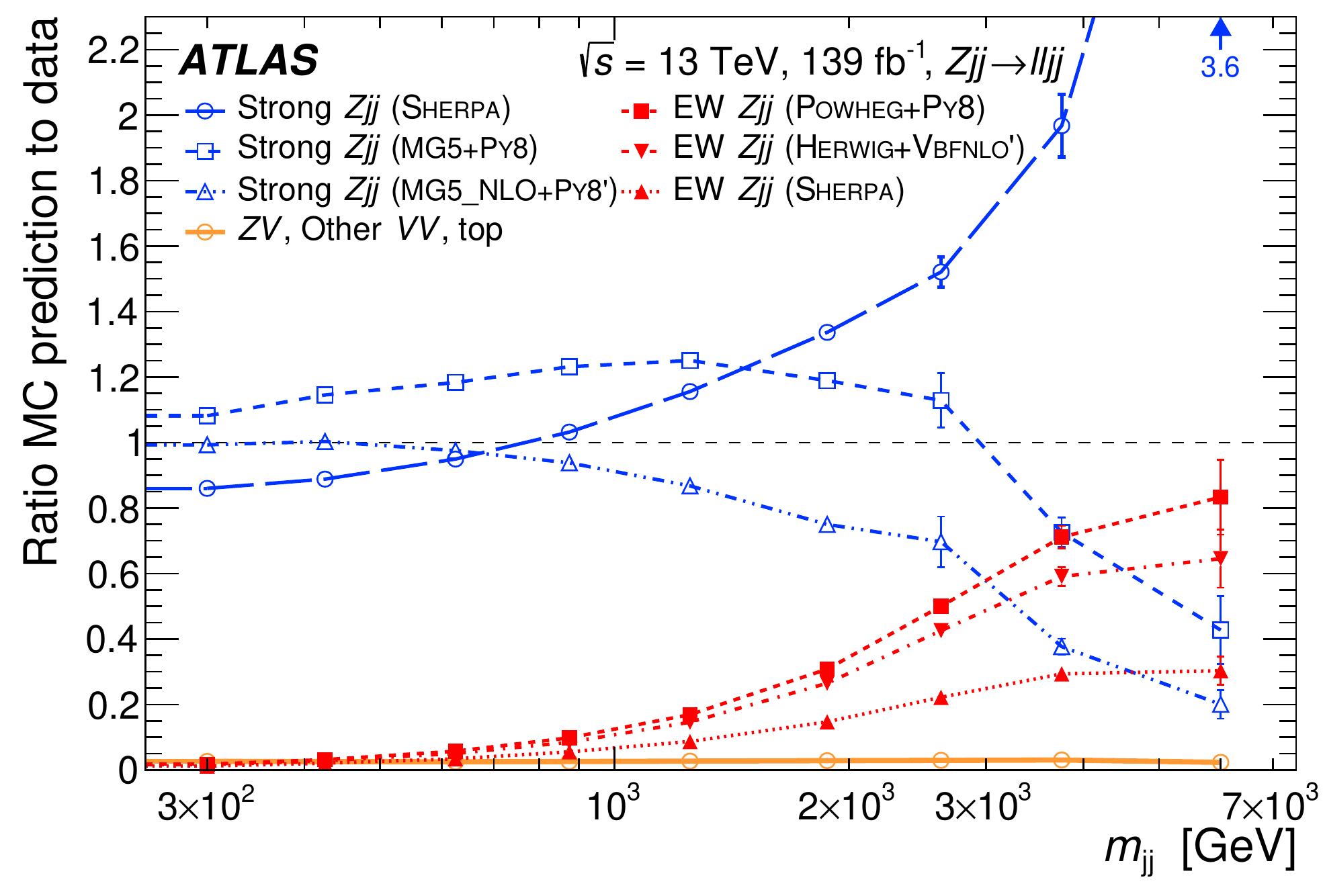}
\includegraphics[width=0.48\textwidth]{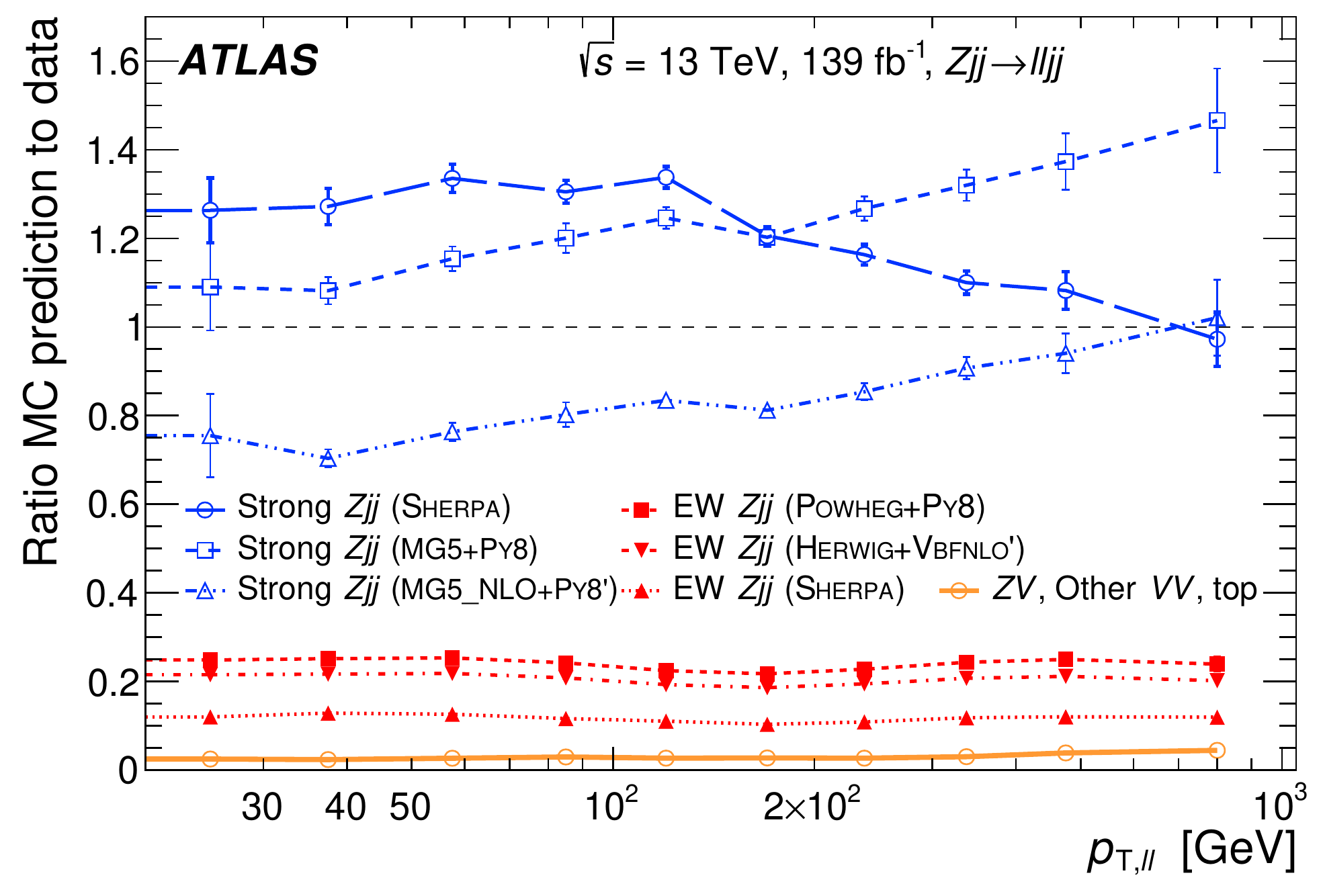}
\caption{Ratio of Monte Carlo prediction to data for different physics processes and generators for the \mjj{} and \pTll{} distributions, following the event selection described in Section~\ref{sec:evsel}.  The data contain all processes that pass the event selection and the ratio demonstrates the contribution to the observed event yield that is predicted by each MC generator. The \mjj{} distribution extends down to \SI{250}{\GeV} and hence includes a larger phase space than the \pTll{} distribution, which requires $\mjj>\SI{1000}{\GeV}$. Only statistical uncertainties are shown. The prediction labelled \mgnloR{} for the strong $Zjj$ prediction is obtained by a particle-level reweighting of the strong $Zjj$ simulation provided by \mglo{}. The EW $Zjj$ prediction labelled \hervbfR{} is also obtained by a particle-level reweighting of the EW $Zjj$ simulation provided by \powpy{}.
}
\label{fig:MCmodelling}
\end{figure}
 
The poor agreement between data and simulation observed in Figure~\ref{fig:dataMCcomparison} implies that the EW $Zjj$ event yield cannot be  extracted by simply subtracting the background simulations from the data. Furthermore, the level of mismodelling in the simulation changes when different strong $Zjj$ simulations are used, as shown in Figure~\ref{fig:MCmodelling} for the \mjj{} and \pTll{} distributions. A data-driven method is therefore used to constrain both the shape and normalisation of the strong $Zjj$ background during the extraction of the EW $Zjj$ event yield.
 
\begin{figure}[t]
\begin{center}
\includegraphics[width=0.5\textwidth]{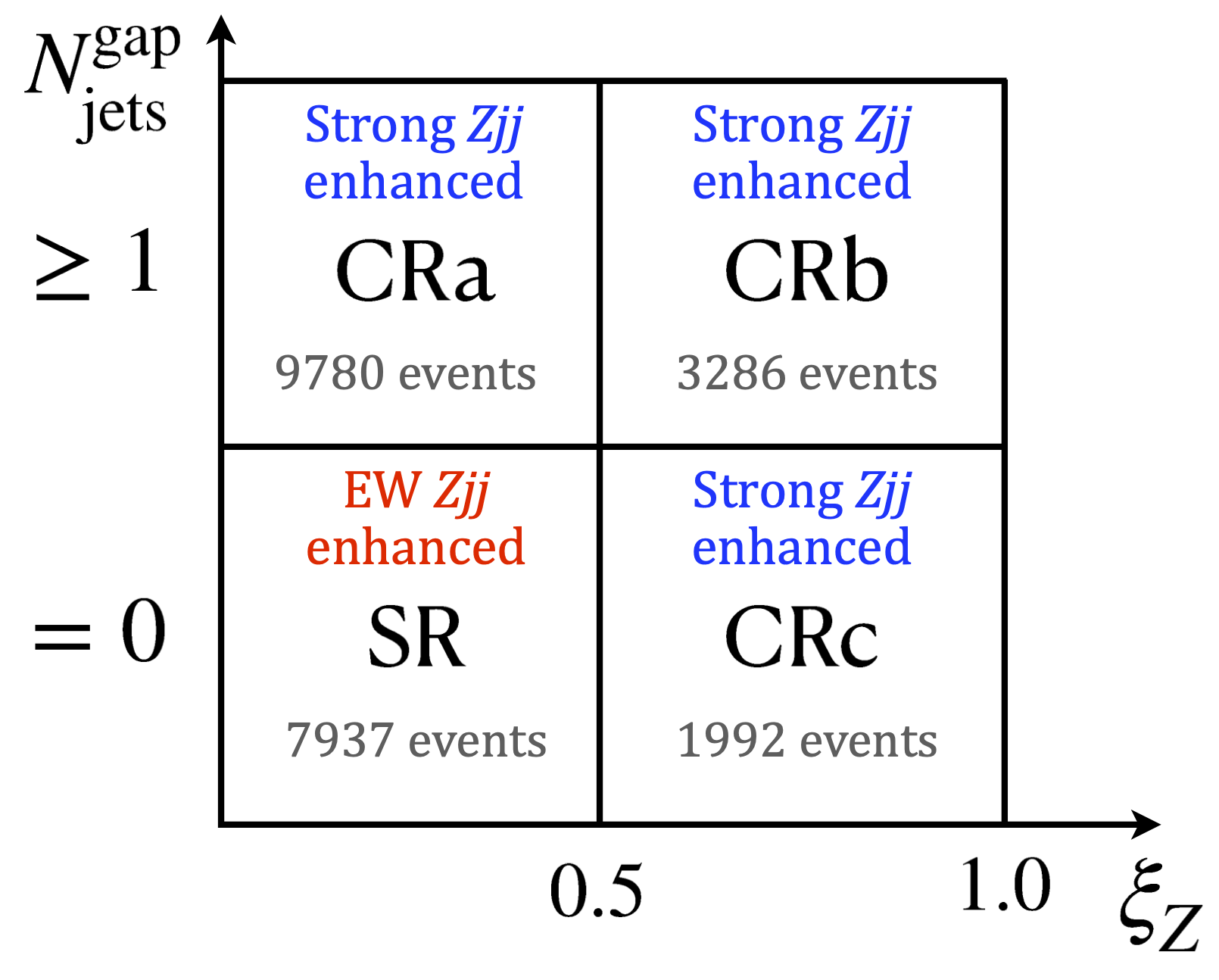}
\end{center}
\caption{Definition of the signal region (SR) and control regions (CRa, CRb, CRc) used in the extraction of the electroweak component.}
\label{fig:SRCR}
\end{figure}

The data are split into four regions by imposing criteria on \Zcent{} as well as on the multiplicity of jets in the rapidity interval between the leading and subleading jets, \Ngapjets{}.
These two variables are chosen because they are almost uncorrelated for both the strong and EW $Zjj$ processes, with calculated correlation coefficients ranging from $-$0.04 to $+$0.02 depending on the event generator and process.  Approximately 80\% of the EW $Zjj$ events are predicted to fall into the EW-enhanced signal region (SR) defined by $\Ngapjets=0$ and $\Zcent<0.5$.
The remaining three regions define EW-suppressed control regions (CR), which can be used to constrain the dominant background from strong $Zjj$ production. These regions are labelled as
CRa ($\Ngapjets \geq 1,~\Zcent<0.5$), CRb ($\Ngapjets \geq 1,~\Zcent>0.5$) and CRc ($\Ngapjets=0,~\Zcent>0.5$) and are depicted in Figure~\ref{fig:SRCR}. All analysis decisions and optimisations were performed with the signal region blinded, to avoid any unintended biases.
 
The EW $Zjj$ event yield is measured in the EW-enhanced SR using a binned maximum-likelihood fit~\cite{Barlow:1990vc,Tanabashi:2018oca}. The log likelihood is defined according to
\begin{equation*}
\label{eq:lnL}
\ln \mathcal{L} = - \sum_{r,i}\nu_{ri}(\boldsymbol{\theta}) + \sum_{r,i} N_{ri}^\mathrm{data}\,\ln{\nu_{ri}(\boldsymbol{\theta})} - \sum_s \frac{\theta_s^2}{2},
\end{equation*}
where $r$ is an index corresponding to the region $r\in \left\{\mathrm{CRa},\mathrm{CRb},\mathrm{CRc},\mathrm{SR}\right\}$, $i$ is the bin of the kinematic observable, $N_{ri}^\mathrm{data}$ is the observed event yield and $\nu_{ri}(\boldsymbol{\theta})$ is the prediction that is dependent on the $s$ sources of experimental systematic uncertainty that are each constrained by nuisance parameters $\boldsymbol{\theta} = (\theta_1,\dots,\theta_s$).\footnote{The dependence of the prediction on the systematic uncertainties is given by $\nu_{ri}(\boldsymbol{\theta})=\nu_{ri}^\mathrm{MC}\,\prod_s(1+\lambda_{ris}\,\theta_s)$, where $s$ is an index for the uncertainty source, $\theta_s$ is the associated nuisance parameter and $\lambda_{ris}$ is the fractional uncertainty amplitude for bin $i$ in region $r$.} The fitted number of events in each region and in each bin of a distribution is given by
\begin{equation}
\label{eq:fittedTotal}
\nu_{ri} = \mu_i\,\nu_{ri}^\mathrm{EW,MC} + \nu_{ri}^\mathrm{strong} + \nu_{ri}^\mathrm{other,MC},
\end{equation}
where $\mu_i$ is the EW $Zjj$ signal strength of bin $i$, $\nu_{ri}^\mathrm{EW,MC}$ and $\nu_{ri}^\mathrm{other,MC}$ are the MC predictions of EW $Zjj$ and contributions from other processes (diboson, \ttbar{} and single top), respectively. The strong $Zjj$ prediction is constrained using the different EW-suppressed control regions according to
\begin{equation}
\label{eq:strongConstraints}
\begin{array}{r@{}lcr@{}l}
\nu_{\mathrm{CRa},i}^\mathrm{strong} &{}= b_{\mathrm{L},i}\,\nu_{\mathrm{CRa},i}^\mathrm{strong,MC}, & ~ &
\nu_{\mathrm{CRb},i}^\mathrm{strong} &{}= b_{\mathrm{H},i}\,\nu_{\mathrm{CRb},i}^\mathrm{strong,MC},\\
\nu_{\mathrm{SR}i}^\mathrm{strong} &{}= b_{\mathrm{L},i}\,f(x_i)\,\nu_{\mathrm{SR},i}^\mathrm{strong,MC}, & ~ &
\nu_{\mathrm{CRc},i}^\mathrm{strong} &{}= b_{\mathrm{H},i}\,f(x_i)\,\nu_{\mathrm{CRc},i}^\mathrm{strong,MC}.
\end{array}
\end{equation}
Here, the $b_{\mathrm{L},i}$ and $b_{\mathrm{H},i}$ are sets of bin-dependent factors that apply to the $\Zcent < 0.5$ and  $\Zcent>0.5$ regions, respectively.
These factors are primarily constrained in CRa and CRb, where they adjust the predicted simulated strong $Zjj$ event yields and bring the total predicted yield ($v_{ri}$ of Equation~\ref{eq:fittedTotal}) into better agreement with data.
The $f(x_i)$ is a two-parameter function of the observable that is being measured and is evaluated at the centre of each bin. This function provides a residual correction to the constrained strong $Zjj$ yield to account for the extrapolation from CRa ($\Ngapjets{}\geq1$) to the SR ($\Ngapjets{}=0$) and is primarily constrained by CRb and CRc. The function is taken to be a first-order polynomial.
 
The free parameters in the binned maximum-likelihood fit are therefore the signal strengths $\mu_i$, the two parameters of the function $f(x_i)$, and the $b_{\mathrm{L},i}$ and $b_{\mathrm{H},i}$ corrections to the strong $Zjj$ process.
In total, this amounts to $3\,N_\mathrm{bins}+2$ parameters that are constrained using $4\,N_\mathrm{bins}$ measurements in data, where $N_\mathrm{bins}$ is the number of bins measured for a specific observable (\mjj{}, \Dyjj{}, \pTll{} and \Dphijj{}).
 
\begin{figure}[t]
\includegraphics[width=0.48\textwidth]{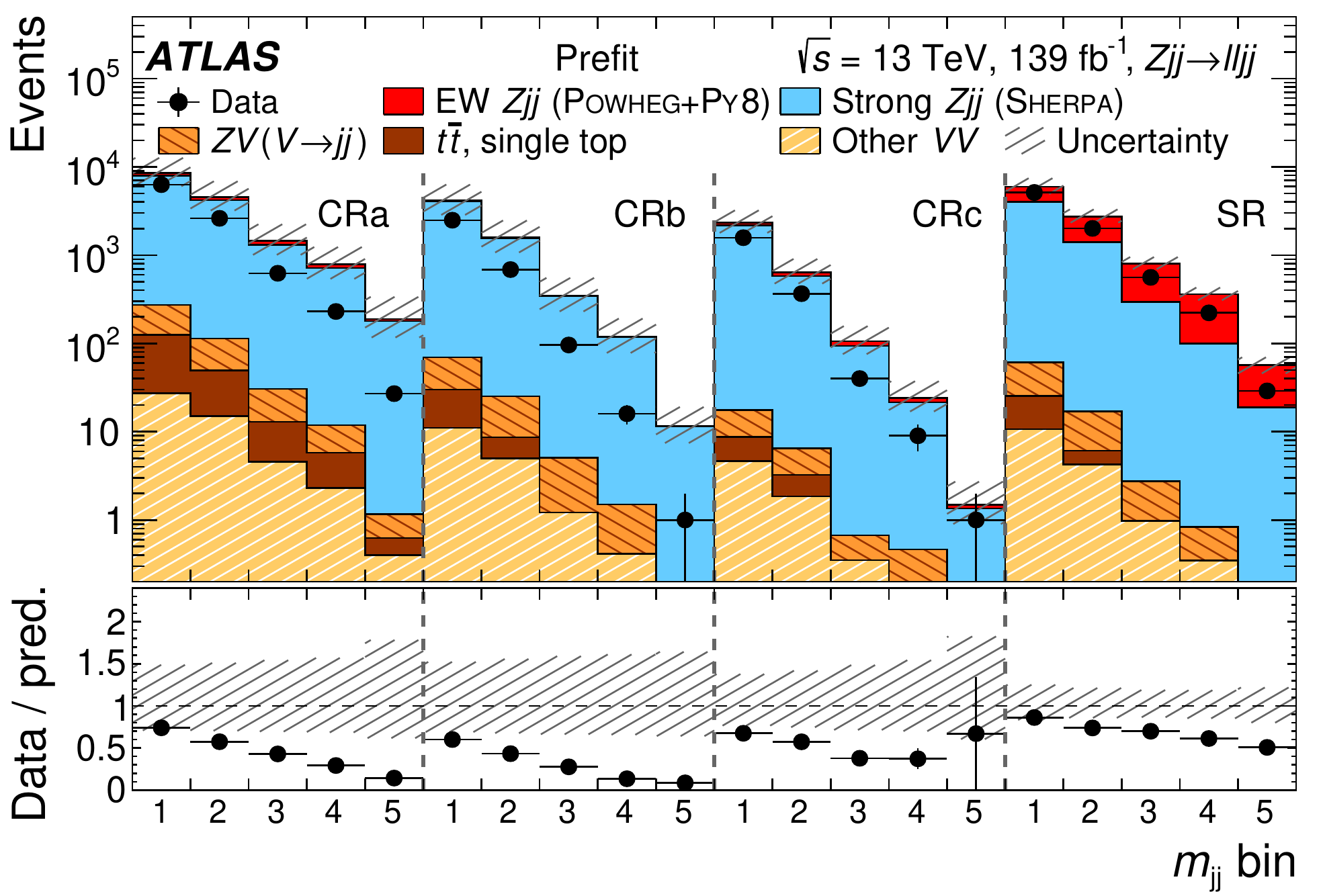}
\includegraphics[width=0.48\textwidth]{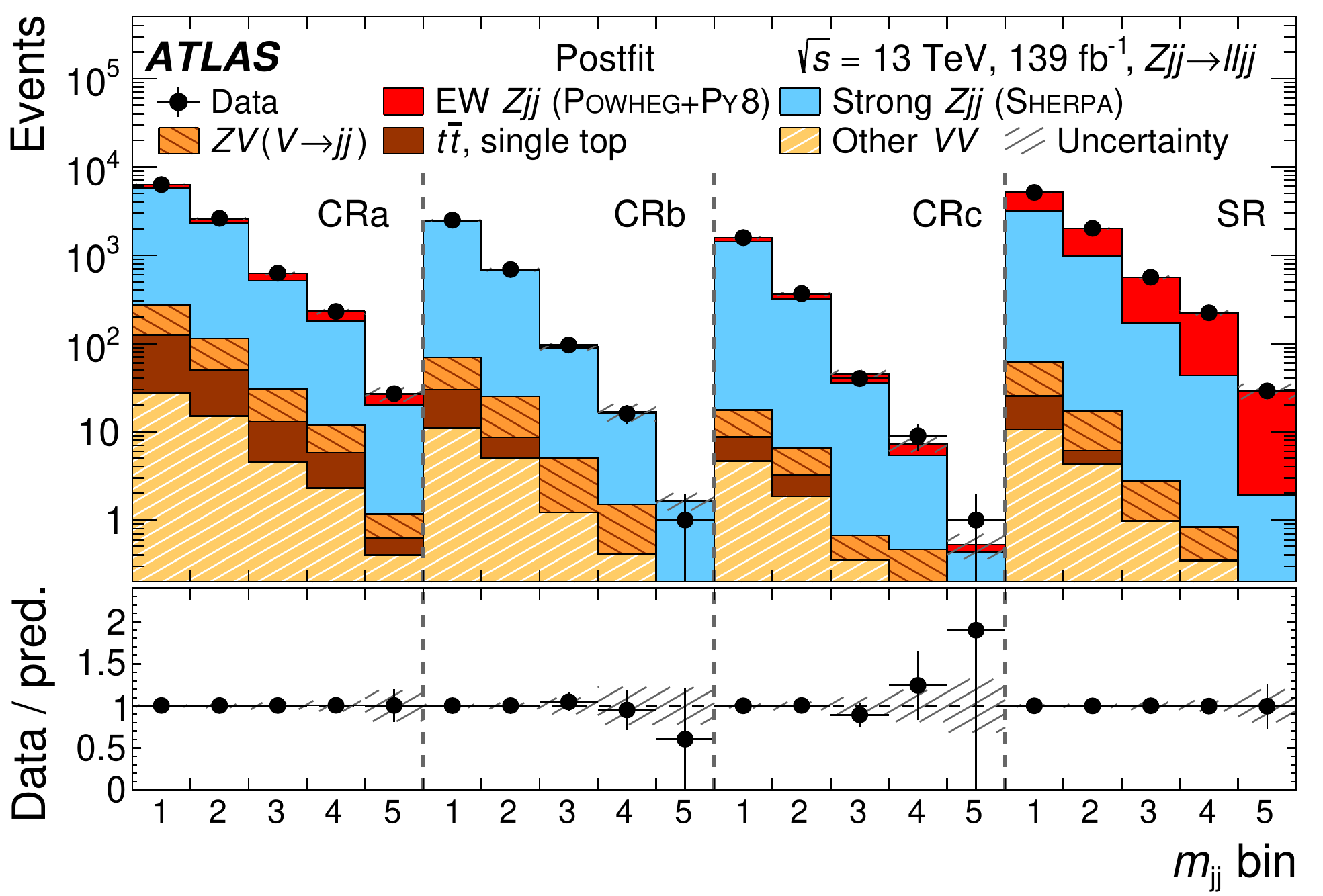}\\
\includegraphics[width=0.48\textwidth]{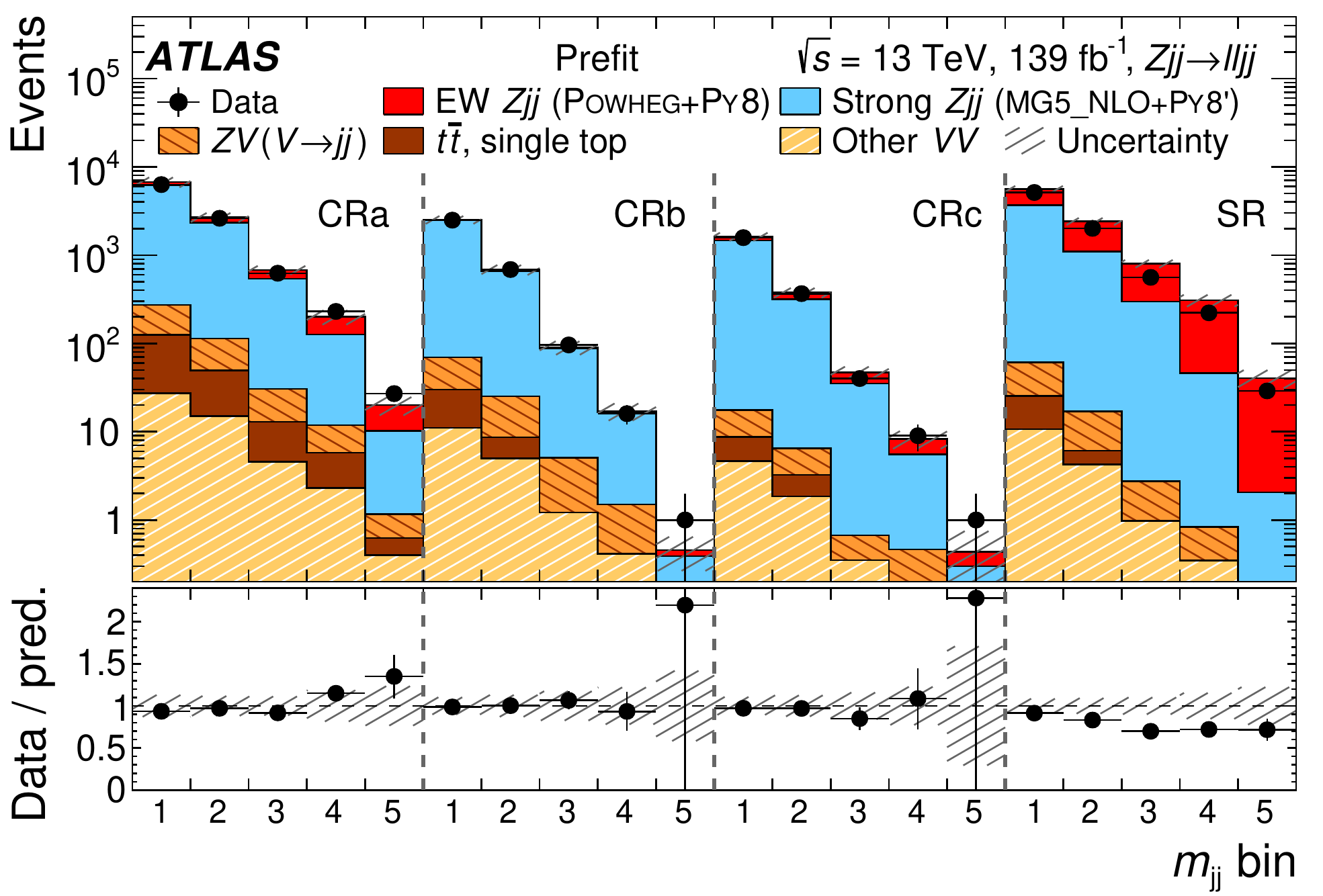}
\includegraphics[width=0.48\textwidth]{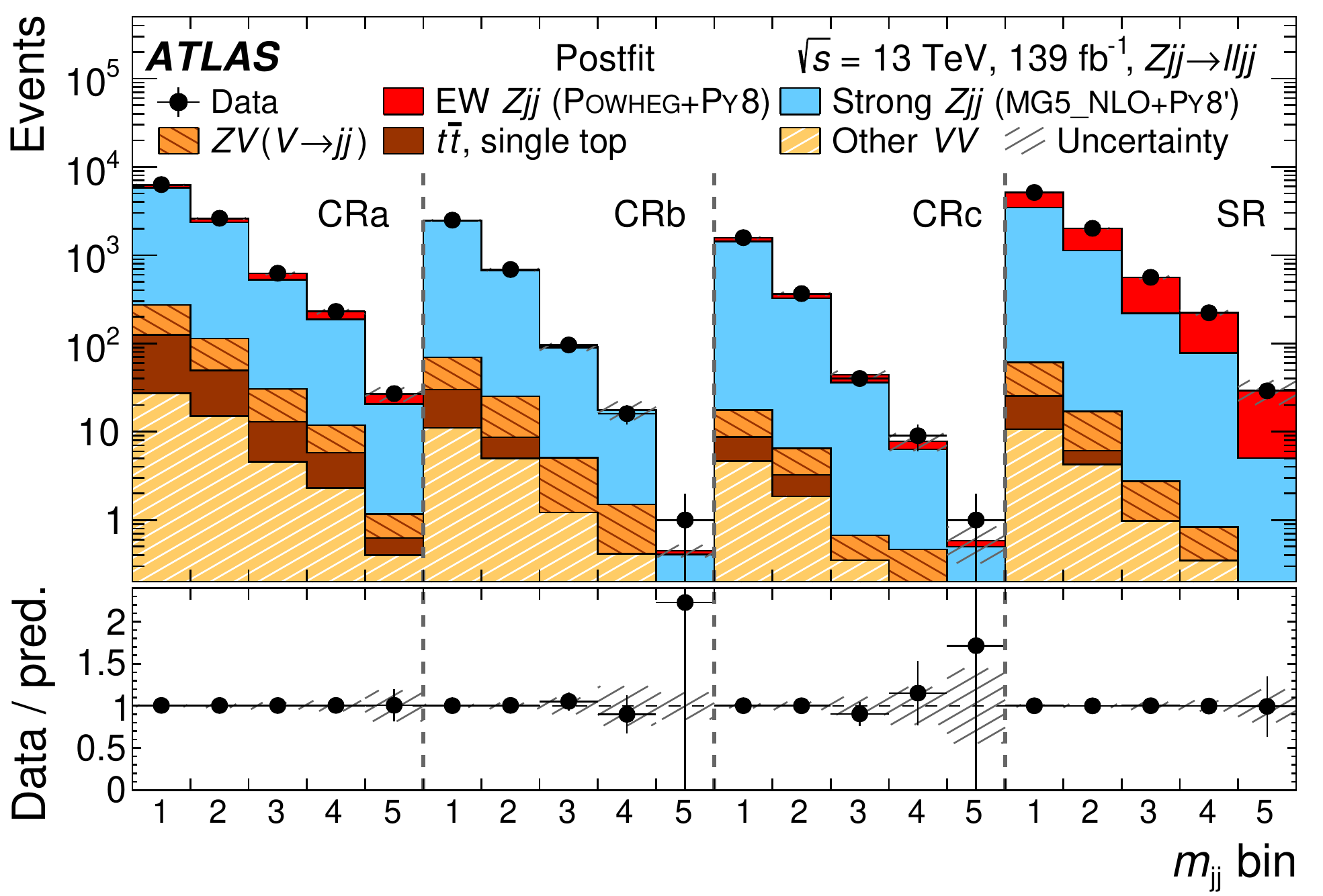}
\caption{Comparison between data and prediction before (left) and after (right) the fit using strong $Zjj$ estimates based on \sherpa{} (top) and \mgnloR{} (bottom) in bins of \mjj{} in the different control and signal regions. The \mgnloR{} prediction is obtained by a particle-level reweighting of the strong $Zjj$ simulation provided by \mglo{}.
The \mjj{} bin edges are defined by $(1.0,1.5,2.25,3.0,4.5,7.5)$~\TeV{}.
}
\label{fig:prePostFit}
\end{figure}

The pre-fit and post-fit agreement between data and simulation is shown in Figure \ref{fig:prePostFit} as a function of \mjj{} in the signal and control regions. Two separate fits are shown, one using the \sherpa{} strong $Zjj$ prediction (top row) and one using the \mgnloR{} prediction (bottom row). These simulations initially have very different mismodelling as a function of \mjj{}, but produce very good agreement with the data following the fitting procedure. The overall scaling factor applied to the strong $Zjj$ prediction from \mgnloR{} in the signal region is 0.93 at low \mjj{} rising to 2.2 at high \mjj{}. For \sherpa{}, the corresponding scaling factors are 0.86 at low \mjj{} and 0.26 at high \mjj{}. The pre-fit systematic uncertainties shown on the plots are derived as outlined in Section~\ref{sec:systematics}.
 
Since there is no \textit{a priori} reason to prefer any strong $Zjj$ generator over another, the EW $Zjj$ component is extracted three times, once using the \sherpa{} strong $Zjj$ prediction, once using the \mgnloR{} strong $Zjj$ prediction, and once using the \mglo{} strong $Zjj$ prediction. The final electroweak signal yield in each bin of the differential distribution is taken to be the midpoint of the envelope of yields obtained using the three different strong $Zjj$ event generators.
The envelope itself is used to define a systematic uncertainty as outlined in Section~\ref{sec:systematics}.
 
The constraints on the strong $Zjj$ simulation in Equation~\ref{eq:strongConstraints} are evaluated independently for each of the measured differential distributions (\mjj{}, \Dyjj{}, \pTll{} and \Dphijj{}). This results in slightly different total EW $Zjj$ and strong $Zjj$ event yields when summed across each differential spectrum. To ensure consistency between the distributions, an additional constraint is applied in the likelihood to ensure that the same integrated strong $Zjj$ yield is obtained for each distribution, i.e.
\begin{equation}
\label{eq:strongyield}
\sum_i \nu_{\mathrm{SR},i}^\mathrm{strong} = \hat{\nu}^\mathrm{strong}_{\mathrm{SR},m_{jj}},
\end{equation}
where $\hat{\nu}^\mathrm{strong}_{\mathrm{SR},m_{jj}}$ is the event yield obtained by integrating the constrained strong $Zjj$ template for the \mjj{} distribution in the SR .
 
The electroweak extraction methodology is validated in four ways. First, a variation of the likelihood method is implemented by switching the control regions used to define the strong $Zjj$ simulation as defined in  Equation~\ref{eq:strongConstraints}, such that the $b_i$ factors are constrained in CRs at high $\xi_Z$ and the $f(x_i)$ function is then defined to correct for non-closure when transferring these corrections to low $\xi_Z$. Second, the constraint on the strong $Zjj$ background includes a function ($f(x_i)$) that is taken to be a first-order polynomial by default. This choice is validated by changing the function to a second-order polynomial. Third, the constraint applied to the integrated strong $Zjj$ event yield (Eq. \ref{eq:strongyield}) is removed.  Finally, a simpler `sequential' method is used to extract the EW $Zjj$ event yields. In this approach, the data-driven correction to the strong $Zjj$ is derived in CRa (assuming the SM prediction for the electroweak process in this region) and directly applied to the strong $Zjj$ simulation in the SR. A transfer factor to account for mismodelling between the SR and CRa is evaluated at low \mjj{} ($250\leq\mjj<500$~\GeV{}). Non-closure of the sequential method is evaluated in CRc using corrections to the strong $Zjj$ process derived in CRb; this non-closure is used as a systematic uncertainty in the sequential method. The extracted electroweak event yields obtained with these four variations are found to be in good agreement with the nominal results and are presented in Appendix \ref{sec:appV}.
 
\section{Correction for detector effects}
\label{sec:unfolding}
Particle-level differential cross-sections are produced by correcting the inclusive $Zjj$ and EW $Zjj$ event yields in each bin for the effects of detector inefficiency and resolution. The EW $Zjj$ event yields are extracted in the signal region using the method outlined in the previous section. The inclusive $Zjj$ event yields are obtained by subtracting, from the data, the small number of events predicted by simulation for processes that do not contain a $Z$ boson and two jets in the final state ($t\bar{t}$, single-top, $VV\not\to Zjj$, and $W$+jets  production). For both inclusive and EW $Zjj$ production, the event yields in the $e^+e^-$ and $\mu^+\mu^-$ decay channels are added together and unfolded in a single step.
 
The particle level is defined using final-state stable particles with mean lifetime satisfying $c\tau> 10$~mm. To reduce model-dependent extrapolations across kinematic phase space, the particle-level event selection is defined to be as close as possible to the detector-level event selection defined in Section~\ref{sec:evsel}. Leptons are defined at the `dressed' level, as the four-momentum combination of a prompt electron or muon (that do not originate from the decay of a hadron) and all nearby prompt photons within $\Delta R < 0.1$. Leptons are required to have $p_{\mathrm{T}}>25$~\GeV{} and have the same acceptance requirement as used at the analysis level, i.e. muons satisfy  $|\eta|<2.4$ and electrons satisfy $|\eta|<2.47$ (but exclude the region $1.37<|\eta|<1.52$).
Jets are reconstructed using the anti-$k_t$ algorithm using all final-state stable particles as input, except those that are part of a dressed-lepton object. Jets are required to have $\pT>\SI{25}{\GeV}$ and $|y|<4.4$. Using these jets and leptons, events are then selected in a VBF topology using   requirements identical to those imposed at detector level. The EW $Zjj$ differential cross-sections are measured in the SR, whereas inclusive $Zjj$ differential cross-sections are measured in the SR and the three CRs. The VBF topology, SR and the three CRs are defined in Table ~\ref{tab:fidDef}.
 
\begin{table}[t]
\centering
\bgroup
\def\arraystretch{1.2}
\begin{tabular}{r|l}\hline\hline
Dressed muons & $p_\mathrm{T}^{}>25$~\GeV{} and $|\eta|<2.4$ \\
Dressed electrons & $p_\mathrm{T}^{}>25$~\GeV{} and $|\eta|<2.47$ (excluding $1.37<|\eta|<1.52$) \\
Jets & $p_\mathrm{T}^{}>25$~\GeV{} and $|y|<4.4$ \\
\hline
VBF topology &
$N_\ell = 2$ (same flavour, opposite charge), $m_{\ell\ell} \in (81,101)$\,\GeV{}\\
& $\Delta R_\text{min}(\ell_1,j)>0.4$,~~$\Delta  R_\text{min}(\ell_2,j)>0.4$\\
& $N_\text{jets} \geq 2$,~~$p_\mathrm{T}^{j1} > 85$~\GeV{},~~$p_\mathrm{T}^{j2} > 80$~\GeV{}\\
& $\pTll > 20$~\GeV{},~~$\pTbal < 0.15$\\
& $\mjj > 1000$~\GeV{},~~$\Dyjj>2$,~~$\Zcent<1$\\
\hline
CRa & VBF topology~$\oplus$~$\Ngapjets \geq 1$  and $\Zcent<0.5$\\[0.5mm]
CRb & VBF topology~$\oplus$~$\Ngapjets \geq 1$ and $\Zcent > 0.5$\\[0.5mm]
CRc & VBF topology~$\oplus$~$\Ngapjets =0$ and $\Zcent > 0.5$\\[0.5mm]
SR & VBF topology~$\oplus$~$\Ngapjets =0$ and $\Zcent<0.5$\\[0.5mm] \hline\hline
\end{tabular}
\egroup
\caption{
Particle-level definition of the measurement.
$\Delta R_\text{min}(\ell_1,j)$ denotes the minimum $\Delta R$ distance between the highest transverse-momentum lepton ($\ell_1$) and any of the jets in the event. $\Delta R_\text{min}(\ell_2,j)$ is similarly defined.
\label{tab:fidDef}
}
\end{table}
 
Each distribution is unfolded separately using the iterative Bayesian method proposed by D'Agostini~\cite{DAGOSTINI1995487,Adye:2011gm} with two iterations. This procedure uses MC simulations to (i) correct for events that pass the detector-level selection but not the particle-level selection, (ii)  invert the migration between bins of the differential distribution, and (iii) correct for events that pass the particle-level selection but not the detector-level selection. For the EW $Zjj$ differential cross-section measurements, the \powpy{} EW $Zjj$ simulation  is used to define the corrections and the response matrices. For the inclusive $Zjj$ differential cross-section measurements, all sources of $Zjj$ production are part of the measurement and the unfolding is carried out using the cross-section weighted sum of the \powpy{} EW $Zjj$ simulation, the \sherpa{} strong $Zjj$ simulation, and the \sherpa{} diboson samples that contain a leptonically decaying $Z$ boson produced in association with a hadronically decaying weak boson.
 
Statistical uncertainties in the data are propagated through the unfolding procedure using the bootstrap method~\cite{PhysRevD.39.274} with 1000 pseudo-experiments. For the EW $Zjj$ measurements, the electroweak extraction is repeated for each pseudo-experiment after fluctuating the event yields, in each bin of the signal and control regions, using a Poisson distribution. For the inclusive $Zjj$ measurements, the background-subtracted event yields are fluctuated using a Gaussian distribution centred on the data-minus-background value and with a width given by the data statistical uncertainty. The statistical uncertainties in the MC simulation are propagated through the unfolding procedure in a similar fashion, by fluctuating each bin of the response matrix using a Gaussian distribution. The unfolding is repeated with the modified distributions (or response matrices) created for each pseudo-experiment. The final statistical uncertainties in the measurement are taken to be the standard deviation of the unfolded values obtained from the ensemble of pseudo-experiments.
 
\section{Systematic uncertainties}
\label{sec:systematics}
 
\subsection*{Experimental systematic uncertainties}
 
Experimental systematic uncertainties arise from jet reconstruction, lepton reconstruction, the pile-up of multiple proton--proton interactions, and the luminosity determination. These uncertainties affect the normalisation and shape of the background simulations used in the extraction of the EW $Zjj$ process, as well as the MC simulations used to unfold the EW $Zjj$ and inclusive $Zjj$ event yields. For the extraction of the electroweak signal, each source of experimental uncertainty is included as a Gaussian-constrained nuisance parameter in the likelihood, as outlined in Section~\ref{sec:ewextraction}. For the unfolding, each source of uncertainty is propagated to the MC simulations and the change in the unfolded event yield is taken as the systematic uncertainty.
 
The luminosity is measured to an accuracy of 1.7\% using van der Meer beam separation scans, as outlined in Refs.~\cite{ATLAS-CONF-2019-021,Avoni:2018iuv}. Uncertainties in the modelling of pile-up interactions are estimated by repeating the analysis after varying the average number of pile-up interactions in the simulation. This variation accounts for the uncertainty in the ratio of the predicted and measured inelastic cross-sections within the ATLAS fiducial volume~\cite{Aaboud:2016mmw}.
 
A variation in the pile-up reweighting of simulated events (referred to as pile-up uncertainty) is included to
account for the uncertainty in the ratio of the predicted and measured inelastic cross-sections.
 
The lepton trigger, reconstruction and isolation efficiencies in simulation are corrected using scale factors derived from data, as outlined in Section~\ref{sec:MC}. Systematic uncertainties associated with this procedure are estimated by varying these scale factors according to their associated uncertainties~\cite{Aaboud:2019ynx,Aad:2016jkr}.  In addition, uncertainties due to differences between data and simulation in the reconstructed lepton momentum~\cite{Aaboud:2018ugz,Aad:2016jkr} are estimated by scaling and smearing the lepton momentum in the simulation. The overall impact on the differential cross-section measurement from systematic uncertainties associated with leptons is typically 1\%, but rises to 2\% at the highest dilepton transverse momentum.
 
The uncertainties associated with jet energy scale and jet energy resolution have a larger impact on the analysis. As discussed in Section~\ref{sec:evsel}, the jets are calibrated in data using a combination of MC-based and data-driven correction factors. The uncertainty in the measurement due to these corrections is estimated by scaling and smearing the jet four-momentum in the simulation by one standard deviation in the associated uncertainties of the calibration procedure~\cite{Aaboud:2017jcu}. The impact on the differential cross-section measurements is between 5\% at low \mjj{} or \pTll{}, but more than 10\% for $m_{jj}>\SI{4}{\TeV}$. An additional uncertainty arises from the use of the jet vertex tagger, which suppresses jets arising from pile-up interactions but is not fully efficient for jets produced in the hard scatter. Uncertainties arising from imperfect modelling of the JVT efficiency are estimated by varying the JVT requirement~\cite{Aad:2015ina} and result in an uncertainty of about 1\%, which is anti-correlated between the $\Ngapjets=0$ and $\Ngapjets\geq 1$ regions.

\subsection*{Theoretical uncertainties in the electroweak signal extraction}
 
Theoretical uncertainties associated with the modelling of the signal and background processes can impact the extraction of the electroweak signal yield. The impact of each source of theory uncertainty on the extracted signal yield is evaluated by repeating the electroweak extraction procedure (outlined in Section~\ref{sec:ewextraction}) after varying the input MC event generator templates in the SR and the CRs. The variation in the extracted signal yield is then propagated through the unfolding procedure.
 
Theoretical uncertainties associated with the modelling of the strong $Zjj$ process are the dominant uncertainties in the extraction of the electroweak signal yield. Three sources of uncertainty in the strong $Zjj$ modelling are investigated, arising from (i) the choice of event generator, (ii) the  renormalisation and factorisation scale dependence in the strong $Zjj$ calculations, and (iii) the parton distribution functions. The systematic uncertainty associated with the choice of event generator is defined by the envelope of electroweak event yields extracted using the \sherpa{}, \mgnloR{} and \mglo{} strong $Zjj$ simulations (the default electroweak event yield defined as the midpoint of this envelope, as discussed in Section~\ref{sec:ewextraction}).  The uncertainty associated with the choice of renormalisation and factorisation scales is assessed by repeating the analysis using new strong $Zjj$ templates for \sherpa{} in which the renormalisation ($\muR$) and factorisation ($\muF$) scales have been varied independently by factors of 0.5 and 2.0. Six variations are considered for each generator corresponding to ($\muR$, $\muF$)~$=$ (0.5,1.0), (2.0,1.0), (1.0, 0.5), (1.0, 2.0), (0.5,0.5) and (2.0,2.0). For each variation, the change in the extracted EW event yield relative to that obtained with the default \sherpa{} strong $Zjj$ sample is evaluated, and the envelope of the variations is then taken to be the relative uncertainty in the extracted electroweak yields.
Finally, the impact of uncertainties associated with the parton distribution functions is estimated using the \sherpa{} generator, by reweighting the nominal strong $Zjj$ sample to reproduce the variations of the NNPDF3.0nnlo PDF set (including the associated \alphas{} variations) and repeating the full analysis chain for each variation. The systematic uncertainty in the extracted EW signal yields due to PDFs is then taken as the RMS of signal yields extracted from the PDF set variations. Of the three sources of uncertainty associated with modelling strong $Zjj$ production, the choice of event generator has the largest impact on the extracted electroweak yields.
 
Theoretical uncertainties associated with the modelling of the EW $Zjj$ process have a much smaller impact on the extraction of the electroweak component because, for each bin of a measured distribution, the only theoretical input is the relative event yields in the SR and CRs. The theoretical uncertainty due to the mismodelling of the EW $Zjj$ process is determined by repeating the analysis after reweighting the default \POWHEGBOXV{} EW $Zjj$ simulation such that it matches the prediction of the \hervbf{} EW $Zjj$ simulation at particle level. The change in extracted EW event yield with respect to the nominal event yield extracted with \powpy{} is taken as a symmetric uncertainty. The signal-modelling dependence is further validated using the leading-order \sherpa{} EW $Zjj$ simulation to extract the electroweak event yield and the results are found to be consistent and within the assigned uncertainty due to electroweak $Zjj$ modelling. Systematic uncertainties associated with the parton distribution functions used in the matrix-element calculation are investigated, by applying the NNPDF3.0nnlo PDF set variations to the \sherpa{} EW $Zjj$ simulation, and found to have a much smaller impact than the choice of event generator. Variations of renomalisation and factorisation scales in the matrix-element calculations are also found to have a negligible impact on the final result. The total systematic uncertainty associated with the signal modelling is typically between 2--3\%.
 
The electroweak extraction methodology assumes that there is no interference between the EW $Zjj$ process and the strong $Zjj$ process. The size of the interference contribution relative to the electroweak signal process is estimated at particle level using \mg{} as a function of the measured kinematic variables in the SR and CRs. The uncertainty associated with the interference is then defined as the change in the extracted electroweak yield induced by reweighting the default \powpy{} EW $Zjj$ sample such that it contains the interference contribution, and is taken to be symmetric. This source of uncertainty is typically a factor of five smaller than the uncertainty associated with the modelling of the strong $Zjj$ process.
 
\begin{figure}[t]
\includegraphics[width=0.48\textwidth]{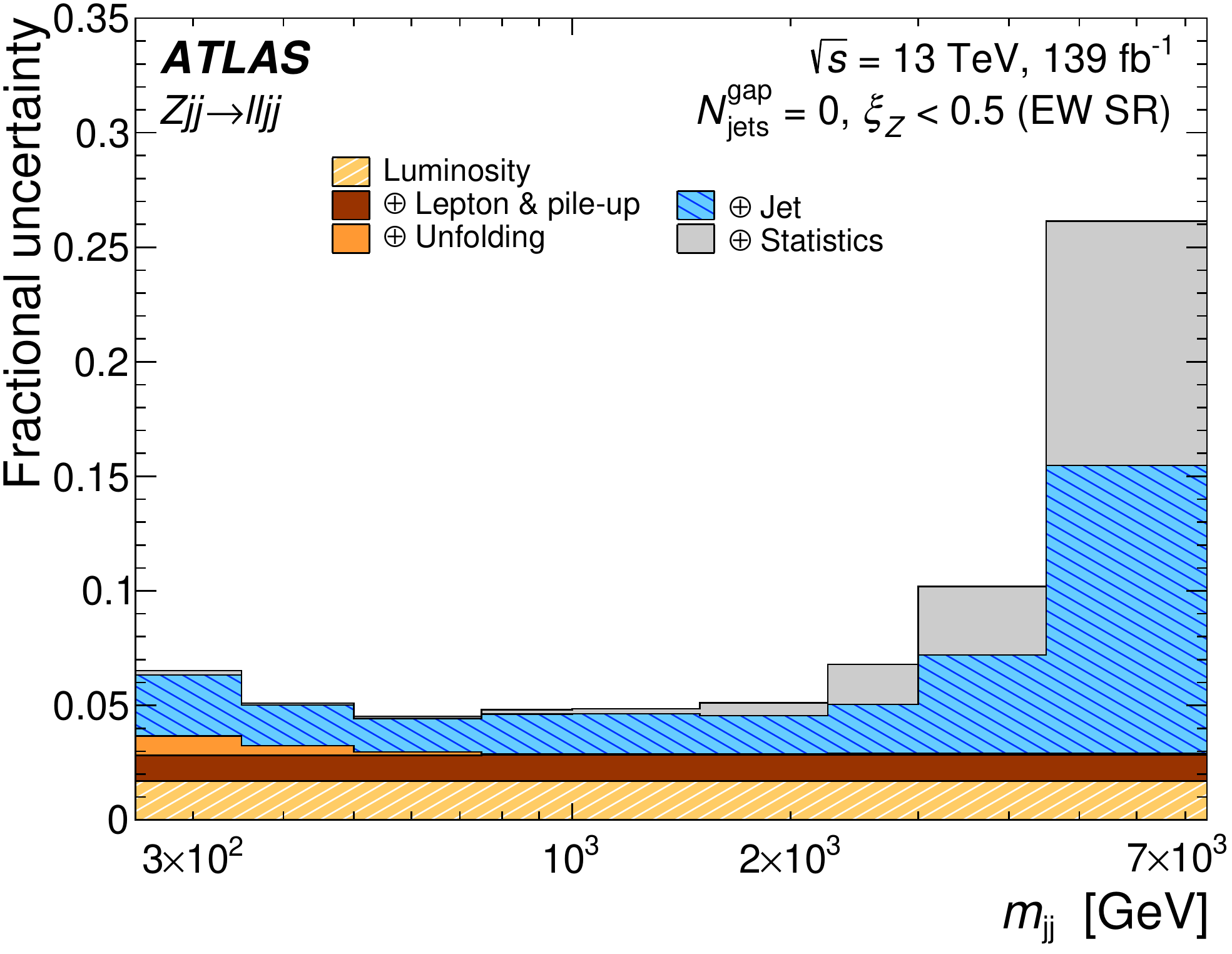}
\includegraphics[width=0.48\textwidth]{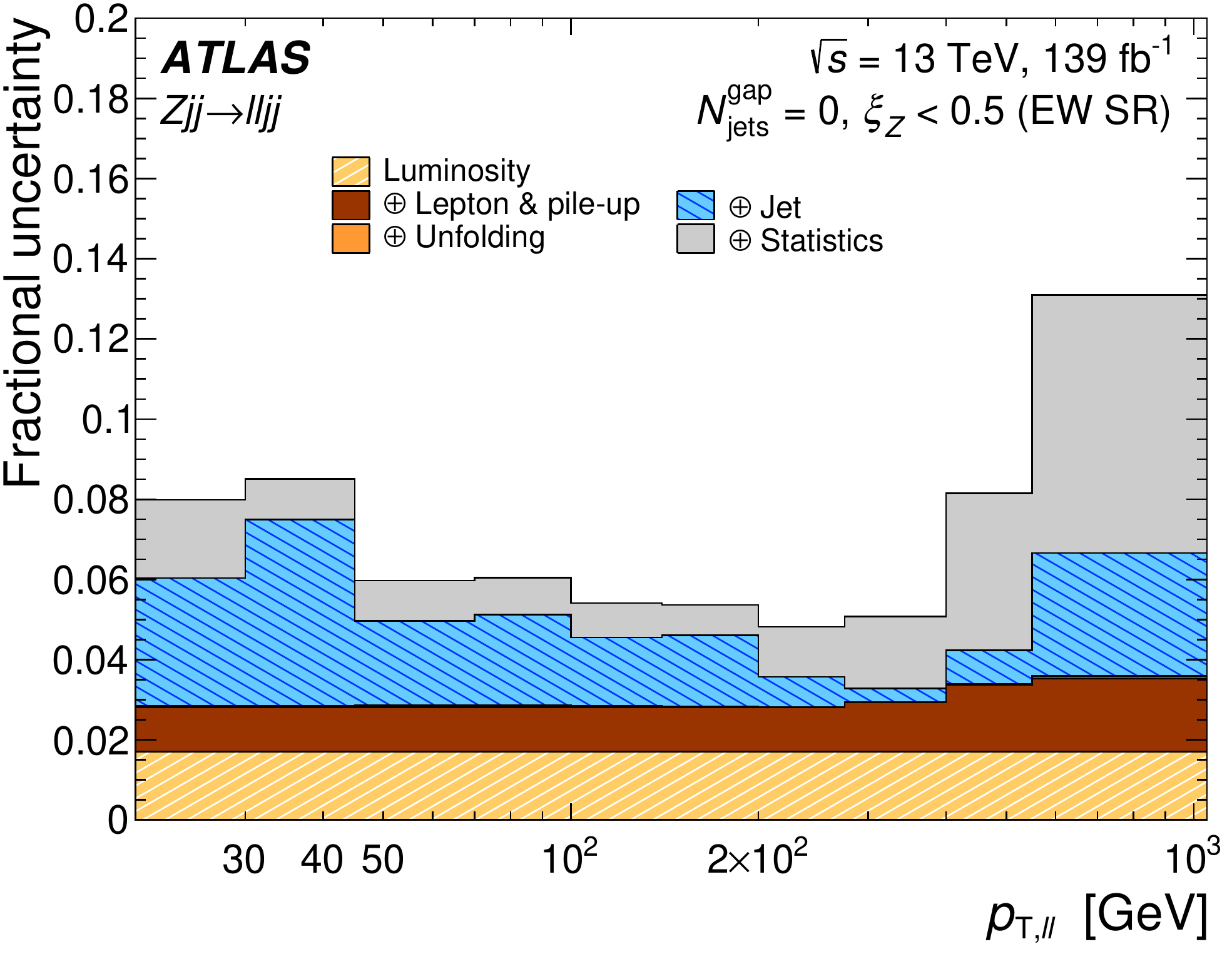}\\
\includegraphics[width=0.48\textwidth]{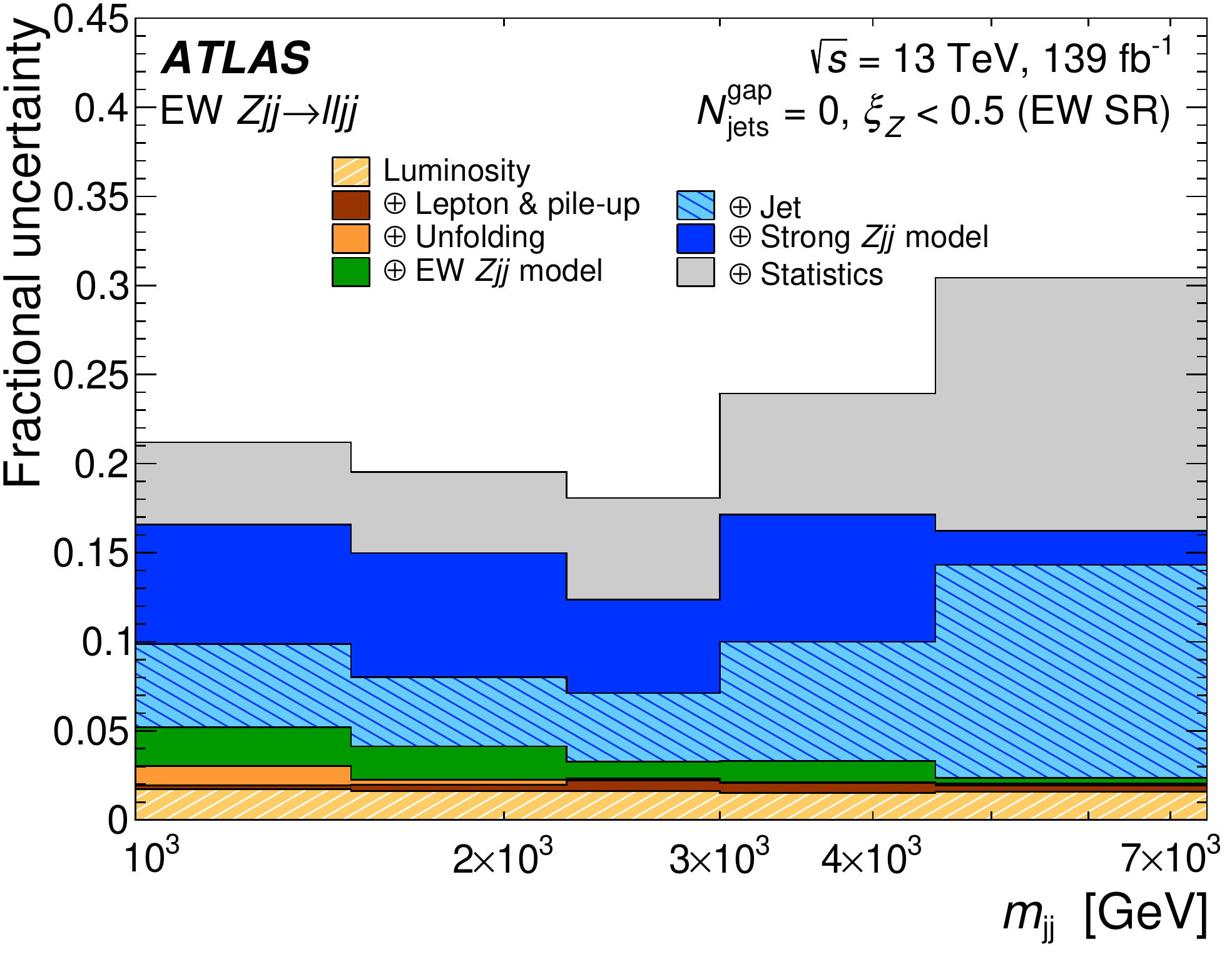}
\includegraphics[width=0.48\textwidth]{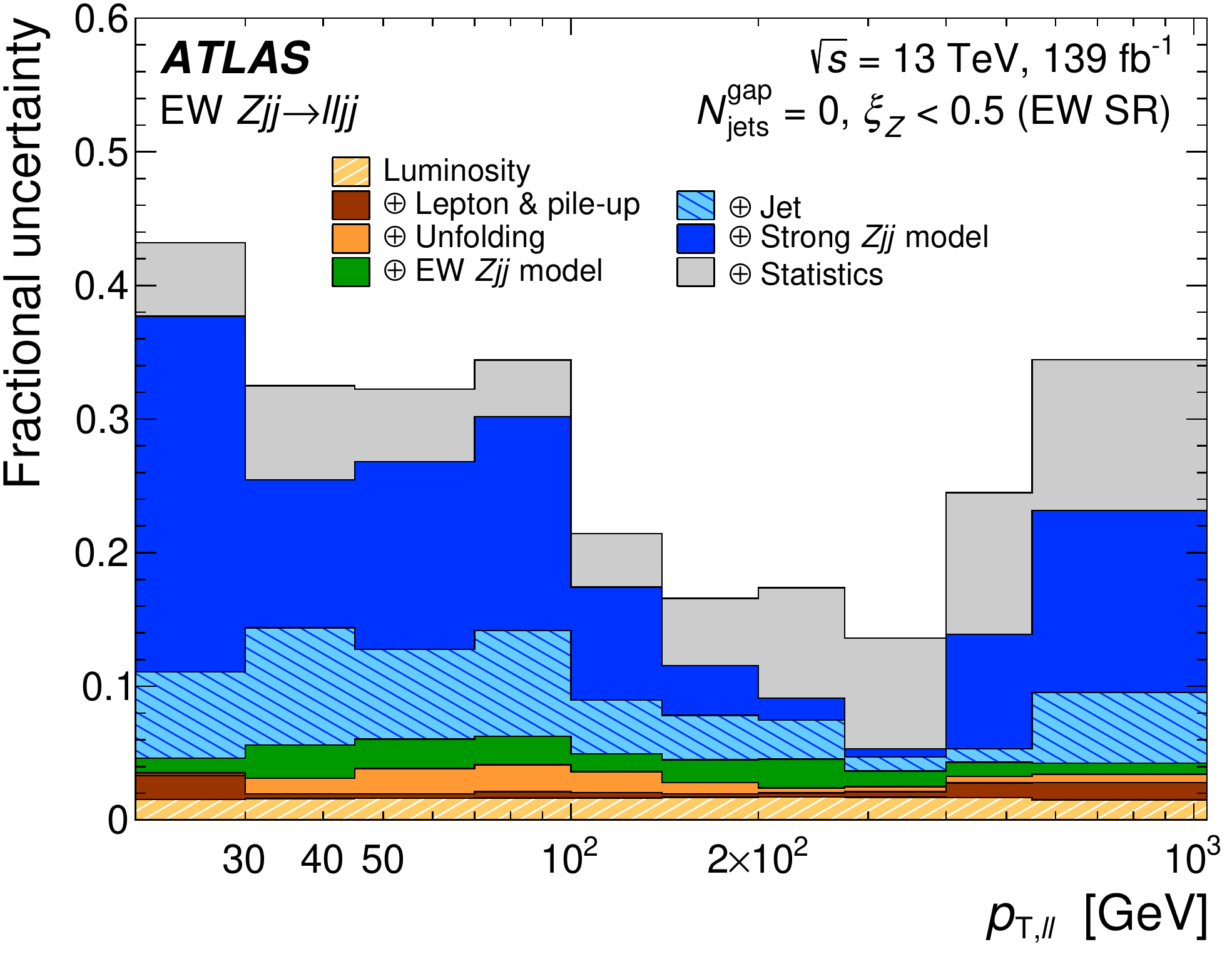}
 
\caption{Fractional uncertainty in the inclusive $Zjj$ measurement (top) and the EW $Zjj$ measurement (bottom) as a function of \mjj{} (left) and \pTll{} (right). Uncertainty sources are grouped in categories that are added in quadrature (denoted $\oplus$) to give the total uncertainty. The `EW $Zjj$ model' component includes the uncertainty on the EW $Zjj$ prediction and the impact of interference between the strong $Zjj$ and EW $Zjj$ processes. The `strong $Zjj$ model' uncertainty is dominated by the choice of generator used for the strong $Zjj$ prediction, but also includes the impact of renormalisation/factorisation scale variations and PDF set variations.}
\label{fig:syst}
\end{figure}
 
\subsection*{Uncertainties in the unfolding procedure}
 
Uncertainties associated with the unfolding procedure are estimated in two ways. First, the data are unfolded using a different simulation and the deviation  from the nominal result is taken as a systematic uncertainty. For the EW $Zjj$ differential cross-section measurements, the \sherpa{} EW $Zjj$ simulation is used in place of the \powpy{} EW $Zjj$ simulation. For the inclusive $Zjj$ differential cross-section measurements, the \mglo{} strong $Zjj$ simulation is used in place of the \sherpa{} strong $Zjj$ simulation. Second, a data-driven closure test is performed separately for each observable, to assess the potential bias in the unfolding method. In this approach, the particle-level distribution is reweighted such that it provides a better description of the data at detector level. The reweighted detector-level prediction is then unfolded using the response matrix and other corrections derived from nominal (unweighted) \powpy{} EW $Zjj$ simulation. The systematic uncertainty associated with the unfolding method is defined as the difference between the unfolded spectrum and the reweighted particle-level prediction; it is taken to be a symmetric uncertainty.
 
\subsection*{Summary of systematic uncertainties}
 
The final uncertainties in the differential cross-section measurements of EW $Zjj$ production and inclusive $Zjj$ production are shown in Figure \ref{fig:syst}. For the inclusive $Zjj$ measurements, the jet energy scale and jet energy resolution uncertainties dominate. However, for the EW $Zjj$ measurements the uncertainties associated with the modelling of the strong $Zjj$ process dominate.
 
\section{Results}
\label{sec:results}
 
\begin{figure}[t]
\includegraphics[width=0.48\textwidth]{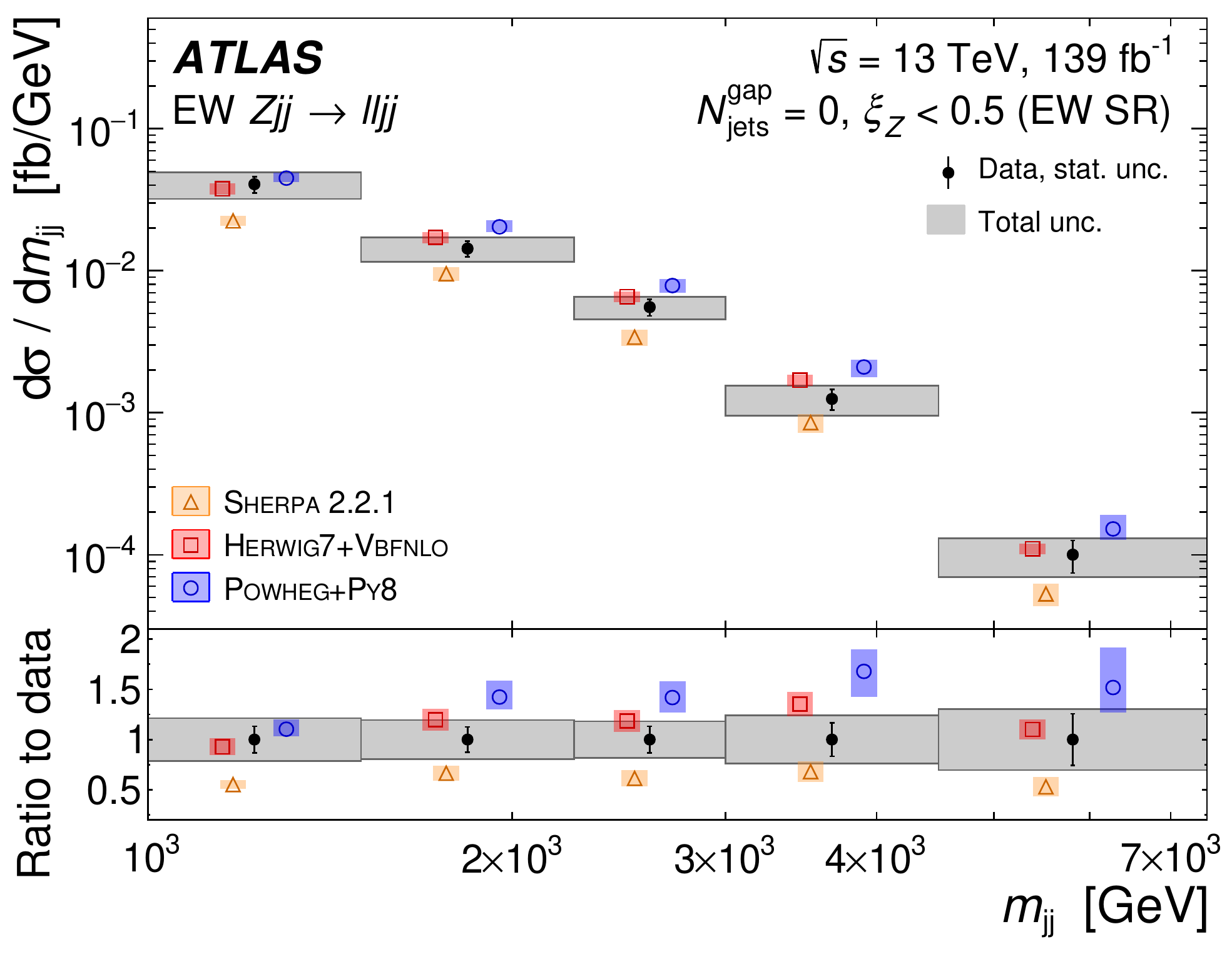}
\includegraphics[width=0.48\textwidth]{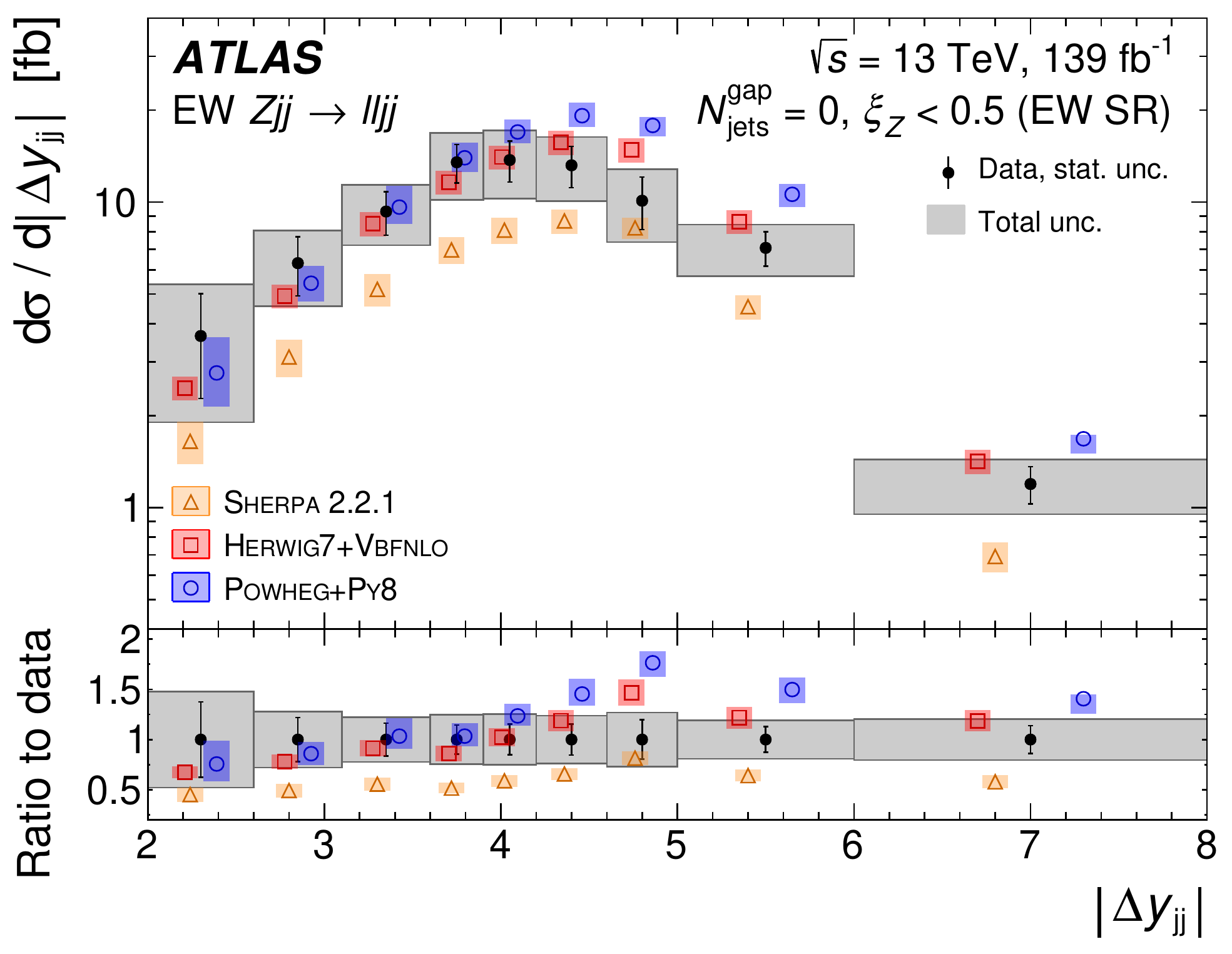}\\
\includegraphics[width=0.48\textwidth]{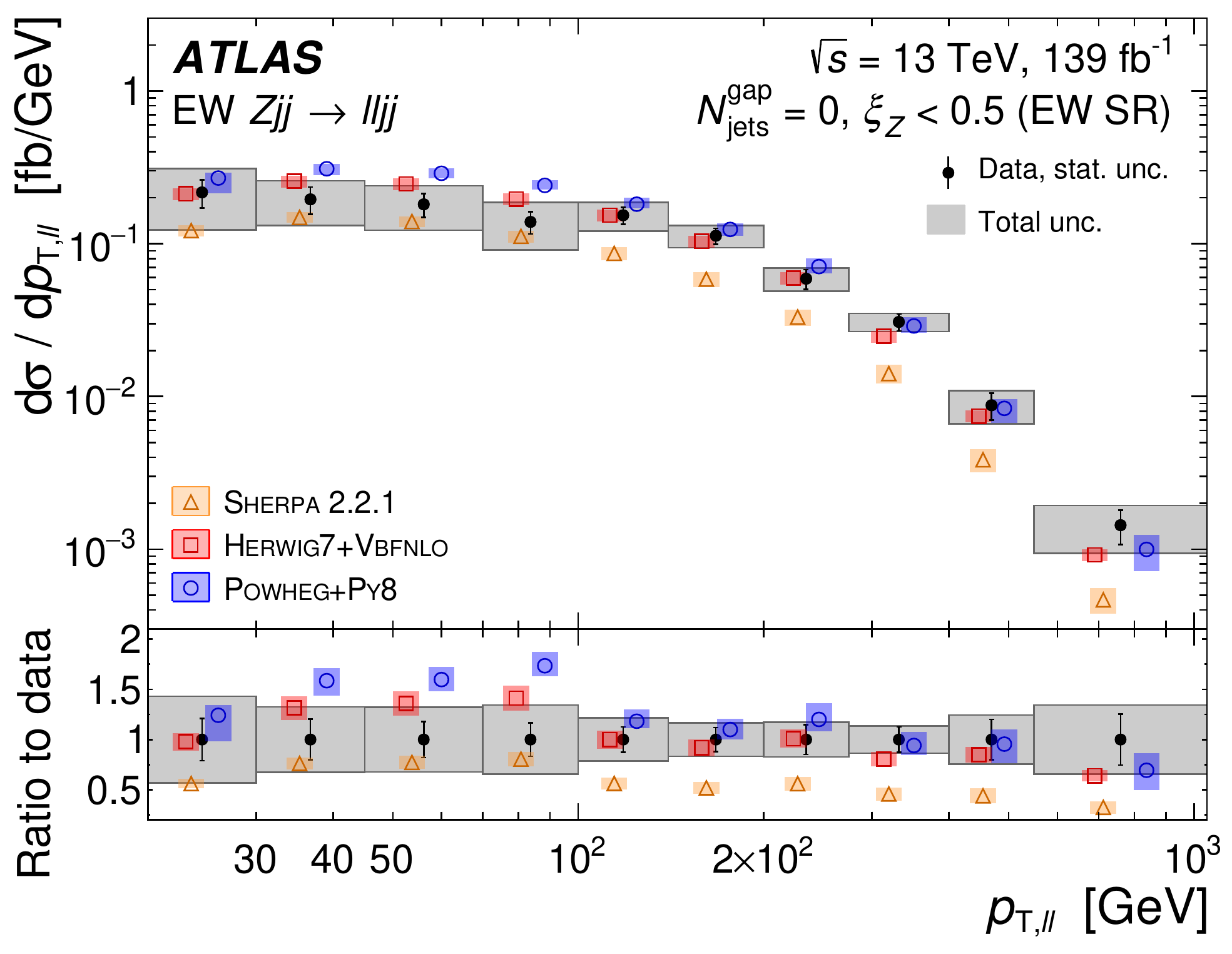}
\includegraphics[width=0.48\textwidth]{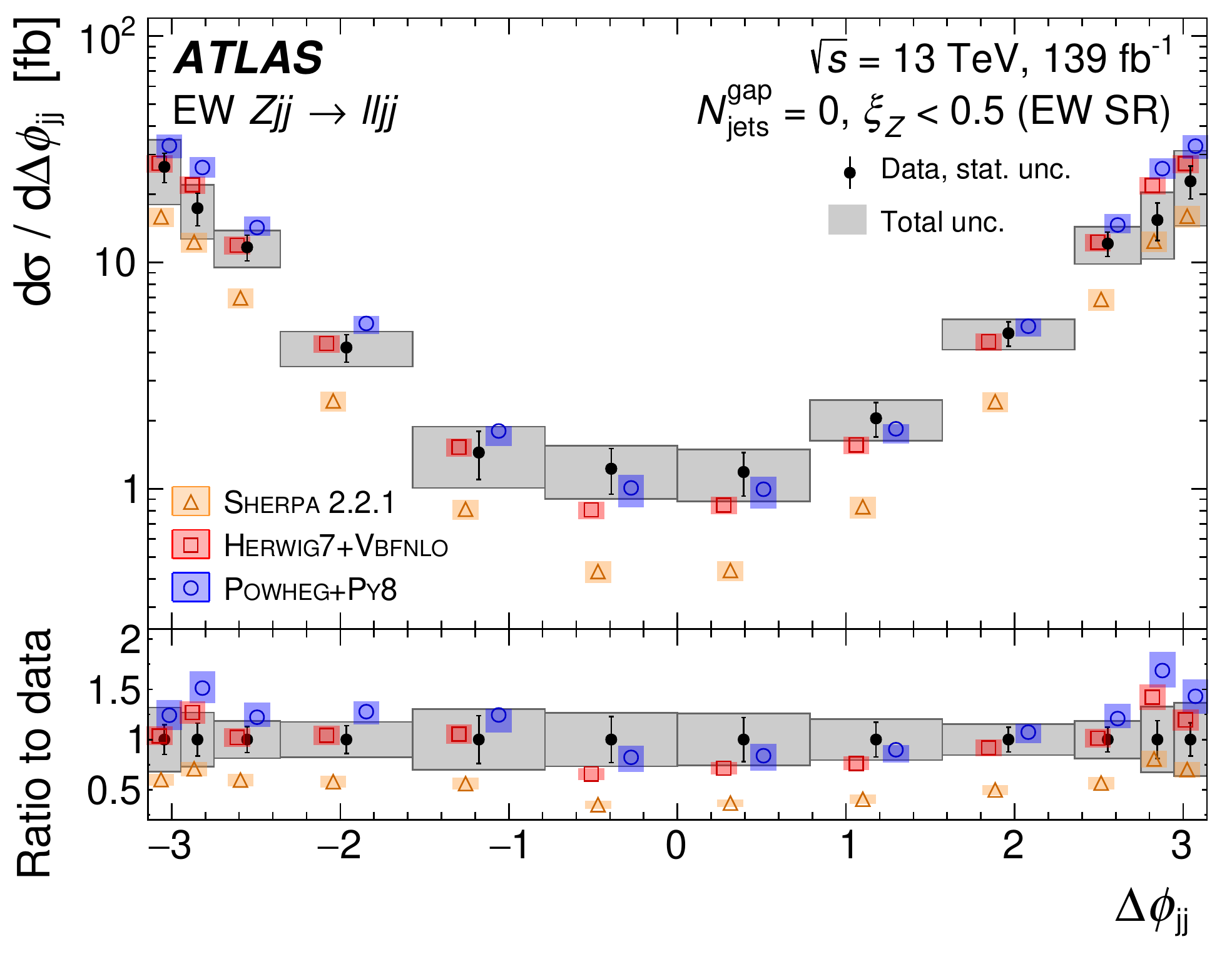}
\caption{Differential cross-sections for EW $Zjj$ production as a function of \mjj{} (top left), \Dyjj{} (top right), \pTll{} (bottom left) and \Dphijj{} (bottom right).
The unfolded data are shown as black points, with the statistical uncertainty represented by an error bar and the total uncertainty represented as a grey band. The data are compared with theoretical predictions produced by \hervbf{} (red points), \powpy{} (blue points) and \sherpa{} 2.2.1 (orange points). Uncertainty bands are shown for the three theoretical predictions. Each theory prediction is slightly offset from the bin center to avoid overlap.}
\label{fig:ewdiffxs}
\end{figure}

The differential cross-sections for EW $Zjj$ production as a function of \mjj{}, \Dyjj{}, \pTll{}, and \Dphijj{} are shown in Figure~\ref{fig:ewdiffxs} and are compared with theoretical predictions produced by \hervbf{}, \powpy{} and \sherpa{}. The set-up of the theoretical predictions is discussed in Section~\ref{sec:MC}.
The effects of scale uncertainties on the \hervbf{} prediction are estimated by independently varying the scale used in the matrix-element calculation and the scale associated with the parton shower by factors of 0.5 or 2.0.
The effects of scale uncertainties on the \sherpa{} prediction are estimated by varying the renormalisation and factorisation scales used in the matrix-element calculation independently by a factor of 0.5 or 2.0. The effects of scale uncertainties on the \powpy{} prediction are evaluated by independently varying the renormalisation, factorisation and resummation scales
by factors of 0.5 or 2.0. Additional uncertainties on the \powpy{} prediction associated with the parton-shower and underlying-event parameters in \pythia{} are evaluated using the AZNLO eigentune variations~\cite{Aad:2014xaa}. PDF uncertainties on the EW $Zjj$ predictions are estimated by reweighting the nominal sample to reproduce the 100 variations of the NNPDF3.0nnlo PDF sets and taking the RMS of these variations; the impact of PDF-related uncertainties on the EW $Zjj$ predictions are found to be much smaller than the impact of scale uncertainties.

In general, the \hervbf{} prediction is found to be in reasonable agreement with the data for all measured distributions. The \powpy{} prediction is found to overestimate the EW $Zjj$ cross-section at high \mjj{}, high \Dyjj{}, and intermediate \pTll{}. Furthermore, the central value of the \powpy{} prediction often does not agree with the \hervbf{} prediction, within the assigned theoretical uncertainties. A similar discrepancy between theoretical predictions was noted for EW $VVjj$ processes in Ref.~\cite{ATL-PHYS-PUB-2019-004} and was attributed to the set-up of the parton shower when matched to the matrix-element calculations. The \sherpa{} prediction significantly underestimates the measured differential cross-sections, due to a non-optimal setting of the colour flow~\cite{ATL-PHYS-PUB-2019-004}. However, despite the offset in normalisation, the shape of the measured distributions is reasonably well produced by \sherpa{}.
Under the assumption that there are no new physics contributions to the EW $Zjj$ process, the measurements presented in this article therefore constrain the choice of theoretical predictions that should be used for signal modelling in future measurements that exploit weak-boson fusion or weak-boson scattering. In particular, the EW $Zjj$ differential cross-section measurements can be used to determine the optimal parameter choices for each event generator, and poor parameter choices can be ruled out entirely.
 
\begin{figure}[t]
\includegraphics[width=0.48\textwidth]{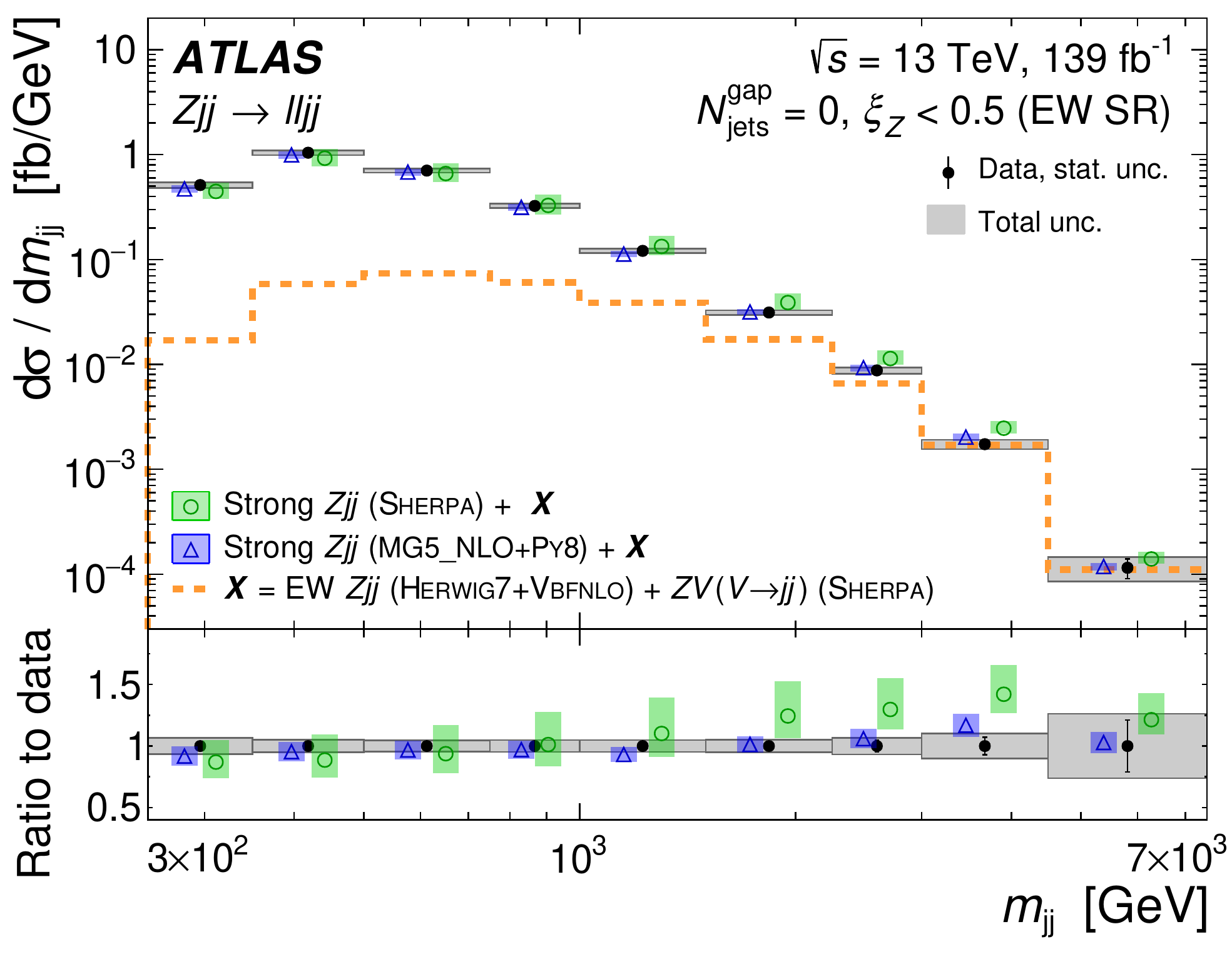}
\includegraphics[width=0.48\textwidth]{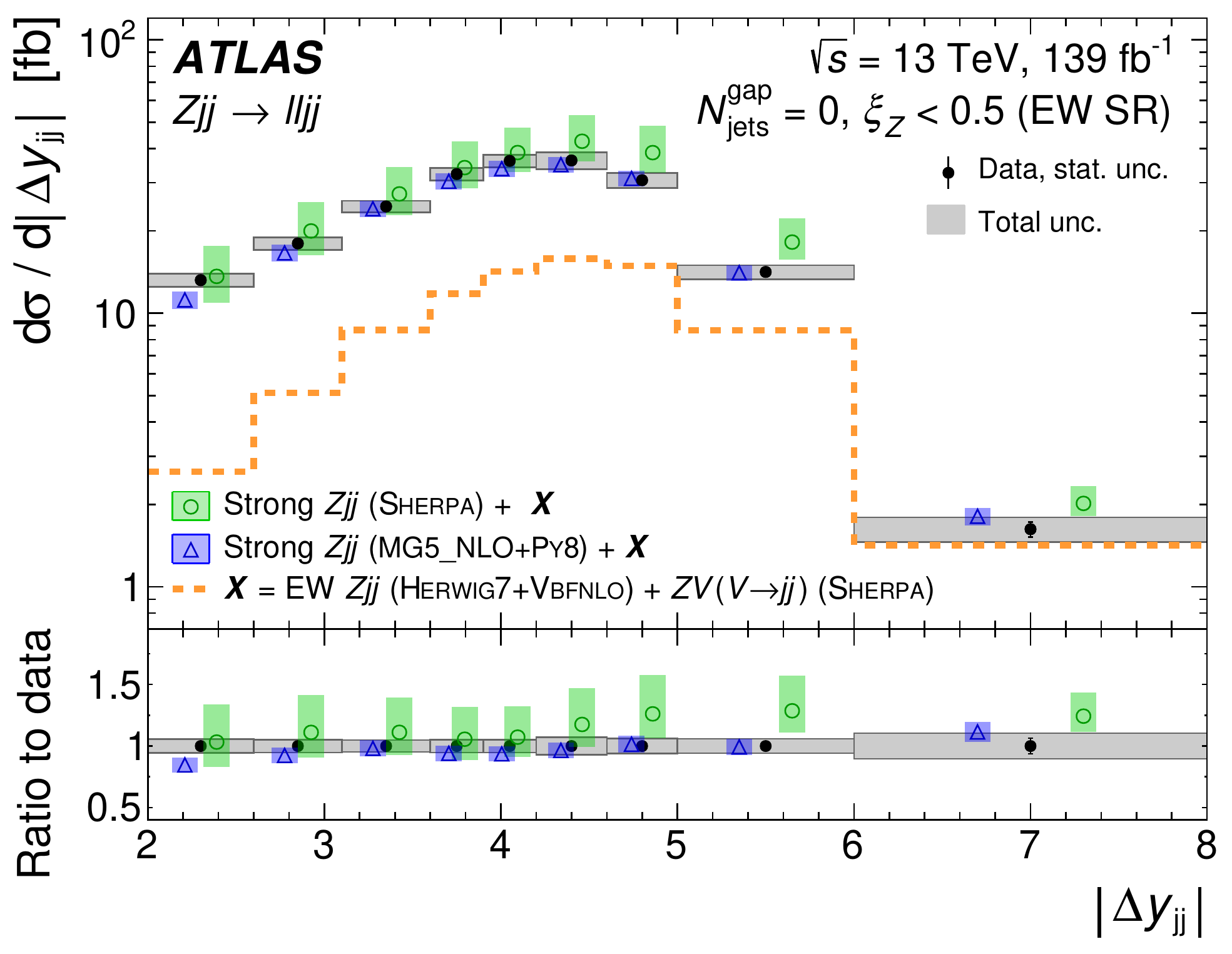}\\
\includegraphics[width=0.48\textwidth]{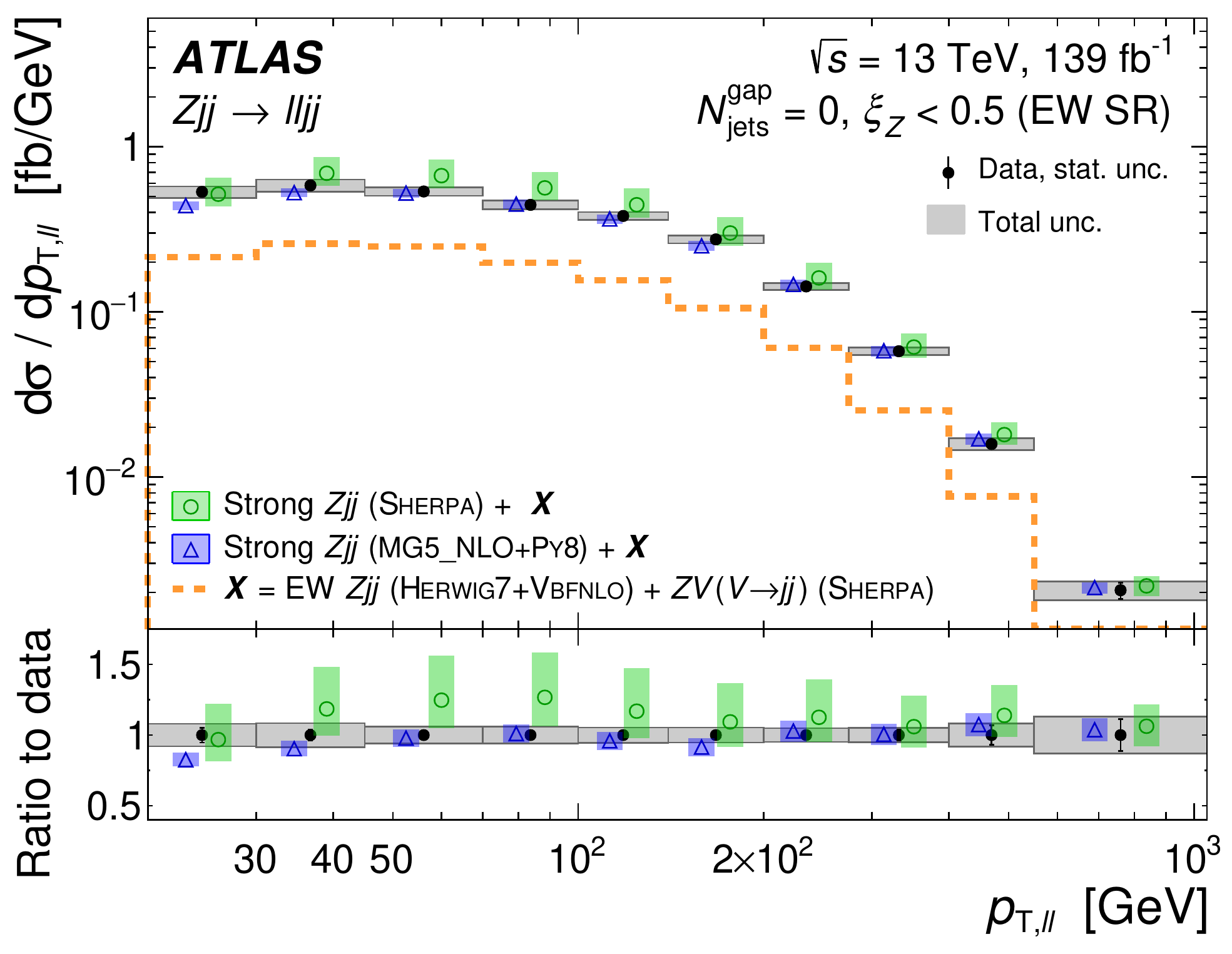}
\includegraphics[width=0.48\textwidth]{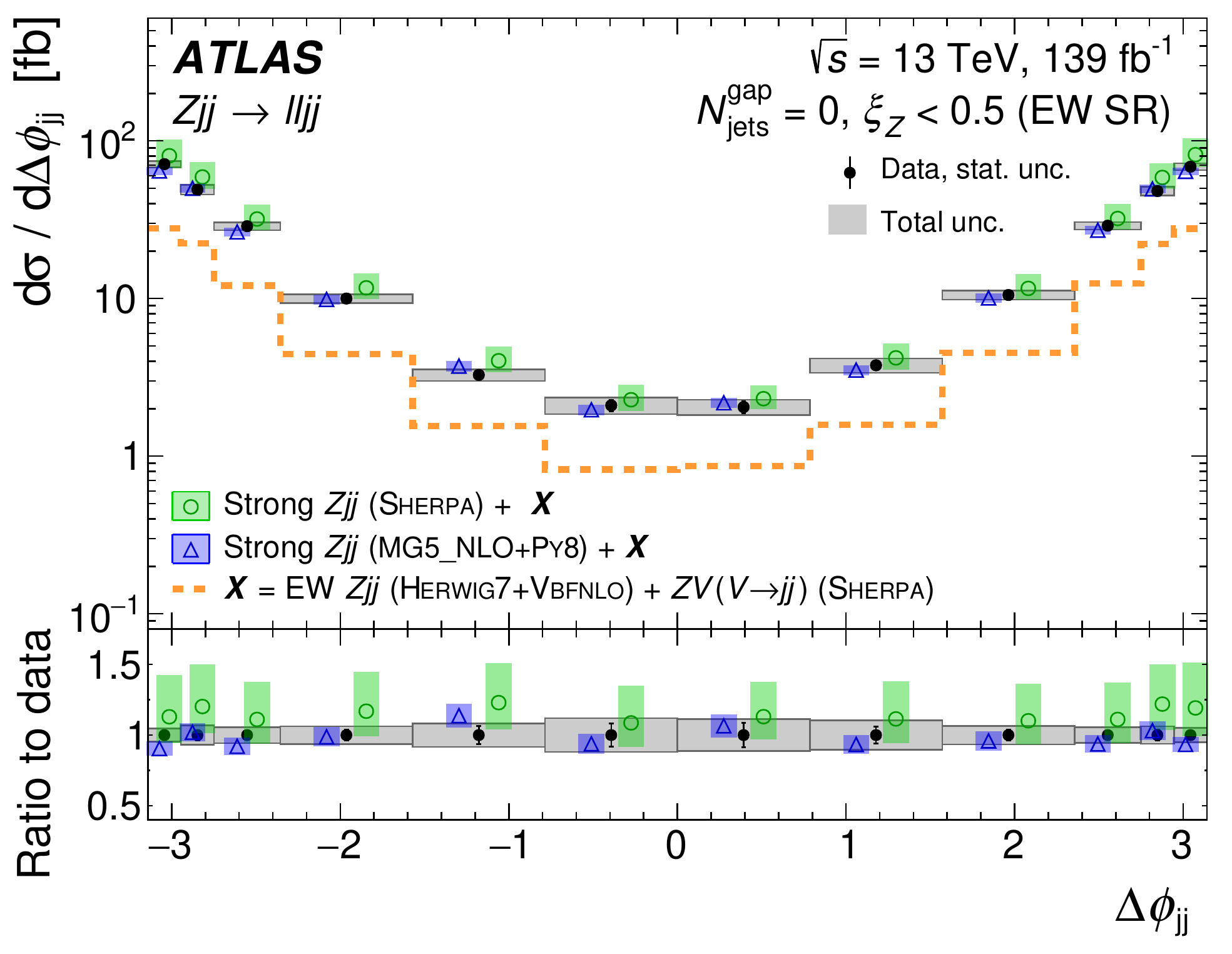}
\caption{Differential cross-sections measured in the SR for inclusive $Zjj$ production as a function of \mjj{} (top left), \Dyjj{} (top right), \pTll{} (bottom left) and \Dphijj{} (bottom right). The unfolded data are shown as black points, with the statistical uncertainty represented by an error bar and the total uncertainty represented as a grey band. The data are compared with theoretical predictions constructed from different strong $Zjj$ predictions provided by \sherpa{} (green) and \mgnlo{} (blue). Uncertainty bands are shown for the two theoretical predictions. Each theory prediction is slightly offset from the bin center to avoid overlap.}
\label{fig:incdiffxs}
\end{figure}
 
A fiducial cross-section for EW $Zjj$ production is calculated, by integrating the differential cross-section as a function of \mjj{}, and found to be
\begin{equation*}
\sigma_{\textrm{EW}} = 37.4 \pm 3.5 \,(\textrm{stat}) \pm 5.5\,(\textrm{syst})\, \,
\textrm{fb}.
\end{equation*}
This is in excellent agreement with the theoretical prediction from \hervbf{}, which is $39.5 \pm 3.4\,(\textrm{scale}) \pm 1.2\,(\textrm{PDF})$~fb.
 
\begin{figure}[t]
\includegraphics[width=0.48\textwidth]{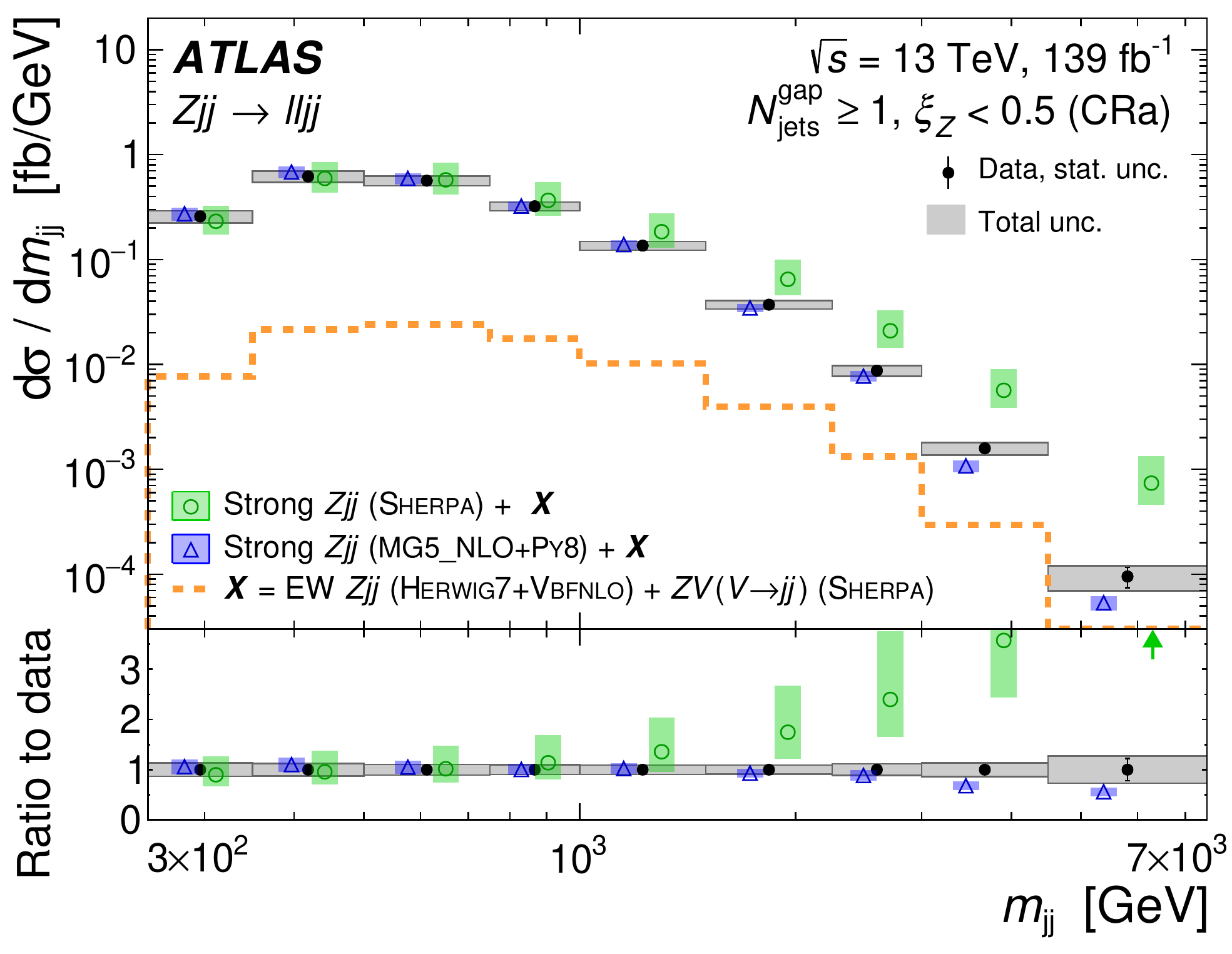}
\includegraphics[width=0.48\textwidth]{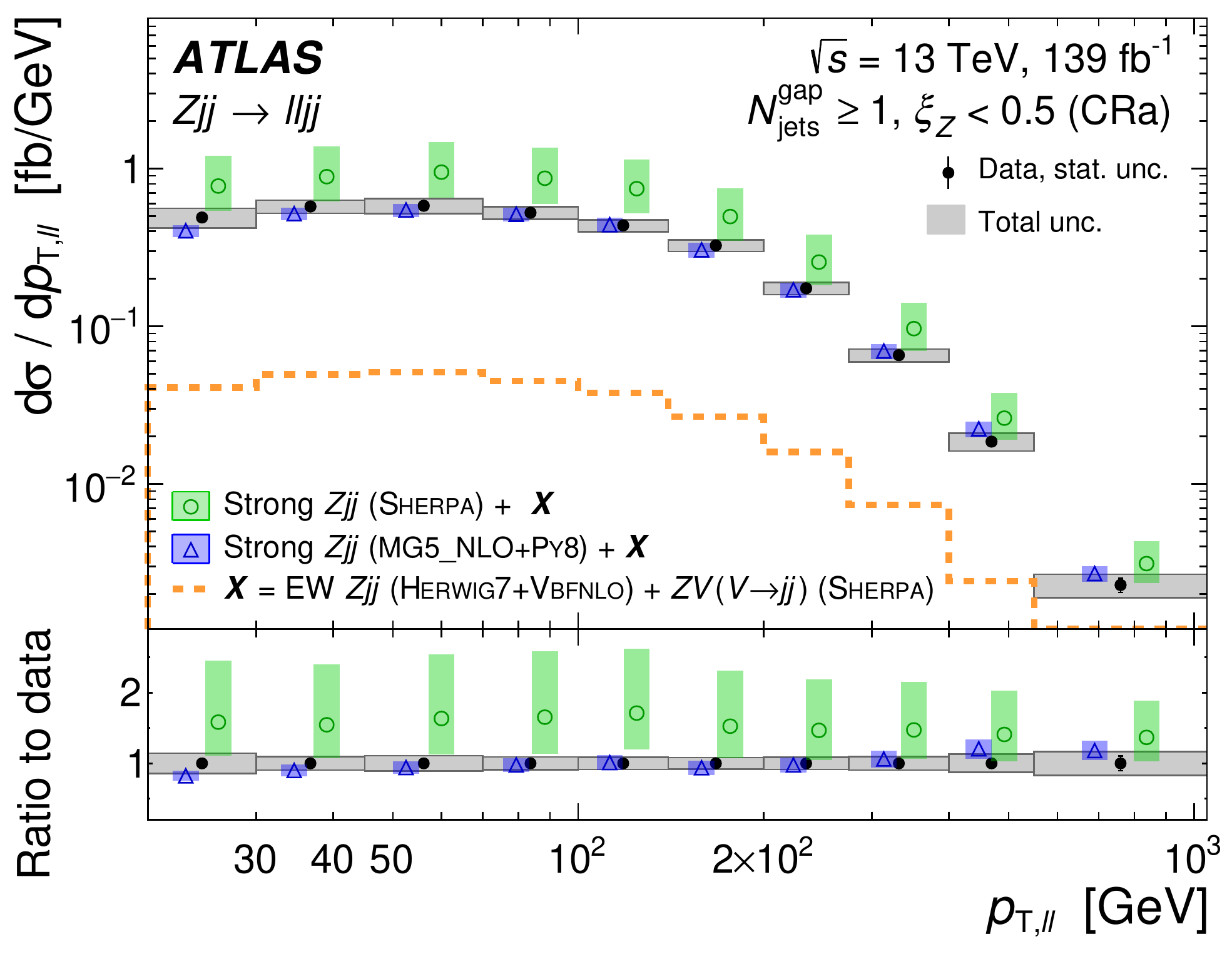}
\caption{Differential cross-sections measured in CRa for inclusive $Zjj$ production as a function of \mjj{} (left) and \pTll{} (right), where CRa is defined by $\Ngapjets \geq 1$ and $\Zcent<0.5$. The unfolded data are shown as black points, with the statistical uncertainty represented by an error bar and the total uncertainty represented as a grey band. The data are compared with theoretical predictions constructed from different strong $Zjj$ predictions provided by \sherpa{} (green) and \mgnlo{} (blue). Uncertainty bands are shown for the two theoretical predictions. Each theory prediction is slightly offset from the bin center to avoid overlap.}
\label{fig:incdiffxs2}
\end{figure}
 
Differential cross-sections for inclusive $Zjj$ production as a function of \mjj{}, \Dyjj{}, \pTll{} and \Dphijj{} are also measured in the signal and control regions that are used to extract the electroweak component. These measurements can be used to re-evaluate the electroweak contribution in the future, when new theoretical predictions for the strong $Zjj$ background presumably will become available. The differential cross-sections for inclusive $Zjj$ production measured in the SR as a function of \mjj{}, \Dyjj{}, \pTll{}, and \Dphijj{} are shown in Figure~\ref{fig:incdiffxs}. The differential cross-sections measured in CRa for inclusive $Zjj$ production as a function of \mjj{} and \pTll{} are shown in Figure~\ref{fig:incdiffxs2}. The data are compared with the strong $Zjj$ predictions provided by \sherpa{} and \mgnlo{}, augmented with the EW $Zjj$ contribution predicted by \hervbf{} and the $VZ$ contribution predicted by \sherpa{}. The effects of scale uncertainties on the strong $Zjj$ predictions dominate the overall uncertainty in each prediction and are estimated by independently varying the renormalisation and factorisation scales by factors of 0.5 and 2.0 (with six  variations considered for each generator). PDF uncertainties on the strong $Zjj$ predictions are estimated using the variations of the NNPDF PDF sets. The total uncertainty on the strong $Zjj$ predictions is taken to be the envelope of the scale variations added in quadrature with the PDF uncertainty. Overall, the data is best described when using the \mgnlo{} prediction for strong $Zjj$ production.
 
The unfolded differential cross-sections for EW $Zjj$ production and inclusive $Zjj$ production are documented in tabular form in Appendix \ref{sec:appA}. The data are also provided in the HEPDATA repository~\cite{Maguire:2017ypu} and a Rivet analysis routine is provided~\cite{Buckley:2010ar,Bierlich:2019rhm}.
 
\section{Constraints on anomalous weak-boson self-interactions}
\label{sec:eft}
In this section, the measured EW $Zjj$ differential cross-sections are used to constrain extensions to the SM that produce anomalous weak-boson self-interactions. The anomalous interactions are introduced using an effective field theory (EFT), for which the effective Lagrangian is given by
\begin{equation}
\label{eq:eft}
\mathcal{L}_{\mathrm{eff}} = \mathcal{L}_{\mathrm{SM}} + \sum_i \frac{c_i}{\Lambda^2} \mathcal{O}_i ,
\end{equation}
where $\mathcal{L}_{\mathrm{SM}}$ is the SM Lagrangian, the $\mathcal{O}_i$ are dimension-six operators in the Warsaw basis \cite{Grzadkowski:2010es}, and the ${c_i}/\Lambda^2$ are Wilson coefficients that describe the strength of the anomalous interactions induced by those operators. Constraints are placed on two CP-even operators ($\mathcal{O}_W$, $\mathcal{O}_{HWB}$) and two CP-odd operators ($\tilde{\mathcal{O}}_W$,  $\tilde{\mathcal{O}}_{HWB}$), which are known to produce anomalous $WWZ$ interactions.
 
Theoretical predictions are constructed for the EW $Zjj$ process using the effective Lagrangian in Equation~\ref{eq:eft}. The amplitude for the EW $Zjj$ process is split into a SM part, $\mathcal{M}_\mathrm{SM}$, and a dimension-six part, $\mathcal{M}_\mathrm{d6}$, which contains the anomalous interactions. The differential cross-section or squared amplitude then has three contributions
\begin{equation}
\label{eq:medim6}
|\mathcal{M}|^2 = |\mathcal{M}_\mathrm{SM}|^2 +  2\,\mathrm{Re}(\mathcal{M}^*_{\mathrm{SM}}\mathcal{M}^{\,}_{\mathrm{d6}})+
|\mathcal{M}_{\mathrm{d6}}|^2,
\end{equation}
namely a pure SM term $|\mathcal{M}_\mathrm{SM}|^2$, a pure dimension-six term $|\mathcal{M}_{\mathrm{d6}}|^2$, and a term that contains the interference between the SM and dimension-six amplitudes, $ 2\,\mathrm{Re}(\mathcal{M}^*_{\mathrm{SM}}\mathcal{M}^{\,}_{\mathrm{d6}})$.
The constraints on the dimension-six operators presented in this section are derived both with and without the pure dimension-six terms included in the theoretical prediction. This tests whether the results are robust against  missing dimension-eight operators in the EFT expansion.
 
The pure-SM contribution to the EW $Zjj$ differential cross-sections in Equation~\ref{eq:medim6} is taken to be the prediction from \hervbf{}. The contributions arising from the interference and pure dimension-six terms are generated at leading order in perturbative QCD using \mg{}$+$\pythia{}, with the interactions from the dimension-six operators provided by the SMEFTSim package \cite{Brivio:2017btx}. The A14 set of tuned parameters is used for parton showering, hadronisation and multiple parton scattering. To account for missing higher-order QCD corrections, the interference and pure dimension-six contributions are scaled using a bin-dependent $K$-factor, which is defined by the ratio of pure-SM EW $Zjj$ differential cross-sections predicted by \hervbf{} and \mg{}$+$\pythia{} in each bin.
 
\begin{figure}[t]
\includegraphics[width=0.9\textwidth]{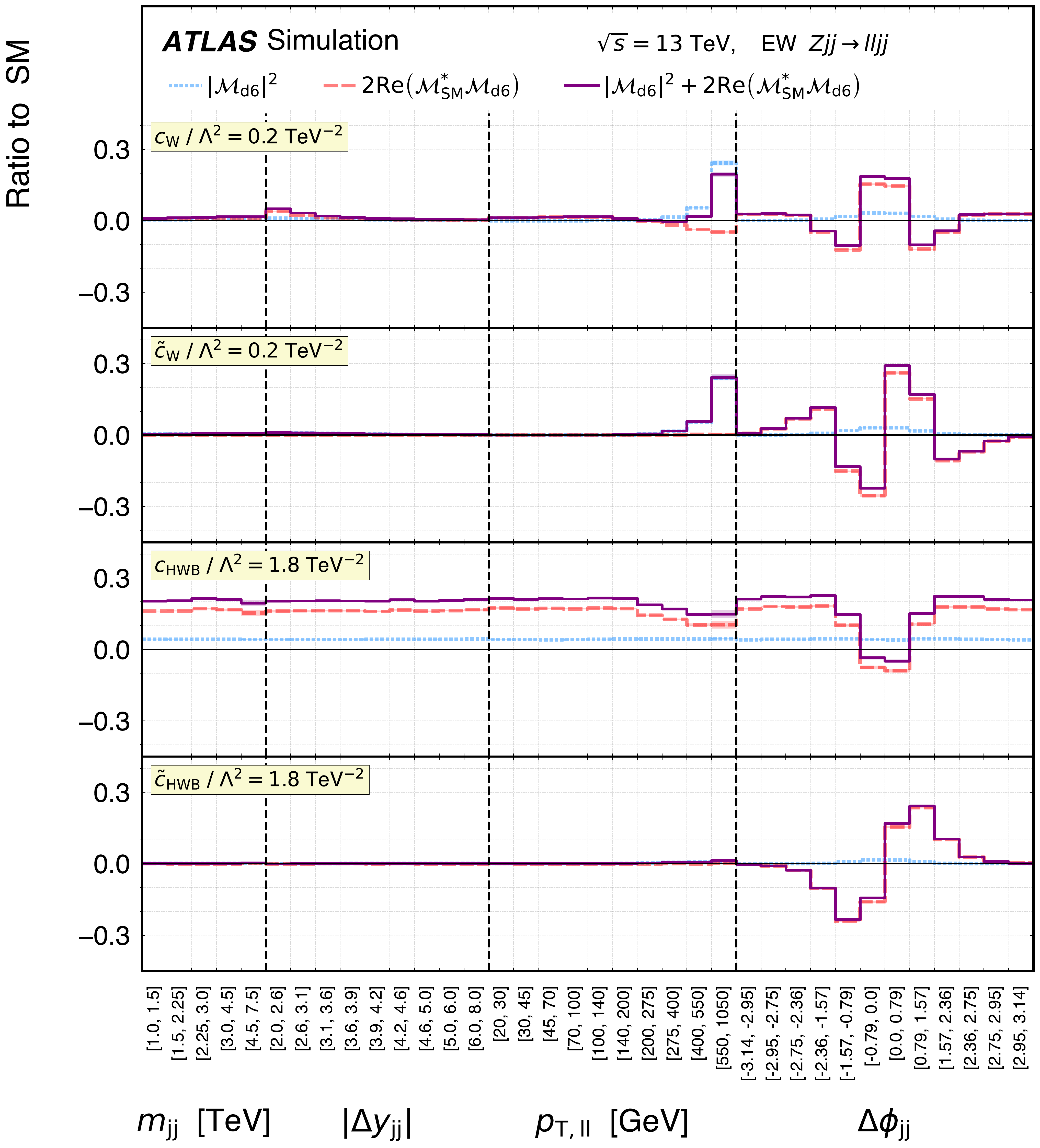}
\caption{Impact of the $\mathcal{O}_W$, $\tilde{\mathcal{O}}_W$, $\mathcal{O}_{HWB}$ and $\tilde{\mathcal{O}}_{HWB}$ operators on the EW $Zjj$ differential cross-sections. The expected contributions from the pure dimension-six term ($|\mathcal{M}_{\mathrm{d6}}|^2$) and from the interference between the SM and dimension-six amplitudes ($2\,\mathrm{Re}(\mathcal{M}^*_{\mathrm{SM}}\mathcal{M}^{\,}_{\mathrm{d6}})$)  are shown relative to the pure-SM prediction and represented as dotted and dashed lines, respectively. The total contribution to the EW $Zjj$ cross-section is shown as a solid line.
}
\label{fig:wilson_coeff_impact}
\end{figure}

The impact of the interference and pure dimension-six contributions to the EW $Zjj$ differential cross-sections is shown relative to the pure SM contribution in Figure \ref{fig:wilson_coeff_impact}. The Wilson coefficients were chosen to be ${c_W}/\Lambda^2 = 0.2$~\TeV{}$^{-2}$,  ${\tilde{c}_W}/\Lambda^2 = 0.2$~\TeV{}$^{-2}$, ${c_{HWB}}/\Lambda^2 = 1.8$~\TeV{}$^{-2}$ and  ${\tilde{c}_{HWB}}/\Lambda^2 = 1.8$~\TeV{}$^{-2}$. For the CP-even $\mathcal{O}_W$ operator, the high-\pTll{} region is particularly sensitive to the anomalous interactions, a feature that was seen in previous studies for EW $Vjj$ production \cite{Aaboud:2017fye,Sirunyan:2017jej}.
The pure dimension-six contributions to the cross-section dominate in this region. The \Dphijj{} observable is also found to be very sensitive to the anomalous interactions induced by the $\mathcal{O}_W$ operator, but in this observable the interference contribution dominates. For the CP-even $\mathcal{O}_{HWB}$ operator, the interference contribution dominates in all distributions, with the \Dphijj{} observable showing the largest kinematic dependences. For the CP-odd operators, the interference contribution is zero in the parity-even observables (\mjj{}, \Dyjj{}, \pTll{}). However, the interference contribution produces large asymmetric effects in the parity-odd \Dphijj{} observable. Constraints are therefore placed on Wilson coefficients using the measured EW $Zjj$ differential cross-section as a function of \Dphijj{}.
 
The measured differential cross-section as a function of \Dphijj{} and the corresponding EFT-dependent theoretical prediction are used to define a likelihood function. Statistical correlations amongst the bins of \Dphijj{} in the EW $Zjj$ measurement are estimated using a bootstrap procedure (as outlined in Section~\ref{sec:unfolding}) and included in the likelihood function. Each source of systematic uncertainty in the measurement is implemented as a Gaussian-constrained nuisance parameter and is hence treated as fully correlated across bins, but uncorrelated with other uncertainty sources. Uncertainties in the theoretical prediction are also implemented as Gaussian-constrained nuisance parameters. These uncertainties include (i) scale and PDF uncertainties in the \hervbf{} prediction, (ii) an additional shape uncertainty defined by the difference between the \hervbf{} and \powpy{} predictions, and (iii) an uncertainty in the bin-dependent $K$-factor that arises from finite statistics in the MC samples.
The confidence level at each value of Wilson coefficient is calculated using the profile-likelihood test statistic~\cite{Feldman:1997qc}, which is assumed to be distributed according to a $\chi^2$ distribution with one degree of freedom following from Wilks' theorem~\cite{wilks1938}.
This allows the 95\% confidence intervals to be constructed for each Wilson coefficient. The expected 95\% coverage is validated by generating pseudo-experiments, both around the SM hypothesis and at various points in the EFT parameter space.
 
The expected and observed 95\% confidence intervals on the  dimension-six operators are shown in Table~\ref{Table::ReInterp::Limits}. For each Wilson coefficient, confidence intervals are shown when including or not-including the pure dimension-six contribution in the theoretical prediction.
As expected from Figure \ref{fig:wilson_coeff_impact}, the 95\% confidence intervals are almost unaffected if the pure dimension-six contributions are excluded from the theoretical prediction. The compatibility with the SM hypothesis is found to be poor for one of the operators ($\mathcal{\tilde{O}}_{HWB}$), with a corresponding $p$-value of 1.6\%. The probability that fluctuations around the SM prediction cause this feature when constraining these four Wilson coefficients is investigated using  pseudo-experiments. For each pseudo-experiment, the $p$-value for the compatibility with the SM hypothesis is calculated for each of the four Wilson coefficients. The fraction of pseudo-experiments that produce a $p$-value lower than 1.6\% for any of the Wilson coefficients is found to be 6.2\%.

The 95\% confidence intervals for the CP-even and CP-odd operators can be translated into the HISZ basis~\cite{Hagiwara:1993ck,Degrande:2012wf,Baglio:2018bkm} and be compared with previous ATLAS and CMS results. The observed and expected 95\% confidence intervals for the $c_{WWW}/\Lambda^2$ Wilson coefficient are [--2.7, 5.8]~\TeV{}$^{-2}$ and [--4.4, 4.1]~\TeV{}$^{-2}$, respectively. The observed and expected 95\% confidence intervals for the $\tilde{c}_{WWW}/\Lambda^2$ Wilson coefficient are [--1.6, 2.0]~\TeV{}$^{-2}$ and [--1.7, 1.7]~\TeV{}$^{-2}$ respectively. These confidence intervals are slightly weaker in sensitivity than the confidence intervals derived using measurements of $W^+W^-$ production at ATLAS ~\cite{Aaboud:2019nkz}, $WZ$ production at CMS \cite{Sirunyan:2019gkh}, and measurements of EW $Zjj$ production at CMS~\cite{Sirunyan:2017jej}. However, the constraints from those previous measurements were obtained with the pure dimension-six terms included in the theoretical prediction and therefore are more sensitive to the impact of missing higher-dimensional operators in the effective field theory expansion. For example, the constraints obtained from measurements of $WW$ and $WZ$ production are shown to weaken by a factor of ten when the pure dimension-six terms are excluded, due to helicity selection rules that suppress the interference contribution in diboson processes~\cite{Azatov:2016sqh,Falkowski:2016cxu}. Similarly, the constraints obtained from EW $Zjj$ production at CMS were obtained from a fit to the \pTll{} distribution, which can be dominated by the pure dimension-six terms as shown in Figure \ref{fig:wilson_coeff_impact}. The results presented in this paper therefore have two novel aspects. First, they constitute the strongest limits when pure dimension-six contributions are excluded from the theoretical prediction. Second, the limits are derived from a parity-odd observable, which is sensitive to the interference between the SM and CP-odd amplitudes and is therefore a direct test of CP invariance in the weak-boson self-interactions~\cite{Aad:2016nal}.
 
\begin{table}[t]
\centering
\begin{tabular}{cccccc}
\hline \hline
Wilson        &  Includes   & \multicolumn{2}{c}{95\% confidence interval [\TeV{}$^{-2}$]} & $p$-value (SM)\\
coefficient  &  $|\mathcal{M}_{\mathrm{d6}}|^2$     &   Expected    &   Observed  & \\
\hline
$c_{W}/ \Lambda^2$ &  no   &  $[-0.30,~0.30]$   &  $[-0.19,~0.41]$  & 45.9\% \\
&  yes   &  $[-0.31,~0.29]$   &  $[-0.19,~0.41]$ & 43.2\% \\
\hline
${\tilde c}_{W} / \Lambda^2$  &  no   &  $[-0.12,~0.12]$  &  $[-0.11,~0.14]$ & 82.0\% \\
&  yes      &  $[-0.12,~0.12]$  &  $[-0.11,~0.14]$  & 81.8\% \\
\hline
$c_{HWB}/ \Lambda^2$  &  no  &  $[-2.45,~2.45]$  &  $[-3.78,~1.13]$ & 29.0\% \\
&  yes     &  $[-3.11,~2.10]$  &  $[-6.31,~1.01]$   & 25.0\%\\
\hline
${\tilde c}_{HWB}/ \Lambda^2$  &  no  &   $[-1.06,~1.06]$  &  $[0.23,~2.34]$ & 1.7\% \\
&  yes     &  $[-1.06,~1.06]$  &  $[0.23,~2.35]$ & 1.6\% \\
\hline \hline
\end{tabular}
\caption{Expected and observed 95\% confidence interval for the four Wilson coefficients, using fits to the EW $Zjj$ differential cross-section measured as a function of \Dphijj{}. Results are presented when including or excluding the pure dimension-six contributions to the EFT prediction. The $p$-value quantifying the compatibility with the SM hypothesis is also shown for each Wilson coefficient. The global $p$-value associated with constraining these four Wilson coefficients is investigated using pseudo-experiments, as outlined in the text.}
\label{Table::ReInterp::Limits}
\end{table}
 
\section{Conclusion}
\label{sec:conclusion}
Differential cross-section measurements for the electroweak production of dijets in association with a $Z$ boson (EW $Zjj$) are presented for the first time, using proton--proton collision data collected by the ATLAS experiment at a centre-of-mass energy of $\sqrt{s}=13$~\TeV{} and with an integrated luminosity of 139~fb$^{-1}$. This process is defined by the $t$-channel exchange of a weak vector boson and is extremely sensitive to the vector-boson fusion process. Measurements of electroweak $Zjj$ production therefore probe the $WWZ$ interaction and provide a fundamental test of the SU(2)$_\text{L} \times$U(1)$_\text{Y}$ gauge symmetry of the Standard Model of particle physics.
 
The differential cross-sections for EW $Zjj$ production are measured in the $Z\rightarrow \ell^+\ell^-$ decay channel ($\ell=e,\mu$) as a function of four observables: the dijet invariant mass, the rapidity interval spanned by the two jets, the signed azimuthal angle between the two jets, and the transverse momentum of the dilepton pair. The data are corrected for detector inefficiency and resolution using an iterative Bayesian method and are compared to state-of-the-art theoretical predictions from \powheg{}+\pythia{}, \hervbf{} and \sherpa{}. The data favour the prediction from \hervbf{}. \powheg{}+\pythia{} predicts too large a cross-section at high values of dijet invariant mass, at large dijet rapidity intervals, and at intermediate values of dilepton transverse momentum. \sherpa{} predicts too small a cross-section across the measured phase space. Differential cross-section measurements for inclusive $Zjj$ production are also provided in the signal and control regions used to extract the electroweak component.
 
The detector-corrected measurements are used to search for signatures of anomalous weak-boson self-interactions using the framework of a dimension-six effective field theory. The signed azimuthal angle between the two jets is found to be the most sensitive observable when examining the impact of both the CP-even and CP-odd dimension-six operators. The dimension-six operators are found to be primarily constrained by the contribution to the cross-section from the interference between the SM and dimension-six scattering amplitudes. This makes the results less sensitive to missing higher-order operators in the effective field theory expansion when compared to previous results that search for anomalous weak-boson self-interactions. Furthermore, all limits are derived from a parity-odd observable, which is sensitive to the interference between the SM and CP-odd amplitudes and is therefore a direct test of charge conjugation and parity invariance in the weak-boson self-interactions.
 
\appendix
 
\section{Validation of electroweak extraction methodology}
\label{sec:appV}
 
The method used to extract the EW $Zjj$ event yield uses three control regions to constrain the modelling of the strong $Zjj$ background process (see Sec.~\ref{sec:ewextraction}). In this appendix, additional details and validations of the method are presented. First, the EW yields extracted with different strong $Zjj$ predictions is presented. The spread of the extracted yields constitutes the dominant modelling uncertainty in the measured EW $Zjj$ differential cross-sections. Second, EW yields are presented for variations of the electroweak extraction method.
 
\subsection*{Impact of strong $Zjj$ generator choice}
 
\begin{figure}[t]
\includegraphics[width=\textwidth]{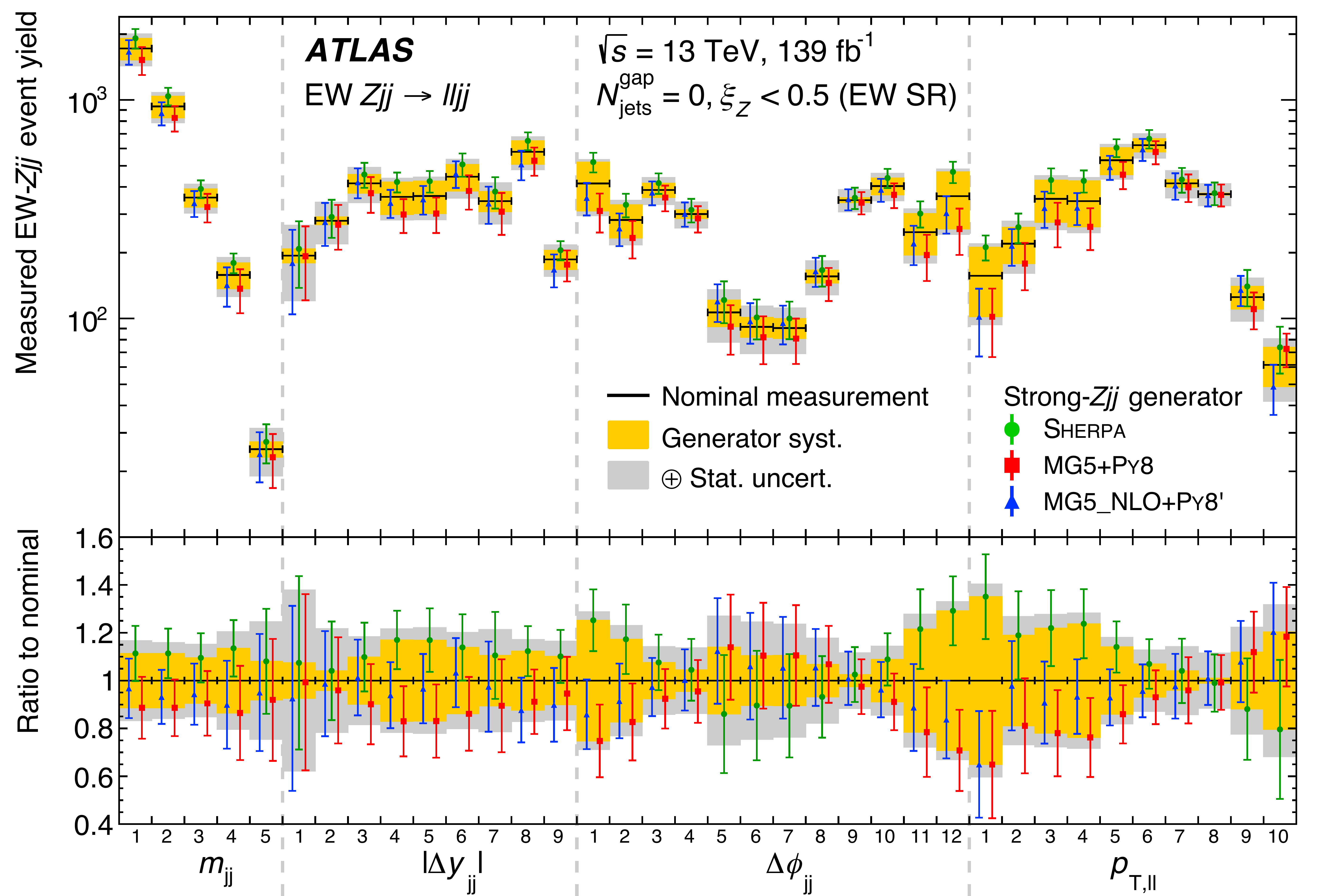}
\caption{ EW $Zjj$ event yields obtained when using the \sherpa{}, \mgnloR{} and \mglo{} predictions for the strong $Zjj$ process. The extracted EW $Zjj$ yields are presented for each bin of the four observables (\mjj{}, \Dyjj{}, \Dphijj{} and \pTll{}). The nominal EW $Zjj$ measurement is defined as the midpoint of the three generator-specific measurements, with a  `generator choice' systematic defined by the envelope. Experimental systematic uncertainties are not shown (they tend to be small as can be seen in Figure~\ref{fig:syst}).}
\label{fig:ewyield_genmodel}
\end{figure}

Figure~\ref{fig:ewyield_genmodel} shows the EW $Zjj$ event yields extracted using the \sherpa{}, \mgnloR{} and \mglo{} event generators to predict the strong $Zjj$ background. The results are presented for each bin of the four measured distributions (\mjj,\Dyjj, \pTll{} and \Dphijj). Also shown is the nominal measurement, the central value of which is taken to be the midpoint of the EW $Zjj$ event yields extracted using the three strong $Zjj$ predictions. The EW $Zjj$ yields obtained using \sherpa{} for the strong $Zjj$ process are typically the largest, while those obtained using \mglo{} for the strong $Zjj$ process are typically the  lowest. The uncertainty on the nominal measurement due to strong $Zjj$ generator choice is defined as the envelope of the EW $Zjj$ event yields obtained using the three strong $Zjj$ predictions.
 
\subsection*{Variations of the electroweak extraction method}
 
The degree to which the measured EW $Zjj$ differential cross-sections depend on the electroweak extraction method is investigated in this section. First, three variations are applied to the nominal likelihood-based extraction method presented in Sec.~\ref{sec:ewextraction}. These are:
\begin{enumerate}
\item \textit{Inverted Control Regions}: In this variation, CRa and CRc are swapped in  Eq.~\ref{eq:strongConstraints}. This means that the $b_i$ factors are constrained in control regions at high $\xi_Z$ and the $f(x_i)$ function is then defined to correct for non-closure when transferring these corrections to low $\xi_Z$.
\item \textit{Function choice}: In this variation, the choice of $f(x_i)$ in  Eq.~\ref{eq:strongConstraints} is changed from a first-order polynomial to a second-order polynomial.
\item \textit{Unconstrained strong $Zjj$ yield}: In this variation, the constraint applied to the integrated strong $Zjj$ event yield (Eq.~\ref{eq:strongyield}) is removed. This allows the integrated strong $Zjj$ yield to be different between different differential distributions.
\end{enumerate}
 
In addition, a simpler `sequential' method is used to extract the EW $Zjj$ yields. In this approach, the EW $Zjj$ yield of bin $i$ in the signal region (SR) is defined by
\begin{equation}
\label{eq:seqmeth}
\nu_{\mathrm{SR},i}^\mathrm{EW} = N_{\mathrm{SR},i}^\mathrm{data} -  k \, r^{\,}_{\mathrm{CRa},i} \, \nu_{SR,i}^\mathrm{strong, MC} - \nu_{\mathrm{SR},i}^\mathrm{other,MC},
\end{equation}
where
$N_{\mathrm{SR},i}^\mathrm{data}$ is the observed event yield in that bin,
$\nu_{\mathrm{SR},i}^\mathrm{strong, MC}$ is the strong $Zjj$ background yield predicted by simulation, and $\nu_{\mathrm{SR},i}^\mathrm{other,MC}$ is the event yield from other simulated background processes. The data-driven constraints on the strong $Zjj$ background, $r^{\,}_{\mathrm{CRa},i}$ and $k$, are defined by
\begin{equation*}
r^{\,}_{\mathrm{CRa}, i} = \left(N_{\mathrm{CRa},i}^\mathrm{data} - \nu_{\mathrm{CRa},i}^\mathrm{EW, MC} - \nu_{\mathrm{CRa},i}^\mathrm{other,MC}\right)/\nu_{\mathrm{CRa},i}^\mathrm{strong, MC} \quad \textrm{and} \quad k= r^{\,}_{\mathrm{SR},0}/r^{\,}_{\mathrm{CRa},0},
\end{equation*}
where $\nu_{\mathrm{CRa},i}^\mathrm{EW, MC}$ is the predicted EW $Zjj$ contamination in the control region  and is estimated using \hervbf.
The index `0' in the definition of $k$ specifies a normalisation region defined by $250\leq\mjj<500$~GeV. The electroweak signal yield in each bin of the differential distribution is taken to be the midpoint of the envelope of yields obtained using the \sherpa{}, \mgnloR{} and \mglo{} predictions for the strong $Zjj$ background. The systematic uncertainties on the sequential method are evaluated in the same way as for the nominal method, using the procedures outlined in Sec.~\ref{sec:systematics}. Non-closure of the sequential method is evaluated by deriving the  constraints on the strong $Zjj$ process using data in CRb and applying them to the strong $Zjj$ prediction in CRc. This non-closure is assigned as an additional systematic uncertainty in the sequential method.
 
\begin{figure}[t]
\includegraphics[width=\textwidth]{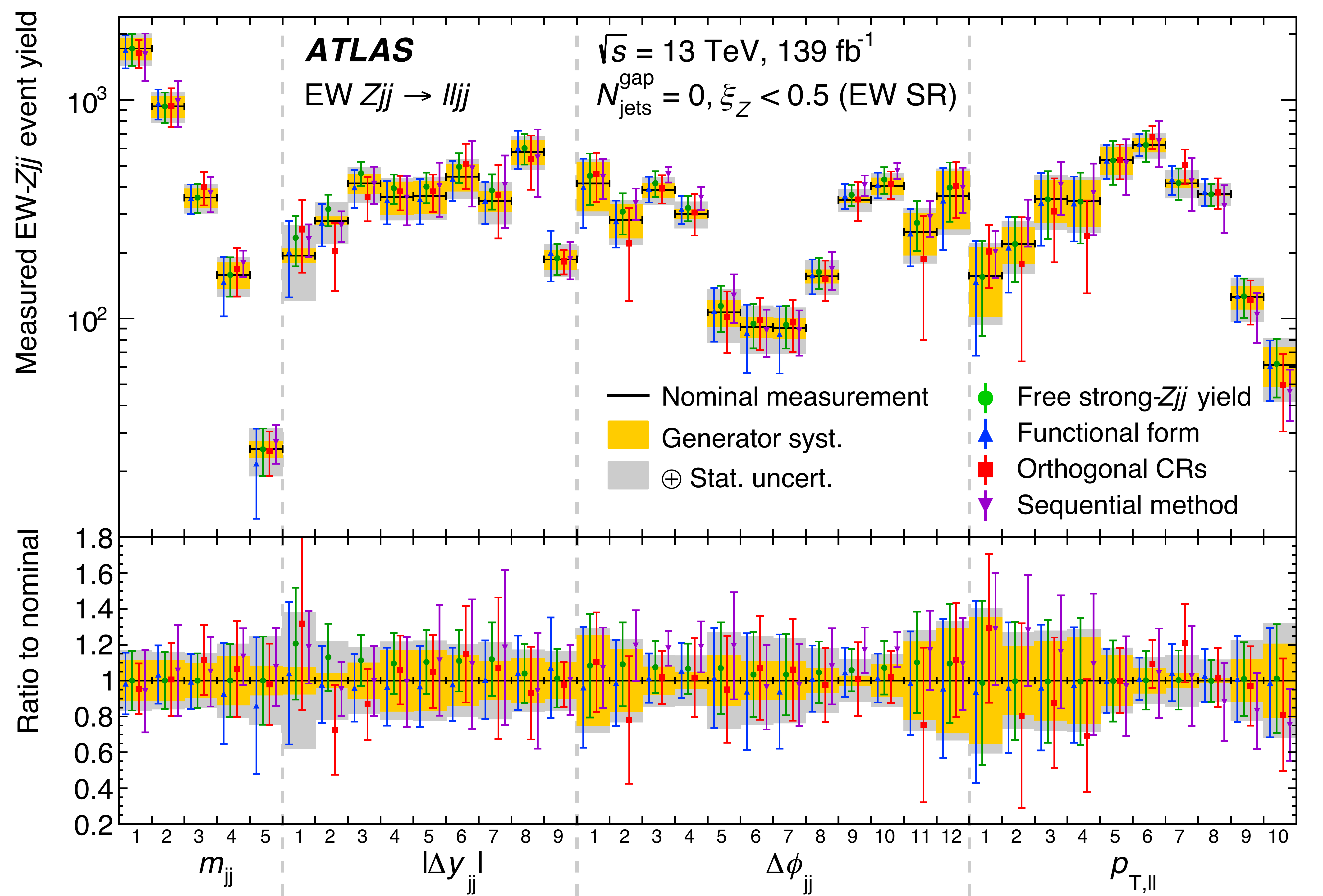}
\caption{Extracted EW $Zjj$ event yields in bins of the four  observables measured using five different methods. The nominal measurements with their associated uncertainties are shown as bands, while the measurements based on the four variations to the EW $Zjj$ extraction methodology are shown as points with error bars corresponding to their associated total uncertainty (excluding the experimental systematics, that tend to be significantly smaller). }
\label{fig:methodcheck}
\end{figure}
 
Figure \ref{fig:methodcheck} shows the EW $Zjj$ event yields measured with the four variations in the electroweak extraction method. The results are found to be in good agreement with the nominal method. In particular, the deviation in the extracted event yields tend to lie within the uncertainty that arises from the strong $Zjj$ generator choice. The agreement between the nominal result and the result obtained with each variation in extraction method is quantified using a $\chi^2$ test for each distribution, with the correlations between measurements determined using the bootstrap method~\cite{PhysRevD.39.274}.
Good agreement is found between the nominal result and each of the results obtained with the different electroweak extraction methods.

\section{Tabulated differential cross-section measurements}
\label{sec:appA}
 
In this section, the measured EW $Zjj$ and inclusive $Zjj$ differential cross-sections are presented in tabular form. The differential cross-sections measured as a function of \mjj{}, \Dyjj{}, \pTll{}, and \Dphijj{} are presented in Tables \ref{tab:mjjMeas}--\ref{tab:DphijjMeas}. The EW $Zjj$ differential cross-sections are measured in the signal region, whereas the inclusive $Zjj$ differential cross-sections are measured in the signal region and the three control regions. The fiducial definition for the signal and control regions are defined in Sec.~\ref{sec:unfolding}.
 
\begin{table}[tbh]
\centering
 
\begin{tabular}{|l|r|r|r|r|r|r|r|r|r|}
\hline \hline
\multicolumn{10}{|l|}{\textbf{EW $Zjj$ SR,~~$m_{jj}$ cross-section measurements}} \\ \hline
$\mathrm{d}\sigma\,/\,\mathrm{d} m_{jj}$ [ab/GeV] & - & - & - & -  & 41 & 14 & 5.5 & 1.3 & 0.10 \\
\hline
Stat.\ unc. [\%]    & - & - & - & -  & $13$ & $13$ & $13$ & $17$ & $26$ \\
Gen.\ choice [\%]   & - & - & - & -  & $11$ & $11$ & $9.4$ & $14$ & $7.6$ \\
Theory syst. [\%]    & - & - & - & -  & $8.1$ & $6.6$ & $4.3$ & $3.1$ & $1.2$ \\
Jet syst. [\%]       & - & - & - & -  & $8.4$ & $6.9$ & $6.3$ & $9.4$ & $14$ \\
Unfolding syst. [\%] & - & - & - & -  & $2.3$ & $1.1$ & $0.7$ & $0.6$ & $0.6$ \\
Other syst. [\%]     & - & - & - & -  & $2.0$ & $2.0$ & $2.3$ & $2.2$ & $3.0$ \\
\hline \hline
\multicolumn{10}{|l|}{\textbf{Inclusive $Zjj$ SR,~~$m_{jj}$ cross-section measurements}} \\ \hline
$\mathrm{d}\sigma\,/\,\mathrm{d} m_{jj}$ [ab/GeV]  & 510 & 1040 & 700 & 320 & 120 & 31 & 8.8 & 1.7 & 0.12 \\
\hline
Stat.\ unc. [\%]  & $1.6$ & $1.0$ & $0.9$ & $1.3$ & $1.5$ & $2.3$ & $4.5$ & $7.2$ & $21$ \\
Jet syst. [\%]  & $5.2$ & $3.8$ & $3.3$ & $3.6$ & $3.6$ & $3.5$ & $4.1$ & $6.6$ & $15$ \\
Unfolding syst. [\%]  & $2.3$ & $1.6$ & $0.9$ & $0.6$ & $0.5$ & $0.4$ & $0.5$ & $0.6$ & $0.6$ \\
Other syst. [\%]  & $2.8$ & $2.8$ & $2.8$ & $2.8$ & $2.8$ & $2.8$ & $2.9$ & $2.9$ & $3.4$ \\
\hline \hline
\multicolumn{10}{|l|}{\textbf{Inclusive $Zjj$ CRa,~~$m_{jj}$ cross-section measurements}} \\ \hline
$\mathrm{d}\sigma\,/\,\mathrm{d} m_{jj}$ [ab/GeV]  & 250 & 610 & 560 & 320 & 130 & 37 & 8.7 & 1.6 & 0.10 \\
\hline
Stat.\ unc. [\%]  & $2.2$ & $1.2$ & $1.0$ & $1.3$ & $1.3$ & $2.1$ & $4.4$ & $7.3$ & $22$ \\
Jet syst. [\%]  & $11$ & $11$ & $9.4$ & $8.6$ & $8.6$ & $8.1$ & $9.9$ & $11$ & $14$ \\
Unfolding syst. [\%]  & $6.7$ & $5.3$ & $4.1$ & $3.3$ & $2.7$ & $2.6$ & $3.0$ & $3.9$ & $5.3$ \\
Other syst. [\%]  & $2.3$ & $2.3$ & $2.3$ & $2.4$ & $2.4$ & $2.5$ & $2.5$ & $2.6$ & $2.8$ \\
\hline \hline
\multicolumn{10}{|l|}{\textbf{Inclusive $Zjj$ CRb,~~$m_{jj}$ cross-section measurements}} \\ \hline
$\mathrm{d}\sigma\,/\,\mathrm{d} m_{jj}$ [ab/GeV]  & 190 & 430 & 330 & 150 & 54 & 10 & 1.4 & 0.11 & -  \\
\hline
Stat.\ unc. [\%]  & $2.5$ & $1.4$ & $1.2$ & $1.8$ & $2.2$ & $4.2$ & $11$ & $28$ & -  \\
Jet syst. [\%]  & $11$ & $9.0$ & $7.6$ & $8.0$ & $7.4$ & $7.9$ & $9.0$ & $8.9$ & -  \\
Unfolding syst. [\%]  & $2.3$ & $2.4$ & $2.4$ & $2.1$ & $1.8$ & $2.1$ & $3.0$ & $3.8$ & -  \\
Other syst. [\%]  & $2.3$ & $2.3$ & $2.3$ & $2.4$ & $2.4$ & $2.5$ & $2.6$ & $2.6$ & -  \\
\hline \hline
\multicolumn{10}{|l|}{\textbf{Inclusive $Zjj$ CRc,~~$m_{jj}$ cross-section measurements}} \\ \hline
$\mathrm{d}\sigma\,/\,\mathrm{d} m_{jj}$ [ab/GeV]  & 350 & 690 & 390 & 140 & 37 & 5.7 & 0.60 & 0.07 & -  \\
\hline
Stat.\ unc. [\%]  & $1.9$ & $1.2$ & $1.2$ & $2.0$ & $2.7$ & $5.8$ & $18$ & $36$ & -  \\
Jet syst. [\%]  & $6.7$ & $3.6$ & $3.3$ & $5.0$ & $2.3$ & $4.7$ & $5.5$ & $4.0$ & -  \\
Unfolding syst. [\%]  & $1.2$ & $1.0$ & $0.8$ & $0.9$ & $1.1$ & $1.6$ & $2.1$ & $2.3$ & -  \\
Other syst. [\%]  & $2.8$ & $2.8$ & $2.8$ & $2.8$ & $2.8$ & $2.9$ & $2.9$ & $3.1$ & -  \\
\hline \hline
Low bin edge [TeV]  & $0.25$ & $0.35$ & $0.50$ & $0.75$ & $1.0$ & $1.5$ & $2.2$ & $3.0$ & $4.5$ \\
High bin edge [TeV]  & $0.35$ & $0.50$ & $0.75$ & $1.0$ & $1.5$ & $2.2$ & $3.0$ & $4.5$ & $7.5$ \\ \hline \hline
\end{tabular}
 
\caption{Differential cross-section measurements for EW $Zjj$ production and inclusive $Zjj$ production as a function of \mjj{}. The EW $Zjj$ measurements are performed in the signal region. The inclusive $Zjj$ measurements are performed in the signal region and three control regions. The different sources of uncertainty are grouped together for each measurement, with a more granular breakdown of each uncertainty available in HEPDATA.
\label{tab:mjjMeas}}
\end{table}
 
\begin{table}[tbh]
\centering
 
\begin{tabular}{|l|r|r|r|r|r|r|r|r|r|}
\hline \hline
\multicolumn{10}{|l|}{\textbf{EW $Zjj$ SR,~~$|\Delta y_{jj}|$ cross-section measurements}} \\ \hline
$\mathrm{d}\sigma\,/\,\mathrm{d} |\Delta y_{jj}|$ [fb]  & 3.6 & 6.3 & 9.3 & 13 & 14 & 13 & 10 & 7.1 & 1.2 \\
\hline
Stat.\ unc. [\%]     & $38$ & $22$ & $16$ & $14$ & $15$ & $15$ & $20$ & $13$ & $14$ \\
Gen.\ choice [\%]    & $7.5$ & $4.0$ & $9.8$ & $17$ & $17$ & $14$ & $10$ & $12$ & $10$ \\
Theory syst. [\%]     & $18$ & $11$ & $8.5$ & $6.7$ & $8.0$ & $7.5$ & $7.5$ & $3.6$ & $1.8$ \\
Jet syst. [\%]        & $21$ & $10.0$ & $6.9$ & $7.0$ & $6.4$ & $8.3$ & $13$ & $6.2$ & $11$ \\
Unfolding syst. [\%]  & $8.4$ & $6.3$ & $3.9$ & $2.8$ & $2.3$ & $1.5$ & $1.1$ & $0.7$ & $0.1$ \\
Other syst. [\%]      & $2.6$ & $2.3$ & $2.2$ & $2.4$ & $2.1$ & $2.0$ & $2.0$ & $1.9$ & $2.2$ \\
\hline \hline
\multicolumn{10}{|l|}{\textbf{Inclusive $Zjj$ SR,~~$|\Delta y_{jj}|$ cross-section measurements}} \\ \hline
$\mathrm{d}\sigma\,/\,\mathrm{d} |\Delta y_{jj}|$ [fb]  & 13 & 18 & 25 & 32 & 36 & 36 & 31 & 14 & 1.6 \\
\hline
Stat.\ unc. [\%]  & $3.9$ & $3.8$ & $3.2$ & $3.5$ & $3.3$ & $2.9$ & $3.2$ & $3.0$ & $6.3$ \\
Jet syst. [\%]  & $2.7$ & $2.1$ & $2.2$ & $2.6$ & $2.9$ & $5.7$ & $4.5$ & $4.2$ & $7.7$ \\
Unfolding syst. [\%]  & $0.7$ & $0.6$ & $0.5$ & $0.4$ & $0.3$ & $0.3$ & $0.4$ & $0.4$ & $0.3$ \\
Other syst. [\%]  & $2.9$ & $2.9$ & $2.9$ & $2.9$ & $2.9$ & $2.8$ & $2.9$ & $2.8$ & $2.9$ \\
\hline \hline
\multicolumn{10}{|l|}{\textbf{Inclusive $Zjj$ CRa,~~$|\Delta y_{jj}|$ cross-section measurements}} \\ \hline
$\mathrm{d}\sigma\,/\,\mathrm{d} |\Delta y_{jj}|$ [fb]  & 13 & 21 & 30 & 36 & 43 & 42 & 37 & 15 & 1.6 \\
\hline
Stat.\ unc. [\%]  & $3.8$ & $3.5$ & $2.9$ & $3.3$ & $3.0$ & $2.6$ & $2.9$ & $2.8$ & $6.2$ \\
Jet syst. [\%]  & $9.3$ & $8.1$ & $8.0$ & $7.5$ & $7.5$ & $8.1$ & $8.7$ & $9.5$ & $13$ \\
Unfolding syst. [\%]  & $2.0$ & $2.1$ & $2.3$ & $2.5$ & $2.7$ & $2.8$ & $2.9$ & $3.2$ & $3.7$ \\
Other syst. [\%]  & $2.5$ & $2.5$ & $2.5$ & $2.5$ & $2.5$ & $2.5$ & $2.5$ & $2.5$ & $2.6$ \\
\hline \hline
\multicolumn{10}{|l|}{\textbf{Inclusive $Zjj$ CRb,~~$|\Delta y_{jj}|$ cross-section measurements}} \\ \hline
$\mathrm{d}\sigma\,/\,\mathrm{d} |\Delta y_{jj}|$ [fb]  & 11 & 14 & 15 & 14 & 13 & 9.6 & 6.1 & 1.1 & -  \\
\hline
Stat.\ unc. [\%]  & $4.1$ & $4.1$ & $4.0$ & $5.4$ & $5.6$ & $5.8$ & $7.1$ & $10$ & -  \\
Jet syst. [\%]  & $6.3$ & $6.1$ & $6.1$ & $7.1$ & $8.3$ & $9.5$ & $8.7$ & $11$ & -  \\
Unfolding syst. [\%]  & $1.1$ & $1.3$ & $1.7$ & $2.0$ & $2.1$ & $2.0$ & $2.0$ & $2.0$ & -  \\
Other syst. [\%]  & $2.4$ & $2.4$ & $2.5$ & $2.5$ & $2.5$ & $2.5$ & $2.5$ & $2.6$ & -  \\
\hline \hline
\multicolumn{10}{|l|}{\textbf{Inclusive $Zjj$ CRc,~~$|\Delta y_{jj}|$ cross-section measurements}} \\ \hline
$\mathrm{d}\sigma\,/\,\mathrm{d} |\Delta y_{jj}|$ [fb]  & 7.9 & 9.2 & 9.6 & 8.2 & 8.2 & 6.1 & 3.7 & 0.70 & -  \\
\hline
Stat.\ unc. [\%]  & $4.9$ & $5.2$ & $5.2$ & $7.3$ & $7.3$ & $7.3$ & $9.4$ & $13$ & -  \\
Jet syst. [\%]  & $1.8$ & $2.6$ & $2.2$ & $3.2$ & $2.5$ & $3.5$ & $10$ & $9.7$ & -  \\
Unfolding syst. [\%]  & $0.5$ & $0.6$ & $0.6$ & $0.6$ & $0.6$ & $0.5$ & $0.4$ & $0.2$ & -  \\
Other syst. [\%]  & $2.9$ & $2.9$ & $2.9$ & $2.9$ & $2.9$ & $2.9$ & $3.1$ & $3.4$ & -  \\
\hline \hline
Low bin edge  & $2.0$ & $2.6$ & $3.1$ & $3.6$ & $3.9$ & $4.2$ & $4.6$ & $5.0$ & $6.0$ \\
High bin edge  & $2.6$ & $3.1$ & $3.6$ & $3.9$ & $4.2$ & $4.6$ & $5.0$ & $6.0$ & $8.0$ \\ \hline \hline
\end{tabular}
 
\caption{Differential cross-section measurements for EW $Zjj$ production and inclusive $Zjj$ production as a function of \Dyjj{}. The EW $Zjj$ measurements are performed in the signal region. The inclusive $Zjj$ measurements are performed in the signal region and three control regions. The different sources of uncertainty are grouped together for each measurement, with a more granular breakdown of each uncertainty available in HEPDATA.\label{tab:DyjjMeas}}
\end{table}

\begin{table}[tbh]
\centering
 
\begin{tabular}{|l|r|r|r|r|r|r|r|r|r|r|}
\hline \hline
\multicolumn{11}{|l|}{\textbf{EW $Zjj$ SR,~~$p_{\mathrm{T},\ell\ell}$ cross-section measurements}} \\ \hline
$\mathrm{d}\sigma\,/\,\mathrm{d} p_{\mathrm{T},\ell\ell}$ [ab/GeV]  & 210 & 190 & 180 & 130 & 150 & 110 & 59 & 31 & 8.8 & 1.4 \\
\hline
Stat.\ unc. [\%]     & $21$ & $20$ & $18$ & $17$ & $12$ & $12$ & $15$ & $13$ & $20$ & $25$ \\
Gen.\ choice [\%]    & $36$ & $19$ & $22$ & $24$ & $14$ & $6.9$ & $4.2$ & $-0.9$ & $-12$ & $-21$ \\
Theory syst. [\%]     & $4.7$ & $11$ & $9.8$ & $12$ & $5.8$ & $6.0$ & $5.0$ & $3.5$ & $5.0$ & $4.1$ \\
Jet syst. [\%]        & $10$ & $13$ & $11$ & $13$ & $7.5$ & $6.4$ & $5.9$ & $2.9$ & $3.1$ & $8.5$ \\
Unfolding syst. [\%]  & $1.2$ & $2.4$ & $3.3$ & $3.5$ & $3.0$ & $2.0$ & $1.3$ & $1.3$ & $1.7$ & $2.0$ \\
Other syst. [\%]      & $3.5$ & $2.1$ & $2.1$ & $2.2$ & $2.1$ & $2.0$ & $2.1$ & $2.2$ & $3.0$ & $3.4$ \\
\hline \hline
\multicolumn{11}{|l|}{\textbf{Inclusive $Zjj$ SR,~~$p_{\mathrm{T},\ell\ell}$ cross-section measurements}} \\ \hline
$\mathrm{d}\sigma\,/\,\mathrm{d} p_{\mathrm{T},\ell\ell}$ [ab/GeV]  & 530 & 580 & 530 & 440 & 380 & 270 & 140 & 58 & 16 & 2.1 \\
\hline
Stat.\ unc. [\%]  & $5.2$ & $4.0$ & $3.3$ & $3.2$ & $2.9$ & $2.8$ & $3.2$ & $3.9$ & $7.0$ & $11$ \\
Jet syst. [\%]  & $5.3$ & $6.9$ & $4.1$ & $4.2$ & $3.6$ & $3.6$ & $2.2$ & $1.4$ & $2.5$ & $5.6$ \\
Unfolding syst. [\%]  & $0.4$ & $0.4$ & $0.5$ & $0.5$ & $0.4$ & $0.3$ & $0.2$ & $0.2$ & $0.3$ & $0.3$ \\
Other syst. [\%]  & $2.9$ & $2.9$ & $2.8$ & $2.8$ & $2.8$ & $2.8$ & $2.8$ & $3.0$ & $3.4$ & $3.6$ \\
\hline \hline
\multicolumn{11}{|l|}{\textbf{Inclusive $Zjj$ CRa,~~$p_{\mathrm{T},\ell\ell}$ cross-section measurements}} \\ \hline
$\mathrm{d}\sigma\,/\,\mathrm{d} p_{\mathrm{T},\ell\ell}$ [ab/GeV]  & 480 & 570 & 580 & 520 & 430 & 320 & 170 & 66 & 19 & 2.3 \\
\hline
Stat.\ unc. [\%]  & $5.4$ & $3.9$ & $3.1$ & $2.9$ & $2.7$ & $2.5$ & $2.9$ & $3.5$ & $6.2$ & $10$ \\
Jet syst. [\%]  & $13$ & $7.7$ & $9.5$ & $7.8$ & $7.4$ & $7.1$ & $7.7$ & $8.2$ & $11$ & $13$ \\
Unfolding syst. [\%]  & $3.4$ & $3.4$ & $3.4$ & $3.2$ & $2.8$ & $2.3$ & $1.6$ & $1.1$ & $0.8$ & $0.7$ \\
Other syst. [\%]  & $2.5$ & $2.5$ & $2.4$ & $2.4$ & $2.4$ & $2.4$ & $2.5$ & $2.6$ & $3.0$ & $3.3$ \\
\hline \hline
\multicolumn{11}{|l|}{\textbf{Inclusive $Zjj$ CRb,~~$p_{\mathrm{T},\ell\ell}$ cross-section measurements}} \\ \hline
$\mathrm{d}\sigma\,/\,\mathrm{d} p_{\mathrm{T},\ell\ell}$ [ab/GeV]  & 190 & 210 & 200 & 190 & 140 & 97 & 56 & 24 & 6.8 & -  \\
\hline
Stat.\ unc. [\%]  & $8.4$ & $6.8$ & $5.3$ & $4.7$ & $4.6$ & $4.3$ & $5.2$ & $5.9$ & $11$ & -  \\
Jet syst. [\%]  & $5.8$ & $7.6$ & $7.3$ & $7.4$ & $6.4$ & $6.5$ & $6.8$ & $6.8$ & $7.0$ & -  \\
Unfolding syst. [\%]  & $1.6$ & $1.8$ & $2.0$ & $2.1$ & $1.9$ & $1.5$ & $1.0$ & $0.7$ & $0.8$ & -  \\
Other syst. [\%]  & $2.5$ & $2.5$ & $2.4$ & $2.4$ & $2.4$ & $2.4$ & $2.4$ & $2.6$ & $3.0$ & -  \\
\hline \hline
\multicolumn{11}{|l|}{\textbf{Inclusive $Zjj$ CRc,~~$p_{\mathrm{T},\ell\ell}$ cross-section measurements}} \\ \hline
$\mathrm{d}\sigma\,/\,\mathrm{d} p_{\mathrm{T},\ell\ell}$ [ab/GeV]  & 100 & 160 & 150 & 120 & 89 & 62 & 32 & 15 & 4.0 & -  \\
\hline
Stat.\ unc. [\%]  & $12$ & $8.0$ & $6.4$ & $5.9$ & $5.9$ & $5.5$ & $7.0$ & $7.8$ & $13$ & -  \\
Jet syst. [\%]  & $3.4$ & $4.3$ & $3.0$ & $4.2$ & $2.5$ & $4.4$ & $1.9$ & $2.2$ & $5.8$ & -  \\
Unfolding syst. [\%]  & $1.1$ & $1.1$ & $0.9$ & $0.7$ & $0.5$ & $0.5$ & $0.5$ & $0.3$ & $0.3$ & -  \\
Other syst. [\%]  & $3.0$ & $2.9$ & $2.9$ & $2.9$ & $2.9$ & $2.9$ & $2.9$ & $3.0$ & $3.4$ & -  \\
\hline \hline
Low bin edge [GeV]  & $20$ & $30$ & $45$ & $70$ & $100$ & $140$ & $200$ & $275$ & $400$ & $550$ \\
High bin edge [GeV]  & $30$ & $45$ & $70$ & $100$ & $140$ & $200$ & $275$ & $400$ & $550$ & $1050$ \\ \hline \hline
\end{tabular}
 
\caption{Differential cross-section measurements for EW $Zjj$ production and inclusive $Zjj$ production as a function of \pTll{}. The EW $Zjj$ measurements are performed in the signal region. The inclusive $Zjj$ measurements are performed in the signal region and three control regions. The different sources of uncertainty are grouped together for each measurement, with a more granular breakdown of each uncertainty available in HEPDATA.\label{tab:pTllMeas}}
\end{table}
 
\begin{sidewaystable}[tbh]
\centering
\begin{small}
\begin{tabular}{|l|r|r|r|r|r|r|r|r|r|r|r|r|}
\hline \hline
\multicolumn{13}{|l|}{\textbf{EW $Zjj$ SR,~~$\Delta\phi_{jj}$ cross-section measurements}} \\ \hline
$\mathrm{d}\sigma\,/\,\mathrm{d} \Delta\phi_{jj}$ [fb]  & 26 & 17 & 12 & 4.2 & 1.4 & 1.2 & 1.2 & 2.0 & 4.9 & 12 & 15 & 23 \\
\hline
Stat.\ unc. [\%]     & $15$ & $16$ & $13$ & $14$ & $24$ & $23$ & $22$ & $17$ & $12$ & $12$ & $19$ & $17$ \\
Gen.\ choice [\%]    & $25$ & $17$ & $7.4$ & $4.5$ & $-14$ & $-10$ & $-11$ & $-6.9$ & $2.5$ & $8.8$ & $22$ & $30$ \\
Theory syst. [\%]     & $7.8$ & $9.5$ & $8.2$ & $5.1$ & $4.6$ & $4.7$ & $4.3$ & $3.2$ & $4.3$ & $8.2$ & $12$ & $8.2$ \\
Jet syst. [\%]        & $8.9$ & $8.0$ & $6.6$ & $7.9$ & $11$ & $7.0$ & $7.6$ & $7.5$ & $7.4$ & $6.3$ & $8.7$ & $9.9$ \\
Unfolding syst. [\%]  & $2.6$ & $2.4$ & $2.2$ & $2.1$ & $1.8$ & $1.1$ & $0.6$ & $0.8$ & $1.7$ & $2.8$ & $3.4$ & $3.9$ \\
Other syst. [\%]      & $2.6$ & $2.1$ & $2.1$ & $2.1$ & $2.3$ & $2.6$ & $2.5$ & $2.3$ & $2.1$ & $2.1$ & $2.2$ & $2.7$ \\
\hline \hline
\multicolumn{13}{|l|}{\textbf{Inclusive $Zjj$ SR,~~$\Delta\phi_{jj}$ cross-section measurements}} \\ \hline
$\mathrm{d}\sigma\,/\,\mathrm{d} \Delta\phi_{jj}$ [fb]  & 71 & 49 & 29 & 10 & 3.3 & 2.1 & 2.0 & 3.8 & 11 & 29 & 48 & 69 \\
\hline
Stat.\ unc. [\%]  & $3.1$ & $3.6$ & $3.3$ & $3.8$ & $6.4$ & $8.1$ & $8.6$ & $6.1$ & $3.7$ & $3.3$ & $3.8$ & $3.3$ \\
Jet syst. [\%]  & $2.0$ & $5.1$ & $3.6$ & $4.2$ & $4.4$ & $8.2$ & $6.6$ & $8.0$ & $4.6$ & $3.4$ & $4.0$ & $2.6$ \\
Unfolding syst. [\%]  & $0.4$ & $0.4$ & $0.3$ & $0.3$ & $0.3$ & $0.1$ & $0.1$ & $0.1$ & $0.2$ & $0.4$ & $0.5$ & $0.6$ \\
Other syst. [\%]  & $2.8$ & $2.8$ & $2.9$ & $2.9$ & $3.0$ & $3.1$ & $3.1$ & $3.0$ & $2.9$ & $2.9$ & $2.9$ & $2.8$ \\
\hline \hline
\multicolumn{13}{|l|}{\textbf{Inclusive $Zjj$ CRa,~~$\Delta\phi_{jj}$ cross-section measurements}} \\ \hline
$\mathrm{d}\sigma\,/\,\mathrm{d} \Delta\phi_{jj}$ [fb]  & 62 & 50 & 33 & 14 & 4.9 & 3.1 & 3.1 & 5.7 & 14 & 33 & 54 & 64 \\
\hline
Stat.\ unc. [\%]  & $3.2$ & $3.5$ & $3.0$ & $3.2$ & $5.2$ & $6.4$ & $6.6$ & $5.1$ & $3.2$ & $3.1$ & $3.5$ & $3.3$ \\
Jet syst. [\%]  & $7.7$ & $7.4$ & $7.5$ & $7.4$ & $8.2$ & $8.6$ & $12$ & $9.6$ & $8.1$ & $8.1$ & $7.6$ & $7.0$ \\
Unfolding syst. [\%]  & $2.0$ & $2.1$ & $2.3$ & $2.4$ & $2.5$ & $2.6$ & $2.5$ & $2.5$ & $2.5$ & $2.5$ & $2.5$ & $2.5$ \\
Other syst. [\%]  & $2.4$ & $2.5$ & $2.5$ & $2.5$ & $2.6$ & $2.6$ & $2.6$ & $2.6$ & $2.5$ & $2.5$ & $2.4$ & $2.4$ \\
\hline \hline
\multicolumn{13}{|l|}{\textbf{Inclusive $Zjj$ CRb,~~$\Delta\phi_{jj}$ cross-section measurements}} \\ \hline
$\mathrm{d}\sigma\,/\,\mathrm{d} \Delta\phi_{jj}$ [fb]  & 28 & 21 & 12 & 3.4 & 0.96 & 0.62 & 0.35 & 0.90 & 3.8 & 12 & 22 & 29 \\
\hline
Stat.\ unc. [\%]  & $4.8$ & $5.4$ & $5.2$ & $6.3$ & $12$ & $14$ & $19$ & $12$ & $6.1$ & $5.2$ & $5.4$ & $4.9$ \\
Jet syst. [\%]  & $5.4$ & $6.5$ & $7.0$ & $8.1$ & $11$ & $12$ & $14$ & $9.7$ & $7.5$ & $7.5$ & $6.1$ & $6.3$ \\
Unfolding syst. [\%]  & $1.1$ & $1.4$ & $1.5$ & $1.4$ & $1.0$ & $0.6$ & $0.6$ & $1.0$ & $1.6$ & $2.0$ & $1.8$ & $1.4$ \\
Other syst. [\%]  & $2.4$ & $2.4$ & $2.5$ & $2.5$ & $2.7$ & $2.9$ & $2.8$ & $2.7$ & $2.5$ & $2.5$ & $2.4$ & $2.4$ \\
\hline \hline
\multicolumn{13}{|l|}{\textbf{Inclusive $Zjj$ CRc,~~$\Delta\phi_{jj}$ cross-section measurements}} \\ \hline
$\mathrm{d}\sigma\,/\,\mathrm{d} \Delta\phi_{jj}$ [fb]  & 25 & 15 & 6.3 & 1.7 & 0.36 & 0.26 & 0.15 & 0.49 & 1.7 & 6.0 & 16 & 22 \\
\hline
Stat.\ unc. [\%]  & $5.2$ & $6.6$ & $7.0$ & $8.9$ & $18$ & $23$ & $31$ & $16$ & $9.1$ & $7.4$ & $6.7$ & $5.8$ \\
Jet syst. [\%]  & $1.8$ & $3.0$ & $2.0$ & $7.3$ & $7.2$ & $10$ & $15$ & $8.0$ & $9.2$ & $3.8$ & $5.1$ & $3.4$ \\
Unfolding syst. [\%]  & $0.8$ & $0.8$ & $0.7$ & $0.4$ & $0.2$ & $0.2$ & $0.2$ & $0.2$ & $0.3$ & $0.6$ & $0.7$ & $0.7$ \\
Other syst. [\%]  & $2.8$ & $2.9$ & $2.9$ & $3.0$ & $3.1$ & $3.9$ & $3.7$ & $3.3$ & $3.0$ & $2.9$ & $2.9$ & $2.9$ \\
\hline \hline
Low bin edge [$\pi/16$]  & $-16$ & $-15$ & $-14$ & $-12$ & $-8.0$ & $-4.0$ & $0.00$ & $4.0$ & $8.0$ & $12$ & $14$ & $15$ \\
High bin edge [$\pi/16$]  & $-15$ & $-14$ & $-12$ & $-8.0$ & $-4.0$ & $0.00$ & $4.0$ & $8.0$ & $12$ & $14$ & $15$ & $16$ \\ \hline \hline
\end{tabular}
 
\caption{Differential cross-section measurements for EW $Zjj$ production and inclusive $Zjj$ production as a function of \Dphijj{}. The EW $Zjj$ measurements are performed in the signal region. The inclusive $Zjj$ measurements are performed in the signal region and three control regions. The different sources of uncertainty are grouped together for each measurement, with a more granular breakdown of each uncertainty available in HEPDATA.\label{tab:DphijjMeas}}
 
\end{small}
\end{sidewaystable}

\FloatBarrier
\section*{Acknowledgements}
 

We thank CERN for the very successful operation of the LHC, as well as the
support staff from our institutions without whom ATLAS could not be
operated efficiently.
 
We acknowledge the support of ANPCyT, Argentina; YerPhI, Armenia; ARC, Australia; BMWFW and FWF, Austria; ANAS, Azerbaijan; SSTC, Belarus; CNPq and FAPESP, Brazil; NSERC, NRC and CFI, Canada; CERN; CONICYT, Chile; CAS, MOST and NSFC, China; COLCIENCIAS, Colombia; MSMT CR, MPO CR and VSC CR, Czech Republic; DNRF and DNSRC, Denmark; IN2P3-CNRS and CEA-DRF/IRFU, France; SRNSFG, Georgia; BMBF, HGF and MPG, Germany; GSRT, Greece; RGC and Hong Kong SAR, China; ISF and Benoziyo Center, Israel; INFN, Italy; MEXT and JSPS, Japan; CNRST, Morocco; NWO, Netherlands; RCN, Norway; MNiSW and NCN, Poland; FCT, Portugal; MNE/IFA, Romania; MES of Russia and NRC KI, Russia Federation; JINR; MESTD, Serbia; MSSR, Slovakia; ARRS and MIZ\v{S}, Slovenia; DST/NRF, South Africa; MINECO, Spain; SRC and Wallenberg Foundation, Sweden; SERI, SNSF and Cantons of Bern and Geneva, Switzerland; MOST, Taiwan; TAEK, Turkey; STFC, United Kingdom; DOE and NSF, United States of America. In addition, individual groups and members have received support from BCKDF, CANARIE, Compute Canada and CRC, Canada; ERC, ERDF, Horizon 2020, Marie Sk{\l}odowska-Curie Actions and COST, European Union; Investissements d'Avenir Labex, Investissements d'Avenir Idex and ANR, France; DFG and AvH Foundation, Germany; Herakleitos, Thales and Aristeia programmes co-financed by EU-ESF and the Greek NSRF, Greece; BSF-NSF and GIF, Israel; CERCA Programme Generalitat de Catalunya and PROMETEO Programme Generalitat Valenciana, Spain; G\"{o}ran Gustafssons Stiftelse, Sweden; The Royal Society and Leverhulme Trust, United Kingdom.
 
The crucial computing support from all WLCG partners is acknowledged gratefully, in particular from CERN, the ATLAS Tier-1 facilities at TRIUMF (Canada), NDGF (Denmark, Norway, Sweden), CC-IN2P3 (France), KIT/GridKA (Germany), INFN-CNAF (Italy), NL-T1 (Netherlands), PIC (Spain), ASGC (Taiwan), RAL (UK) and BNL (USA), the Tier-2 facilities worldwide and large non-WLCG resource providers. Major contributors of computing resources are listed in Ref.~\cite{ATL-SOFT-PUB-2020-001}.
 

\printbibliography

@misc{Aad:2020kub,
      author         = "{ATLAS Collaboration}",
      title          = "{Search for the $HH \rightarrow b \bar{b} b \bar{b}$
                        process via vector-boson fusion production using
                        proton-proton collisions at $\sqrt{s} = 13$ TeV with the
                        ATLAS detector}",
      collaboration  = "ATLAS",
    %  year           = "2020",
      eprint         = "2001.05178",
      archivePrefix  = "arXiv",
      primaryClass   = "hep-ex",
      reportNumber   = "CERN-EP-2019-267",
      SLACcitation   = "%%CITATION = ARXIV:2001.05178;%%"
}

@article{wilks1938,
author = "Wilks, S. S.",
doi = "10.1214/aoms/1177732360",
fjournal = "The Annals of Mathematical Statistics",
journal = "Ann. Math. Statist.",
month = "03",
number = "1",
pages = "60--62",
publisher = "The Institute of Mathematical Statistics",
title = "The Large-Sample Distribution of the Likelihood Ratio for Testing Composite Hypotheses",
url = "https://doi.org/10.1214/aoms/1177732360",
volume = "9",
year = "1938"
}

@article{Chatrchyan:2013jya,
      author         = "{CMS Collaboration}",
%      author         = "Chatrchyan, Serguei and others",
      title          = "{Measurement of the hadronic activity in events with a Z
                        and two jets and extraction of the cross section for the
                        electroweak production of a Z with two jets in $pp$
                        collisions at $\sqrt{s}$ = 7 TeV}",
      collaboration  = "CMS",
      journal        = "JHEP",
      volume         = "10",
      year           = "2013",
      pages          = "062",
      doi            = "10.1007/JHEP10(2013)062",
      eprint         = "1305.7389",
      archivePrefix  = "arXiv",
      primaryClass   = "hep-ex",
      reportNumber   = "CMS-FSQ-12-019, CERN-PH-EP-2013-060",
      SLACcitation   = "%%CITATION = ARXIV:1305.7389;%%"
}

@article{Frixione:2007nw,
      author         = "Frixione, Stefano and Nason, Paolo and Ridolfi, Giovanni",
      title          = "{A Positive-weight next-to-leading-order Monte Carlo for
                        heavy flavour hadroproduction}",
      journal        = "JHEP",
      volume         = "09",
      year           = "2007",
      pages          = "126",
      doi            = "10.1088/1126-6708/2007/09/126",
      eprint         = "0707.3088",
      archivePrefix  = "arXiv",
      primaryClass   = "hep-ph",
      reportNumber   = "BICOCCA-FT-07-12, GEF-TH-19-2007",
      SLACcitation   = "%%CITATION = ARXIV:0707.3088;%%"
}

@article{Jager:2012xk,
      author         = "Jager, Barbara and Schneider, Steven and Zanderighi,
                        Giulia",
      title          = "{Next-to-leading order QCD corrections to electroweak $Zjj$
                        production in the POWHEG BOX}",
      journal        = "JHEP",
      volume         = "09",
      year           = "2012",
      pages          = "083",
      doi            = "10.1007/JHEP09(2012)083",
      eprint         = "1207.2626",
      archivePrefix  = "arXiv",
      primaryClass   = "hep-ph",
      reportNumber   = "MZ-TH-12-25",
      SLACcitation   = "%%CITATION = ARXIV:1207.2626;%%"
}

@article{Aad:2019hmz,
      author         = "{ATLAS Collaboration}",
      title          = "{ATLAS data quality operations and performance for
                        2015-2018 data-taking}",
      collaboration  = "ATLAS",
      journal        = "JINST",
      volume         = "15",
      year           = "2020",
      pages          = "P04003",
      eprint         = "1911.04632",
      archivePrefix  = "arXiv",
      primaryClass   = "physics.ins-det",
      reportNumber   = "CERN-EP-2019-207",
      SLACcitation   = "%%CITATION = ARXIV:1911.04632;%%"
}

@article{Aaboud:2019nmv,
      author         = "{ATLAS Collaboration}",
      title          = "{Observation of Electroweak Production of a Same-Sign $W$
                        Boson Pair in Association with Two Jets in $pp$ Collisions
                        at $\sqrt{s}=13$ TeV with the ATLAS Detector}",
      collaboration  = "ATLAS",
      journal        = "Phys. Rev. Lett.",
      volume         = "123",
      year           = "2019",
      number         = "16",
      pages          = "161801",
      doi            = "10.1103/PhysRevLett.123.161801",
      eprint         = "1906.03203",
      archivePrefix  = "arXiv",
      primaryClass   = "hep-ex",
      reportNumber   = "CERN-EP-2019-008",
      SLACcitation   = "%%CITATION = ARXIV:1906.03203;%%"
}

@article{Aaboud:2018ddq,
      author         = "{ATLAS Collaboration}",
      title          = "{Observation of electroweak $W^{\pm}Z$ boson pair
                        production in association with two jets in $pp$ collisions
                        at $\sqrt{s} =$ 13 TeV with the ATLAS detector}",
      collaboration  = "ATLAS",
      journal        = "Phys. Lett. B",
      volume         = "793",
      year           = "2019",
      pages          = "469-492",
      doi            = "10.1016/j.physletb.2019.05.012",
      eprint         = "1812.09740",
      archivePrefix  = "arXiv",
      primaryClass   = "hep-ex",
      reportNumber   = "CERN-EP-2018-286",
      SLACcitation   = "%%CITATION = ARXIV:1812.09740;%%"
}

@article{Aaboud:2018sfi,
      author         = "{ATLAS Collaboration}",
      title          = "{Search for invisible Higgs boson decays in vector boson
                        fusion at $\sqrt{s} = 13$ TeV with the ATLAS detector}",
      collaboration  = "ATLAS",
      journal        = "Phys. Lett. B",
      volume         = "793",
      year           = "2019",
      pages          = "499-519",
      doi            = "10.1016/j.physletb.2019.04.024",
      eprint         = "1809.06682",
      archivePrefix  = "arXiv",
      primaryClass   = "hep-ex",
      reportNumber   = "CERN-EP-2018-184",
      SLACcitation   = "%%CITATION = ARXIV:1809.06682;%%"
}

@article{Aaboud:2018bun,
      author         = "{ATLAS Collaboration}",
      title          = "{Combination of searches for heavy resonances decaying
                        into bosonic and leptonic final states using 36  fb$^{-1}$
                        of proton-proton collision data at $\sqrt{s} = 13$ TeV
                        with the ATLAS detector}",
      collaboration  = "ATLAS",
      journal        = "Phys. Rev. D",
      volume         = "98",
      year           = "2018",
      number         = "5",
      pages          = "052008",
      doi            = "10.1103/PhysRevD.98.052008",
      eprint         = "1808.02380",
      archivePrefix  = "arXiv",
      primaryClass   = "hep-ex",
      reportNumber   = "CERN-EP-2018-179",
      SLACcitation   = "%%CITATION = ARXIV:1808.02380;%%"
}

@article{Aad:2016nal,
      author         = "{ATLAS Collaboration}",
      title          = "{Test of CP invariance in vector-boson fusion production
                        of the Higgs boson using the Optimal Observable method in
                        the ditau decay channel with the ATLAS detector}",
      collaboration  = "ATLAS",
      journal        = "Eur. Phys. J. C",
      volume         = "76",
      year           = "2016",
      number         = "12",
      pages          = "658",
      doi            = "10.1140/epjc/s10052-016-4499-5",
      eprint         = "1602.04516",
      archivePrefix  = "arXiv",
      primaryClass   = "hep-ex",
      reportNumber   = "CERN-EP-2016-002",
      SLACcitation   = "%%CITATION = ARXIV:1602.04516;%%"
}

@article{Aaboud:2018xdt,
      author         = "{ATLAS Collaboration}",
      title          = "{Measurements of Higgs boson properties in the diphoton
                        decay channel with 36 fb$^{-1}$ of $pp$ collision data at
                        $\sqrt{s} = 13$ TeV with the ATLAS detector}",
      collaboration  = "ATLAS",
      journal        = "Phys. Rev. D",
      volume         = "98",
      year           = "2018",
      pages          = "052005",
      doi            = "10.1103/PhysRevD.98.052005",
      eprint         = "1802.04146",
      archivePrefix  = "arXiv",
      primaryClass   = "hep-ex",
      reportNumber   = "CERN-EP-2017-288",
      SLACcitation   = "%%CITATION = ARXIV:1802.04146;%%"
}

@article{Aaboud:2017vzb,
      author         = "{ATLAS Collaboration}",
      title          = "{Measurement of the Higgs boson coupling properties in
                        the $H\rightarrow ZZ^{*} \rightarrow 4\ell$ decay channel
                        at $\sqrt{s}$ = 13 TeV with the ATLAS detector}",
      collaboration  = "ATLAS",
      journal        = "JHEP",
      volume         = "03",
      year           = "2018",
      pages          = "095",
      doi            = "10.1007/JHEP03(2018)095",
      eprint         = "1712.02304",
      archivePrefix  = "arXiv",
      primaryClass   = "hep-ex",
      reportNumber   = "CERN-EP-2017-206",
      SLACcitation   = "%%CITATION = ARXIV:1712.02304;%%"
}

@techreport{ATL-PHYS-PUB-2019-004,
      author         = "{ATLAS Collaboration}",
      title         = "{Modelling of the vector boson scattering process
                       $pp\rightarrow W^\pm W^\pm jj$ in Monte Carlo generators in
                       ATLAS}",
      institution   = "CERN",
      collaboration = "ATLAS Collaboration",
      address       = "Geneva",
      number        = "ATL-PHYS-PUB-2019-004",
      month         = "1",
      year          = "2019",
      reportNumber  = "ATL-PHYS-PUB-2019-004",
      url           = "https://cds.cern.ch/record/2655303",
}

@techreport{ATLAS-CONF-2019-021,
      author         = "{ATLAS Collaboration}",
      title         = "{Luminosity determination in $pp$ collisions at
                       $\sqrt{s}=13$ TeV using the ATLAS detector at the LHC}",
      institution   = "CERN",
      collaboration = "ATLAS Collaboration",
      address       = "Geneva",
      number        = "ATLAS-CONF-2019-021",
      month         = "6",
      year          = "2019",
      reportNumber  = "ATLAS-CONF-2019-021",
      url = "https://cds.cern.ch/record/2677054",
}

@article{Sirunyan:2020gyx,
    author = "{CMS Collaboration}",
    collaboration = "CMS",
    title = "{Measurements of production cross sections of WZ and same-sign WW boson pairs in association with two jets in proton-proton collisions at $\sqrt{s}=$ 13 TeV}",
    eprint = "2005.01173",
    archivePrefix = "arXiv",
    primaryClass = "hep-ex",
    reportNumber = "FERMILAB-PUB-20-189-CMS, CMS-SMP-19-012, CERN-EP-2020-064",
    month = "5",
    year = "2020"
}

@article{Avoni:2018iuv,
    author = "{ATLAS Collaboration}",
    title = "{The new LUCID-2 detector for luminosity measurement and monitoring in ATLAS}",
    doi = "10.1088/1748-0221/13/07/P07017",
    journal = "JINST",
    volume = "13",
    number = "07",
    pages = "P07017",
    year = "2018"
}

@article{Tanabashi:2018oca,
    author = "Tanabashi, M. and others",
    collaboration = "Particle Data Group",
    title = "{Review of Particle Physics}",
    doi = "10.1103/PhysRevD.98.030001",
    journal = "Phys. Rev. D",
    volume = "98",
    number = "3",
    pages = "030001",
    year = "2018"
}

@article{Barlow:1990vc,
    author = "Barlow, Roger J.",
    title = "{Extended maximum likelihood}",
    reportNumber = "M-C-EXP-90-1",
    doi = "10.1016/0168-9002(90)91334-8",
    journal = "Nucl. Instrum. Meth. A",
    volume = "297",
    pages = "496--506",
    year = "1990"
}

@article{Aaboud:2016mmw,
    author = "{ATLAS Collaboration}",
    title = "{Measurement of the Inelastic Proton-Proton Cross Section at $\sqrt{s} = 13$  TeV with the ATLAS Detector at the LHC}",
    eprint = "1606.02625",
    archivePrefix = "arXiv",
    primaryClass = "hep-ex",
    reportNumber = "CERN-EP-2016-140",
    doi = "10.1103/PhysRevLett.117.182002",
    journal = "Phys. Rev. Lett.",
    volume = "117",
    number = "18",
    pages = "182002",
    year = "2016"
}

@article{Alwall:2014hca,
    author = "Alwall, J. and Frederix, R. and Frixione, S. and Hirschi, V. and Maltoni, F. and Mattelaer, O. and Shao, H. -S. and Stelzer, T. and Torrielli, P. and Zaro, M.",
    title = "{The automated computation of tree-level and next-to-leading order differential cross sections, and their matching to parton shower simulations}",
    eprint = "1405.0301",
    archivePrefix = "arXiv",
    primaryClass = "hep-ph",
    reportNumber = "CERN-PH-TH-2014-064, CP3-14-18, LPN14-066, MCNET-14-09, ZU-TH-14-14",
    doi = "10.1007/JHEP07(2014)079",
    journal = "JHEP",
    volume = "07",
    pages = "079",
    year = "2014"
}

@article{Aaboud:2017fye,
      author         = "{ATLAS Collaboration}",
      title          = "{Measurements of electroweak $Wjj$ production and
                        constraints on anomalous gauge couplings with the ATLAS
                        detector}",
      collaboration  = "ATLAS",
      journal        = "Eur. Phys. J. C",
      volume         = "77",
      year           = "2017",
      number         = "7",
      pages          = "474",
      doi            = "10.1140/epjc/s10052-017-5007-2",
      eprint         = "1703.04362",
      archivePrefix  = "arXiv",
      primaryClass   = "hep-ex",
      reportNumber   = "CERN-EP-2017-008",
      SLACcitation   = "%%CITATION = ARXIV:1703.04362;%%"
}

@article{Aad:2014dta,
      author         = "{ATLAS Collaboration}",
      title          = "{Measurement of the electroweak production of dijets in
                        association with a Z-boson and distributions sensitive to
                        vector boson fusion in proton-proton collisions at
                        $\sqrt{s} =$ 8 TeV using the ATLAS detector}",
      collaboration  = "ATLAS",
      journal        = "JHEP",
      volume         = "04",
      year           = "2014",
      pages          = "031",
      doi            = "10.1007/JHEP04(2014)031",
      eprint         = "1401.7610",
      archivePrefix  = "arXiv",
      primaryClass   = "hep-ex",
      reportNumber   = "CERN-PH-EP-2013-227",
      SLACcitation   = "%%CITATION = ARXIV:1401.7610;%%"
}

@article{Aaboud:2017emo,
      author         = "{ATLAS Collaboration}",
%      author         = "Aaboud, M. and others",
      title          = "{Measurement of the cross-section for electroweak
                        production of dijets in association with a Z boson in pp
                        collisions at $\sqrt {s}$ = 13 TeV with the ATLAS
                        detector}",
      collaboration  = "ATLAS",
      journal        = "Phys. Lett. B",
      volume         = "775",
      year           = "2017",
      pages          = "206-228",
      doi            = "10.1016/j.physletb.2017.10.040",
      eprint         = "1709.10264",
      archivePrefix  = "arXiv",
      primaryClass   = "hep-ex",
      reportNumber   = "CERN-EP-2017-115",
      SLACcitation   = "%%CITATION = ARXIV:1709.10264;%%"
}

@article{Khachatryan:2014dea,
      author         = "{CMS Collaboration}",
%      author         = "Khachatryan, Vardan and others",
      title          = "{Measurement of electroweak production of two jets in
                        association with a Z boson in proton-proton collisions at
                        $\sqrt{s}=8\,\text {TeV}$}",
      collaboration  = "CMS",
      journal        = "Eur. Phys. J. C",
      volume         = "75",
      year           = "2015",
      number         = "2",
      pages          = "66",
      doi            = "10.1140/epjc/s10052-014-3232-5",
      eprint         = "1410.3153",
      archivePrefix  = "arXiv",
      primaryClass   = "hep-ex",
      reportNumber   = "CMS-FSQ-12-035, CERN-PH-EP-2014-234",
      SLACcitation   = "%%CITATION = ARXIV:1410.3153;%%"
}

@article{Sirunyan:2017jej,
      author         = "{CMS Collaboration}",
%      author         = "Sirunyan, Albert M and others",
      title          = "{Electroweak production of two jets in association with a
                        Z boson in proton–proton collisions at $\sqrt{s}= $ 13
                        $\,\text {TeV}$}",
      collaboration  = "CMS",
      journal        = "Eur. Phys. J. C",
      volume         = "78",
      year           = "2018",
      number         = "7",
      pages          = "589",
      doi            = "10.1140/epjc/s10052-018-6049-9",
      eprint         = "1712.09814",
      archivePrefix  = "arXiv",
      primaryClass   = "hep-ex",
      reportNumber   = "CMS-SMP-16-018, CERN-EP-2017-328",
      SLACcitation   = "%%CITATION = ARXIV:1712.09814;%%"
}

@article{Hankele:2006ma,
      author         = "Hankele, V. and Klamke, G. and Zeppenfeld, D. and Figy,
                        T.",
      title          = "{Anomalous Higgs boson couplings in vector boson fusion
                        at the CERN LHC}",
      journal        = "Phys. Rev. D",
      volume         = "74",
      year           = "2006",
      pages          = "095001",
      doi            = "10.1103/PhysRevD.74.095001",
      eprint         = "hep-ph/0609075",
      archivePrefix  = "arXiv",
      primaryClass   = "hep-ph",
      reportNumber   = "KA-TP-09-2006",
      SLACcitation   = "%%CITATION = HEP-PH/0609075;%%"
}

@article{Aad:2016jkr,
      author         = "{ATLAS Collaboration}",
%      author         = "Aad, Georges and others",
      title          = "{Muon reconstruction performance of the ATLAS detector in
                        proton–proton collision data at $\sqrt{s}$ =13 TeV}",
      collaboration  = "ATLAS",
      journal        = "Eur. Phys. J. C",
      volume         = "76",
      year           = "2016",
      number         = "5",
      pages          = "292",
      doi            = "10.1140/epjc/s10052-016-4120-y",
      eprint         = "1603.05598",
      archivePrefix  = "arXiv",
      primaryClass   = "hep-ex",
      reportNumber   = "CERN-EP-2016-033",
      SLACcitation   = "%%CITATION = ARXIV:1603.05598;%%"
}

@article{Gleisberg:2008fv,
      author         = "Gleisberg, Tanju and Hoeche, Stefan",
      title          = "{Comix, a new matrix element generator}",
      journal        = "JHEP",
      volume         = "12",
      year           = "2008",
      pages          = "039",
      doi            = "10.1088/1126-6708/2008/12/039",
      eprint         = "0808.3674",
      archivePrefix  = "arXiv",
      primaryClass   = "hep-ph",
      reportNumber   = "SLAC-PUB-13232, IPPP-08-31, DCPT-08-62, MCNET-08-08",
      SLACcitation   = "%%CITATION = ARXIV:0808.3674;%%"
}

@article{Hoeche:2012yf,
      author         = "Hoeche, Stefan and others",
      title          = "{QCD matrix elements + parton showers: The NLO case}",
      journal        = "JHEP",
      volume         = "04",
      year           = "2013",
      pages          = "027",
      doi            = "10.1007/JHEP04(2013)027",
      eprint         = "1207.5030",
      archivePrefix  = "arXiv",
      primaryClass   = "hep-ph",
      reportNumber   = "SLAC-PUB-15191, IPPP-12-52, DCPT-12-104, LPN12-081,
                        FR-PHENO-2012-017, MCNET-12-09, --FR-PHENO-2012-017",
      SLACcitation   = "%%CITATION = ARXIV:1207.5030;%%"
}

@article{Ball:2012cx,
      author         = "Ball, Richard D. and others",
      title          = "{Parton distributions with LHC data}",
      journal        = "Nucl. Phys. B",
      volume         = "867",
      year           = "2013",
      pages          = "244-289",
      doi            = "10.1016/j.nuclphysb.2012.10.003",
      eprint         = "1207.1303",
      archivePrefix  = "arXiv",
      primaryClass   = "hep-ph",
      reportNumber   = "EDINBURGH-2012-08, IFUM-FT-997, FR-PHENO-2012-014,
                        RWTH-TTK-12-25, CERN-PH-TH-2012-037, SFB-CPP-12-47",
      SLACcitation   = "%%CITATION = ARXIV:1207.1303;%%"
}

@article{powheg1,
      author         = "Nason, Paolo",
      title          = "{A New method for combining NLO QCD with shower Monte
                        Carlo algorithms}",
      journal        = "JHEP",
      volume         = "11",
      year           = "2004",
      pages          = "040",
      doi            = "10.1088/1126-6708/2004/11/040",
      eprint         = "hep-ph/0409146",
      archivePrefix  = "arXiv",
      primaryClass   = "hep-ph",
      reportNumber   = "BICOCCA-FT-04-11",
      SLACcitation   = "%%CITATION = HEP-PH/0409146;%%"
}

@article{powheg2,
      author         = "Frixione, Stefano and Nason, Paolo and Oleari, Carlo",
      title          = "{Matching NLO QCD computations with Parton Shower
                        simulations: the POWHEG method}",
      journal        = "JHEP",
      volume         = "11",
      year           = "2007",
      pages          = "070",
      doi            = "10.1088/1126-6708/2007/11/070",
      eprint         = "0709.2092",
      archivePrefix  = "arXiv",
      primaryClass   = "hep-ph",
      reportNumber   = "BICOCCA-FT-07-9, GEF-TH-21-2007",
      SLACcitation   = "%%CITATION = ARXIV:0709.2092;%%"
}

@article{powheg3,
      author         = "Alioli, Simone and others",
      title          = "{A general framework for implementing NLO calculations in
                        shower Monte Carlo programs: the POWHEG BOX}",
      journal        = "JHEP",
      volume         = "06",
      year           = "2010",
      pages          = "043",
      doi            = "10.1007/JHEP06(2010)043",
      eprint         = "1002.2581",
      archivePrefix  = "arXiv",
      primaryClass   = "hep-ph",
      reportNumber   = "DESY-10-018, SFB-CPP-10-22, IPPP-10-11, DCPT-10-22",
      SLACcitation   = "%%CITATION = ARXIV:1002.2581;%%"
}

@article{Agostinelli2003250,
      author         = "Agostinelli, S. and others",
      title          = "{GEANT4: a simulation toolkit}",
      collaboration  = "GEANT4",
      journal        = "Nucl. Instrum. Meth. A",
      volume         = "506",
      year           = "2003",
      pages          = "250-303",
      doi            = "10.1016/S0168-9002(03)01368-8",
      reportNumber   = "SLAC-PUB-9350, FERMILAB-PUB-03-339",
      SLACcitation   = "%%CITATION = NUIMA,A506,250;%%"
}

@article{Aad:2010ah,
      author         = "{ATLAS Collaboration}",
%      author         = "Aad, G. and others",
      title          = "{The ATLAS Simulation Infrastructure}",
      collaboration  = "ATLAS",
      journal        = "Eur. Phys. J. C",
      volume         = "70",
      year           = "2010",
      pages          = "823-874",
      doi            = "10.1140/epjc/s10052-010-1429-9",
      eprint         = "1005.4568",
      archivePrefix  = "arXiv",
      primaryClass   = "physics.ins-det",
      SLACcitation   = "%%CITATION = ARXIV:1005.4568;%%"
}

@article{Sjostrand:2007gs,
      author         = "Sjostrand, Torbjorn and Mrenna, Stephen and Skands, Peter
                        Z.",
      title          = "{A brief introduction to PYTHIA 8.1}",
      journal        = "Comput. Phys. Commun.",
      volume         = "178",
      year           = "2008",
      pages          = "852-867",
      doi            = "10.1016/j.cpc.2008.01.036",
      eprint         = "0710.3820",
      archivePrefix  = "arXiv",
      primaryClass   = "hep-ph",
      reportNumber   = "CERN-LCGAPP-2007-04, LU-TP-07-28,
                        FERMILAB-PUB-07-512-CD-T",
      SLACcitation   = "%%CITATION = ARXIV:0710.3820;%%"
}

@techreport{ATL-PHYS-PUB-2014-021,
      author         = "{ATLAS Collaboration}",
      title         = "{ATLAS Pythia8 tunes to 7TeV data}",
      institution   = "CERN",
      address       = "Geneva",
      number        = "ATL-PHYS-PUB-2014-021",
      month         = "11",
      year          = "2014",
      reportNumber  = "ATL-PHYS-PUB-2014-021",
      url           = "https://cds.cern.ch/record/1966419",
}

@techreport{ATL-PHYS-PUB-2016-017,
      author         = "{ATLAS Collaboration}",
      title         = "{The Pythia 8 A3 tune description of ATLAS minimum bias and inelastic measurements incorporating the Donnachie-Landshoff diffractive mode}",
      institution   = "CERN",
      collaboration = "ATLAS Collaboration",
      address       = "Geneva",
      number        = "ATL-PHYS-PUB-2016-017",
      month         = "8",
      year          = "2016",
      reportNumber  = "ATL-PHYS-PUB-2016-017",
      url           = "https://cds.cern.ch/record/2206965",
}

@article{Aad:2014xaa,
      author         = "{ATLAS Collaboration}",
%      author         = "Aad, Georges and others",
      title          = "{Measurement of the $Z/\gamma^*$ boson transverse
                        momentum distribution in $pp$ collisions at $\sqrt{s}$ = 7
                        TeV with the ATLAS detector}",
      collaboration  = "ATLAS",
      journal        = "JHEP",
      volume         = "09",
      year           = "2014",
      pages          = "145",
      doi            = "10.1007/JHEP09(2014)145",
      eprint         = "1406.3660",
      archivePrefix  = "arXiv",
      primaryClass   = "hep-ex",
      reportNumber   = "CERN-PH-EP-2014-075",
      SLACcitation   = "%%CITATION = ARXIV:1406.3660;%%"
}

@article{Bothmann:2019yzt,
      author         = "Bothmann, Enrico and others",
      title          = "{Event generation with Sherpa 2.2}",
      journal        = "SciPost Phys.",
      volume         = "7",
      year           = "2019",
      number         = "3",
      pages          = "034",
      doi            = "10.21468/SciPostPhys.7.3.034",
      eprint         = "1905.09127",
      archivePrefix  = "arXiv",
      primaryClass   = "hep-ph",
      reportNumber   = "FERMILAB-PUB-19-218-T, SLAC-PUB-17433, IPPP/19/42,
                        MCNET-19-11",
      SLACcitation   = "%%CITATION = ARXIV:1905.09127;%%"
}

@article{Cascioli:2011va,
      author         = "Cascioli, Fabio and Maierhofer, Philipp and Pozzorini,
                        Stefano",
      title          = "{Scattering Amplitudes with Open Loops}",
      journal        = "Phys. Rev. Lett.",
      volume         = "108",
      year           = "2012",
      pages          = "111601",
      doi            = "10.1103/PhysRevLett.108.111601",
      eprint         = "1111.5206",
      archivePrefix  = "arXiv",
      primaryClass   = "hep-ph",
      reportNumber   = "ZU-TH-23-11, LPN11-66",
      SLACcitation   = "%%CITATION = ARXIV:1111.5206;%%"
}

@article{Denner:2016kdg,
      author         = "Denner, Ansgar and Dittmaier, Stefan and Hofer, Lars",
      title          = "{Collier: a fortran-based complex one-loop library in
                        extended regularizations}",
      journal        = "Comput. Phys. Commun.",
      volume         = "212",
      year           = "2017",
      pages          = "220-238",
      doi            = "10.1016/j.cpc.2016.10.013",
      eprint         = "1604.06792",
      archivePrefix  = "arXiv",
      primaryClass   = "hep-ph",
      reportNumber   = "FR-PHENO-2016-003, ICCUB-16-016",
      SLACcitation   = "%%CITATION = ARXIV:1604.06792;%%"
}

@article{Ball:2014uwa,
      author         = "Ball, Richard D. and others",
      title          = "{Parton distributions for the LHC run II}",
      collaboration  = "NNPDF",
      journal        = "JHEP",
      volume         = "04",
      year           = "2015",
      pages          = "040",
      doi            = "10.1007/JHEP04(2015)040",
      eprint         = "1410.8849",
      archivePrefix  = "arXiv",
      primaryClass   = "hep-ph",
      reportNumber   = "EDINBURGH-2014-15, IFUM-1034-FT, CERN-PH-TH-2013-253,
                        OUTP-14-11P, CAVENDISH-HEP-14-11",
      SLACcitation   = "%%CITATION = ARXIV:1410.8849;%%"
}

@article{Catani:2001cc,
      author         = "Catani, S. and Krauss, F. and Kuhn, R. and Webber, B. R.",
      title          = "{QCD Matrix Elements + Parton Showers}",
      journal        = "JHEP",
      volume         = "11",
      year           = "2001",
      pages          = "063",
      doi            = "10.1088/1126-6708/2001/11/063",
      eprint         = "hep-ph/0109231",
      archivePrefix  = "arXiv",
      primaryClass   = "hep-ph",
      reportNumber   = "CERN-TH-2000-367, CAVENDISH-HEP-00-03",
      SLACcitation   = "%%CITATION = HEP-PH/0109231;%%"
}

@article{Grzadkowski:2010es,
      author         = "Grzadkowski, B. and Iskrzynski, M. and Misiak, M. and
                        Rosiek, J.",
      title          = "{Dimension-six terms in the Standard Model Lagrangian}",
      journal        = "JHEP",
      volume         = "10",
      year           = "2010",
      pages          = "085",
      doi            = "10.1007/JHEP10(2010)085",
      eprint         = "1008.4884",
      archivePrefix  = "arXiv",
      primaryClass   = "hep-ph",
      reportNumber   = "IFT-9-2010, TTP10-35",
      SLACcitation   = "%%CITATION = ARXIV:1008.4884;%%"
}

@article{Brivio:2017btx,
      author         = "Brivio, Ilaria and Jiang, Yun and Trott, Michael",
      title          = "{The SMEFTsim package, theory and tools}",
      journal        = "JHEP",
      volume         = "12",
      year           = "2017",
      pages          = "070",
      doi            = "10.1007/JHEP12(2017)070",
      eprint         = "1709.06492",
      archivePrefix  = "arXiv",
      primaryClass   = "hep-ph",
      SLACcitation   = "%%CITATION = ARXIV:1709.06492;%%"
}

@article{Anastasiou:2003ds,
      author         = "Anastasiou, Charalampos and Dixon, Lance J. and Melnikov,
                        Kirill and Petriello, Frank",
      title          = "{High precision QCD at hadron colliders: Electroweak
                        gauge boson rapidity distributions at NNLO}",
      journal        = "Phys. Rev. D",
      volume         = "69",
      year           = "2004",
      pages          = "094008",
      doi            = "10.1103/PhysRevD.69.094008",
      eprint         = "hep-ph/0312266",
      archivePrefix  = "arXiv",
      primaryClass   = "hep-ph",
      reportNumber   = "SLAC-PUB-10288, UH-511-1042-03",
      SLACcitation   = "%%CITATION = HEP-PH/0312266;%%"
}

@article{Hoeche:2009rj,
      author         = "Hoeche, Stefan and Krauss, Frank and Schumann, Steffen
                        and Siegert, Frank",
      title          = "{QCD matrix elements and truncated showers}",
      journal        = "JHEP",
      volume         = "05",
      year           = "2009",
      pages          = "053",
      doi            = "10.1088/1126-6708/2009/05/053",
      eprint         = "0903.1219",
      archivePrefix  = "arXiv",
      primaryClass   = "hep-ph",
      reportNumber   = "ZU-TH-02-09, IPPP-09-14, DCPT-09-28, HD-THEP-09-2,
                        MCNET-09-04",
      SLACcitation   = "%%CITATION = ARXIV:0903.1219;%%"
}

@article{Lonnblad:2001iq,
      author         = "Lonnblad, Leif",
      title          = "{Correcting the Color Dipole Cascade Model with Fixed
                        Order Matrix Elements}",
      journal        = "JHEP",
      volume         = "05",
      year           = "2002",
      pages          = "046",
      doi            = "10.1088/1126-6708/2002/05/046",
      eprint         = "hep-ph/0112284",
      archivePrefix  = "arXiv",
      primaryClass   = "hep-ph",
      reportNumber   = "LU-TP-01-38",
      SLACcitation   = "%%CITATION = HEP-PH/0112284;%%"
}

@article{Lonnblad:2011xx,
      author         = "Lonnblad, Leif and Prestel, Stefan",
      title          = "{Matching tree-level matrix elements with interleaved
                        showers}",
      journal        = "JHEP",
      volume         = "03",
      year           = "2012",
      pages          = "019",
      doi            = "10.1007/JHEP03(2012)019",
      eprint         = "1109.4829",
      archivePrefix  = "arXiv",
      primaryClass   = "hep-ph",
      SLACcitation   = "%%CITATION = ARXIV:1109.4829;%%"
}

@article{LANGE2001152, author = "Lange, D.J.", editor = "Erhan, S. and Schlein, P. and Rozen, Y.", title = "{The EvtGen particle decay simulation package}", doi = "10.1016/S0168-9002(01)00089-4", journal = "Nucl. Instrum. Meth. A", volume = "462", pages = "152--155", year = "2001" }

@article{DAGOSTINI1995487, author = "D Agostini, G.", title = "{A Multidimensional unfolding method based on Bayes  theorem}", reportNumber = "DESY-94-099", doi = "10.1016/0168-9002(95)00274-X", journal = "Nucl. Instrum. Meth. A", volume = "362", pages = "487--498", year = "1995" }

@article{Aaboud:2019ynx,
      author         = "{ATLAS Collaboration}",
%      author         = "Aaboud, Morad and others",
      title          = "{Electron reconstruction and identification in the ATLAS
                        experiment using the 2015 and 2016 LHC proton-proton
                        collision data at $\sqrt{s}$ = 13 TeV}",
      collaboration  = "ATLAS",
      journal        = "Eur. Phys. J. C",
      volume         = "79",
      year           = "2019",
      number         = "8",
      pages          = "639",
      doi            = "10.1140/epjc/s10052-019-7140-6",
      eprint         = "1902.04655",
      archivePrefix  = "arXiv",
      primaryClass   = "physics.ins-det",
      reportNumber   = "CERN-EP-2018-273",
      SLACcitation   = "%%CITATION = ARXIV:1902.04655;%%"
}

@article{Cacciari:2008gp,
      author         = "Cacciari, Matteo and Salam, Gavin P. and Soyez, Gregory",
      title          = "{The anti-$k_t$ jet clustering algorithm}",
      journal        = "JHEP",
      volume         = "04",
      year           = "2008",
      pages          = "063",
      doi            = "10.1088/1126-6708/2008/04/063",
      eprint         = "0802.1189",
      archivePrefix  = "arXiv",
      primaryClass   = "hep-ph",
      reportNumber   = "LPTHE-07-03",
      SLACcitation   = "%%CITATION = ARXIV:0802.1189;%%"
}

@article{Cacciari:2011ma,
      author         = "Cacciari, Matteo and Salam, Gavin P. and Soyez, Gregory",
      title          = "{FastJet user manual}",
      journal        = "Eur. Phys. J. C",
      volume         = "72",
      year           = "2012",
      pages          = "1896",
      doi            = "10.1140/epjc/s10052-012-1896-2",
      eprint         = "1111.6097",
      archivePrefix  = "arXiv",
      primaryClass   = "hep-ph",
      reportNumber   = "CERN-PH-TH-2011-297",
      SLACcitation   = "%%CITATION = ARXIV:1111.6097;%%"
}

@article{Falkowski:2016cxu,
      author         = "Falkowski, Adam and Gonzalez-Alonso, Martin and Greljo,
                        Admir and Marzocca, David and Son, Minho",
      title          = "{Anomalous triple gauge couplings in the effective field
                        theory approach at the LHC}",
      journal        = "JHEP",
      volume         = "02",
      year           = "2017",
      pages          = "115",
      doi            = "10.1007/JHEP02(2017)115",
      eprint         = "1609.06312",
      archivePrefix  = "arXiv",
      primaryClass   = "hep-ph",
      reportNumber   = "ZU-TH-34-16",
      SLACcitation   = "%%CITATION = ARXIV:1609.06312;%%"
}

@misc{Aad:2020ddw,
    author         = "{ATLAS Collaboration}",
%    author = "Aad, Georges and others",
    archivePrefix = "arXiv",
    collaboration = "ATLAS",
    eprint = "2004.14636",
    month = "4",
    primaryClass = "hep-ex",
    reportNumber = "CERN-EP-2020-049",
    title = "{Search for heavy diboson resonances in semileptonic final states in $pp$ collisions at $\sqrt{s}=13$ TeV with the ATLAS detector}",
    %year = "2020"
}

@article{Khachatryan:2016vau,
    author         = "{ATLAS and CMS Collaborations}",
%    author = "Aad, Georges and others",
    archivePrefix = "arXiv",
    collaboration = "ATLAS and CMS",
    doi = "10.1007/JHEP08(2016)045",
    eprint = "1606.02266",
    journal = "JHEP",
    pages = "045",
    primaryClass = "hep-ex",
    reportNumber = "CERN-EP-2016-100, ATLAS-HIGG-2015-07, CMS-HIG-15-002",
    title = "{Measurements of the Higgs boson production and decay rates and constraints on its couplings from a combined ATLAS and CMS analysis of the LHC pp collision data at $ \sqrt{s}=7 $ and 8 TeV}",
    volume = "08",
    year = "2016"
}

@article{Sirunyan:2018owy,
    author         = "{CMS Collaboration}",
%    author = "Sirunyan, Albert M and others",
    archivePrefix = "arXiv",
    collaboration = "CMS",
    doi = "10.1016/j.physletb.2019.04.025",
    eprint = "1809.05937",
    journal = "Phys. Lett. B",
    pages = "520--551",
    primaryClass = "hep-ex",
    reportNumber = "CMS-HIG-17-023, CERN-EP-2018-139",
    title = "{Search for invisible decays of a Higgs boson produced through vector boson fusion in proton-proton collisions at $\sqrt{s} =$ 13 TeV}",
    volume = "793",
    year = "2019"
}

@article{Sirunyan:2018koj,
    author         = "{CMS Collaboration}",
%    author = "Sirunyan, Albert M and others",
    archivePrefix = "arXiv",
    collaboration = "CMS",
    doi = "10.1140/epjc/s10052-019-6909-y",
    eprint = "1809.10733",
    journal = "Eur. Phys. J. C",
    number = "5",
    pages = "421",
    primaryClass = "hep-ex",
    reportNumber = "CMS-HIG-17-031, CERN-EP-2018-263",
    title = "{Combined measurements of Higgs boson couplings in proton--proton collisions at $\sqrt{s}=13\,\text {Te}\text {V} $}",
    volume = "79",
    year = "2019"
}

@article{Sirunyan:2019twz,
%    author = "Sirunyan, Albert M and others",
    author         = "{CMS Collaboration}",
    archivePrefix = "arXiv",
    collaboration = "CMS",
    doi = "10.1103/PhysRevD.99.112003",
    eprint = "1901.00174",
    journal = "Phys. Rev. D",
    number = "11",
    pages = "112003",
    primaryClass = "hep-ex",
    reportNumber = "CMS-HIG-18-002, CERN-EP-2018-329",
    title = "{Measurements of the Higgs boson width and anomalous $HVV$ couplings from on-shell and off-shell production in the four-lepton final state}",
    volume = "99",
    year = "2019"
}

@article{Sirunyan:2019nbs,
    %author = "Sirunyan, Albert M and others",
    author         = "{CMS Collaboration}",
    archivePrefix = "arXiv",
    collaboration = "CMS",
    doi = "10.1103/PhysRevD.100.112002",
    eprint = "1903.06973",
    journal = "Phys. Rev. D",
    number = "11",
    pages = "112002",
    primaryClass = "hep-ex",
    reportNumber = "CMS-HIG-17-034, CERN-EP-2019-029",
    title = "{Constraints on anomalous $HVV$ couplings from the production of Higgs bosons decaying to $\tau$ lepton pairs}",
    volume = "100",
    year = "2019"
}

@misc{Aad:2020zbq,
    author         = "{ATLAS Collaboration}",
%    author = "Aad, Georges and others",
    archivePrefix = "arXiv",
    collaboration = "ATLAS",
    eprint = "2004.10612",
    month = "4",
    primaryClass = "hep-ex",
    reportNumber = "CERN-EP-2020-016",
    title = "{Observation of electroweak production of two jets and a $Z$-boson pair with the ATLAS detector at the LHC}",
    %year = "2020"
}

@article{Sirunyan:2017ret,
    author         = "{CMS Collaboration}",
    %author = "Sirunyan, Albert M and others",
    archivePrefix = "arXiv",
    collaboration = "CMS",
    doi = "10.1103/PhysRevLett.120.081801",
    eprint = "1709.05822",
    journal = "Phys. Rev. Lett.",
    number = "8",
    pages = "081801",
    primaryClass = "hep-ex",
    reportNumber = "CMS-SMP-17-004, CERN-EP-2017-232",
    title = "{Observation of Electroweak Production of Same-Sign W boson Pairs in the Two Jet and Two Same-Sign Lepton Final State in Proton-Proton Collisions at $\sqrt{s} = $ 13 TeV}",
    volume = "120",
    year = "2018"
}

@article{Maguire:2017ypu,
      author         = "Maguire, Eamonn and Heinrich, Lukas and Watt, Graeme",
      title          = "{HEPData: a repository for high energy physics data}",
      booktitle      = "{Proceedings, 22nd International Conference on Computing
                        in High Energy and Nuclear Physics (CHEP2016): San
                        Francisco, CA, October 14-16, 2016}",
      journal        = "J. Phys. Conf. Ser.",
      volume         = "898",
      year           = "2017",
      number         = "10",
      pages          = "102006",
      doi            = "10.1088/1742-6596/898/10/102006",
      eprint         = "1704.05473",
      archivePrefix  = "arXiv",
      primaryClass   = "hep-ex",
      reportNumber   = "IPPP-17-31",
      SLACcitation   = "%%CITATION = ARXIV:1704.05473;%%"
}

@article{Bierlich:2019rhm,
      author         = "Bierlich, C. and others",
      title          = "{Robust Independent Validation of Experiment and Theory:
                        Rivet version 3}",
      journal        = "SciPost Phys.",
      volume         = "8",
      year           = "2020",
      pages          = "026",
      doi            = "10.21468/SciPostPhys.8.2.026",
      eprint         = "1912.05451",
      archivePrefix  = "arXiv",
      primaryClass   = "hep-ph",
      reportNumber   = "MCnet-19-26",
      SLACcitation   = "%%CITATION = ARXIV:1912.05451;%%"
}

@article{Buckley:2010ar,
      author         = "Buckley, Andy and Butterworth, Jonathan and Lonnblad,
                        Leif and Grellscheid, David and Hoeth, Hendrik and Monk,
                        James and Schulz, Holger and Siegert, Frank",
      title          = "{Rivet user manual}",
      journal        = "Comput. Phys. Commun.",
      volume         = "184",
      year           = "2013",
      pages          = "2803-2819",
      doi            = "10.1016/j.cpc.2013.05.021",
      eprint         = "1003.0694",
      archivePrefix  = "arXiv",
      primaryClass   = "hep-ph",
      reportNumber   = "MCNET-10-03",
      SLACcitation   = "%%CITATION = ARXIV:1003.0694;%%"
}

@article{PhysRevD.39.274,
  title = {Application of the bootstrap statistical method to the tau-decay-mode problem},
  author = {Hayes, Kenneth G. and Perl, Martin L. and Efron, Bradley},
  journal = {Phys. Rev. D},
  volume = {39},
  issue = {1},
  pages = {274--279},
  numpages = {0},
  year = {1989},
  publisher = {American Physical Society},
  doi = {10.1103/PhysRevD.39.274},
}

@article{Azatov:2016sqh,
      author         = "Azatov, Aleksandr and Contino, Roberto and Machado,
                        Camila S. and Riva, Francesco",
      title          = "{Helicity selection rules and noninterference for BSM
                        amplitudes}",
      journal        = "Phys. Rev. D",
      volume         = "95",
      year           = "2017",
      number         = "6",
      pages          = "065014",
      doi            = "10.1103/PhysRevD.95.065014",
      eprint         = "1607.05236",
      archivePrefix  = "arXiv",
      primaryClass   = "hep-ph",
      reportNumber   = "CERN-TH-2016-165",
      SLACcitation   = "%%CITATION = ARXIV:1607.05236;%%"
}

@article{Sirunyan:2019gkh,
      author         = "{CMS Collaboration}",
%      author         = "Sirunyan, Albert M and others",
      title          = "{Search for anomalous triple gauge couplings in WW and WZ
                        production in lepton+jet events in proton-proton
                        collisions at $\sqrt{s} =$ 13 TeV}",
      collaboration  = "CMS",
      journal        = "JHEP",
      volume         = "12",
      year           = "2019",
      pages          = "062",
      doi            = "10.1007/JHEP12(2019)062",
      eprint         = "1907.08354",
      archivePrefix  = "arXiv",
      primaryClass   = "hep-ex",
      reportNumber   = "CMS-SMP-18-008, CERN-EP-2019-137",
      SLACcitation   = "%%CITATION = ARXIV:1907.08354;%%"
}

@article{Baglio:2018bkm,
      author         = "Baglio, Julien and Dawson, Sally and Lewis, Ian M.",
      title          = "{NLO effects in EFT fits to $W^+W^-$ production at the
                        LHC}",
      journal        = "Phys. Rev. D",
      volume         = "99",
      year           = "2019",
      number         = "3",
      pages          = "035029",
      doi            = "10.1103/PhysRevD.99.035029",
      eprint         = "1812.00214",
      archivePrefix  = "arXiv",
      primaryClass   = "hep-ph",
      SLACcitation   = "%%CITATION = ARXIV:1812.00214;%%"
}

@article{Hagiwara:1993ck,
      author         = "Hagiwara, Kaoru and Ishihara, S. and Szalapski, R. and
                        Zeppenfeld, D.",
      title          = "{Low-energy effects of new interactions in the
                        electroweak boson sector}",
      journal        = "Phys. Rev. D",
      volume         = "48",
      year           = "1993",
      pages          = "2182-2203",
      doi            = "10.1103/PhysRevD.48.2182",
      reportNumber   = "MAD-PH-737, UT-635, KEK-TH-356, KEK-PREPRINT-92-214",
      SLACcitation   = "%%CITATION = PHRVA,D48,2182;%%"
}

@article{Degrande:2012wf,
      author         = "Degrande, Celine and Greiner, Nicolas and Kilian,
                        Wolfgang and Mattelaer, Olivier and Mebane, Harrison and
                        Stelzer, Tim and Willenbrock, Scott and Zhang, Cen",
      title          = "{Effective field theory: A modern approach to anomalous
                        couplings}",
      journal        = "Annals Phys.",
      volume         = "335",
      year           = "2013",
      pages          = "21-32",
      doi            = "10.1016/j.aop.2013.04.016",
      eprint         = "1205.4231",
      archivePrefix  = "arXiv",
      primaryClass   = "hep-ph",
      reportNumber   = "MPP-2011-149, SI-HEP-2011-17, CP3-12-25",
      SLACcitation   = "%%CITATION = ARXIV:1205.4231;%%"
}

@article{Feldman:1997qc,
      author         = "Feldman, Gary J. and Cousins, Robert D.",
      title          = "{A Unified approach to the classical statistical analysis
                        of small signals}",
      journal        = "Phys. Rev. D",
      volume         = "57",
      year           = "1998",
      pages          = "3873-3889",
      doi            = "10.1103/PhysRevD.57.3873",
      eprint         = "physics/9711021",
      archivePrefix  = "arXiv",
      primaryClass   = "physics.data-an",
      reportNumber   = "HUTP-97-A096",
      SLACcitation   = "%%CITATION = PHYSICS/9711021;%%"
}

@article{Aaboud:2019nkz,
      author         = "{ATLAS Collaboration}",
%      author         = "Aaboud, Morad and others",
      title          = "{Measurement of fiducial and differential $W^+W^-$
                        production cross-sections at $\sqrt{s}=13$  TeV with the
                        ATLAS detector}",
      collaboration  = "ATLAS",
      journal        = "Eur. Phys. J. C",
      volume         = "79",
      year           = "2019",
      number         = "10",
      pages          = "884",
      doi            = "10.1140/epjc/s10052-019-7371-6",
      eprint         = "1905.04242",
      archivePrefix  = "arXiv",
      primaryClass   = "hep-ex",
      reportNumber   = "CERN-EP-2019-055",
      SLACcitation   = "%%CITATION = ARXIV:1905.04242;%%"
}

@article{Aaboud:2017jcu,
      author         = "{ATLAS Collaboration}",
%      author         = "Aaboud, M. and others",
      title          = "{Jet energy scale measurements and their systematic
                        uncertainties in proton-proton collisions at $\sqrt{s} =
                        13$ TeV with the ATLAS detector}",
      collaboration  = "ATLAS",
      journal        = "Phys. Rev. D",
      volume         = "96",
      year           = "2017",
      number         = "7",
      pages          = "072002",
      doi            = "10.1103/PhysRevD.96.072002",
      eprint         = "1703.09665",
      archivePrefix  = "arXiv",
      primaryClass   = "hep-ex",
      reportNumber   = "CERN-EP-2017-038",
      SLACcitation   = "%%CITATION = ARXIV:1703.09665;%%"
}

@article{Aad:2015ina,
      author         = "{ATLAS Collaboration}",
%      author         = "Aad, Georges and others",
      title          = "{Performance of pile-up mitigation techniques for jets in
                        $pp$ collisions at $\sqrt{s}=8$  TeV using the ATLAS
                        detector}",
      collaboration  = "ATLAS",
      journal        = "Eur. Phys. J. C",
      volume         = "76",
      year           = "2016",
      number         = "11",
      pages          = "581",
      doi            = "10.1140/epjc/s10052-016-4395-z",
      eprint         = "1510.03823",
      archivePrefix  = "arXiv",
      primaryClass   = "hep-ex",
      reportNumber   = "CERN-PH-EP-2015-206",
      SLACcitation   = "%%CITATION = ARXIV:1510.03823;%%"
}

@inproceedings{Adye:2011gm, author = "Adye, Tim", title = "{Unfolding algorithms and tests using RooUnfold}", booktitle = "{PHYSTAT 2011}", eprint = "1105.1160", archivePrefix = "arXiv", primaryClass = "physics.data-an", doi = "10.5170/CERN-2011-006.313", publisher = "CERN", address = "Geneva", pages = "313--318", year = "2011" }

@article{Aaboud:2018ugz,
      author         = "{ATLAS Collaboration}",
%      author         = "Aaboud, Morad and others",
      title          = "{Electron and photon energy calibration with the ATLAS
                        detector using 2015–2016 LHC proton-proton collision
                        data}",
      collaboration  = "ATLAS",
      journal        = "JINST",
      volume         = "14",
      year           = "2019",
      number         = "03",
      pages          = "P03017",
      doi            = "10.1088/1748-0221/14/03/P03017",
      eprint         = "1812.03848",
      archivePrefix  = "arXiv",
      primaryClass   = "hep-ex",
      reportNumber   = "CERN-EP-2018-296",
      SLACcitation   = "%%CITATION = ARXIV:1812.03848;%%"
}

@article{Bahr:2008pv,
      author         = "Bahr, M. and others",
      title          = "{Herwig++ physics and manual}",
      journal        = "Eur. Phys. J. C",
      volume         = "58",
      year           = "2008",
      pages          = "639-707",
      doi            = "10.1140/epjc/s10052-008-0798-9",
      eprint         = "0803.0883",
      archivePrefix  = "arXiv",
      primaryClass   = "hep-ph",
      reportNumber   = "CERN-PH-TH-2008-038, CAVENDISH-HEP-08-03, KA-TP-05-2008,
                        DCPT-08-22, IPPP-08-11, CP3-08-05",
      SLACcitation   = "%%CITATION = ARXIV:0803.0883;%%"
}

@article{Frederix:2012ps,
      author         = "Frederix, Rikkert and Frixione, Stefano",
      title          = "{Merging meets matching in MC@NLO}",
      journal        = "JHEP",
      volume         = "12",
      year           = "2012",
      pages          = "061",
      doi            = "10.1007/JHEP12(2012)061",
      eprint         = "1209.6215",
      archivePrefix  = "arXiv",
      primaryClass   = "hep-ph",
      reportNumber   = "CERN-PH-TH-2012-247, ZU-TH-21-12",
      SLACcitation   = "%%CITATION = ARXIV:1209.6215;%%"
}

@article{Arnold:2011wj,
      author         = "Baglio, J. and others",
      title          = "{VBFNLO: A parton level Monte Carlo for processes with
                        electroweak bosons -- Manual for Version 2.7.0}",
      year           = "2011",
      eprint         = "1107.4038",
      archivePrefix  = "arXiv",
      primaryClass   = "hep-ph",
      reportNumber   = "KA-TP-15-2011, FTUV-11-0730, WUB-11-06,
                        FR-PHENO-2011-012, MZ-TH-11-19, CERN-PH-TH-2011-173,
                        LPN11-40, DESY-11-125, TTK-10-51, MAN-HEP-2014-02",
      SLACcitation   = "%%CITATION = ARXIV:1107.4038;%%"
}

@article{Harland-Lang:2014zoa,
      author         = "Harland-Lang, L. A. and Martin, A. D. and Motylinski, P.
                        and Thorne, R. S.",
      title          = "{Parton distributions in the LHC era: MMHT 2014 PDFs}",
      journal        = "Eur. Phys. J. C",
      volume         = "75",
      year           = "2015",
      number         = "5",
      pages          = "204",
      doi            = "10.1140/epjc/s10052-015-3397-6",
      eprint         = "1412.3989",
      archivePrefix  = "arXiv",
      primaryClass   = "hep-ph",
      reportNumber   = "LCTS-2014-47, IPPP-14-97, DCPT-14-194",
      SLACcitation   = "%%CITATION = ARXIV:1412.3989;%%"
}

@article{Bellm:2015jjp,
      author         = "Bellm, Johannes and others",
      title          = "{Herwig 7.0/Herwig++ 3.0 release note}",
      journal        = "Eur. Phys. J. C",
      volume         = "76",
      year           = "2016",
      number         = "4",
      pages          = "196",
      doi            = "10.1140/epjc/s10052-016-4018-8",
      eprint         = "1512.01178",
      archivePrefix  = "arXiv",
      primaryClass   = "hep-ph",
      reportNumber   = "CERN-PH-TH-2015-289, MAN-HEP-2015-15, IFJPAN-IV-2015-13,
                        KA-TP-18-2015, DCPT-15-142, MCNET-15-28, IPPP-15-71,
                        HERWIG-2015-01",
      SLACcitation   = "%%CITATION = ARXIV:1512.01178;%%"
}

@Article{PERF-2007-01,
    author         = "{ATLAS Collaboration}",
    title          = "{The ATLAS Experiment at the CERN Large Hadron Collider}",
    journal        = "JINST",
    volume         = "3",
    year           = "2008",
    pages          = "S08003",
    doi            = "10.1088/1748-0221/3/08/S08003",
    primaryClass   = "hep-ex",
}

@Article{TRIG-2016-01,
    author         = "{ATLAS Collaboration}",
    title          = "{Performance of the ATLAS trigger system in 2015}",
    journal        = "Eur. Phys. J. C",
    volume         = "77",
    year           = "2017",
    pages          = "317",
    doi            = "10.1140/epjc/s10052-017-4852-3",
    reportNumber   = "CERN-EP-2016-241",
    eprint         = "1611.09661",
    archivePrefix  = "arXiv",
    primaryClass   = "hep-ex",
}

@Article{PIX-2018-001,
    author         = "Abbott, B. and others",
    title          = "{Production and integration of the ATLAS Insertable B-Layer}",
    journal        = "JINST",
    volume         = "13",
    year           = "2018",
    pages          = "T05008",
    doi            = "10.1088/1748-0221/13/05/T05008",
    eprint         = "1803.00844",
    archivePrefix  = "arXiv",
    primaryClass   = "physics.ins-det",
}

@Booklet{ATL-SOFT-PUB-2020-001,
    author         = "{ATLAS Collaboration}",
    title          = "{ATLAS Computing Acknowledgements}",
    howpublished   = "{ATL-SOFT-PUB-2020-001}",
    url            = "https://cds.cern.ch/record/2717821",
}

@Booklet{ATLAS-TDR-2010-19,
    author         = "{ATLAS Collaboration}",
    title          = "{ATLAS Insertable B-Layer Technical Design Report}",
    howpublished   = "ATLAS-TDR-19",
    url            = "https://cds.cern.ch/record/1291633",
    year           = "2010",
    related        = "ATLAS-TDR-2010-19-add",
    relatedstring  = "Addendum:",
}

\clearpage 
 
\begin{flushleft}
\hypersetup{urlcolor=black}
{\Large The ATLAS Collaboration}

\bigskip

\AtlasOrcid[0000-0002-6665-4934]{G.~Aad}$^\textrm{\scriptsize 102}$,    
\AtlasOrcid[0000-0002-5888-2734]{B.~Abbott}$^\textrm{\scriptsize 128}$,    
\AtlasOrcid{D.C.~Abbott}$^\textrm{\scriptsize 103}$,    
\AtlasOrcid[0000-0002-2788-3822]{A.~Abed~Abud}$^\textrm{\scriptsize 36}$,    
\AtlasOrcid[0000-0002-1002-1652]{K.~Abeling}$^\textrm{\scriptsize 53}$,    
\AtlasOrcid[0000-0002-2987-4006]{D.K.~Abhayasinghe}$^\textrm{\scriptsize 94}$,    
\AtlasOrcid[0000-0002-8496-9294]{S.H.~Abidi}$^\textrm{\scriptsize 167}$,    
\AtlasOrcid[0000-0002-8279-9324]{O.S.~AbouZeid}$^\textrm{\scriptsize 40}$,    
\AtlasOrcid{N.L.~Abraham}$^\textrm{\scriptsize 156}$,    
\AtlasOrcid[0000-0001-5329-6640]{H.~Abramowicz}$^\textrm{\scriptsize 161}$
% ...
\bigskip
\\

% ...
$^{36}$CERN, Geneva; Switzerland.\\
% ...
$^{40}$Niels Bohr Institute, University of Copenhagen, Copenhagen; Denmark.\\
% ...
$^{53}$II. Physikalisches Institut, Georg-August-Universit\"{a}t G\"ottingen, G\"ottingen; Germany.\\
% ...
$^{94}$Department of Physics, Royal Holloway University of London, Egham; United Kingdom.\\
% ...
$^{102}$CPPM, Aix-Marseille Universit\'e, CNRS/IN2P3, Marseille; France.\\
$^{103}$Department of Physics, University of Massachusetts, Amherst MA; United States of America.\\
% ...
$^{128}$Homer L. Dodge Department of Physics and Astronomy, University of Oklahoma, Norman OK; United States of America.\\
% ...
$^{156}$Department of Physics and Astronomy, University of Sussex, Brighton; United Kingdom.\\
% ...
$^{161}$Raymond and Beverly Sackler School of Physics and Astronomy, Tel Aviv University, Tel Aviv; Israel.\\
% ...
$^{167}$Department of Physics, University of Toronto, Toronto ON; Canada.\\
% ...

\end{flushleft}


\end{document}